\documentclass[11pt,journal,a4paper,onecolumn,draftclsnofoot]{IEEEtran}
\usepackage[ruled,vlined]{algorithm2e}
\usepackage{amsfonts}
\usepackage{amssymb}

\usepackage{cancel}
\usepackage{ClearSans}

\usepackage{epsfig}

\usepackage{glossaries}
\usepackage{mathrsfs}
\usepackage{mathtools}

\usepackage{scalerel}
\usepackage{subcaption}

\usepackage{tikz,pgfplots}
\pgfplotsset{compat=newest}

\usepackage[normalem]{ulem}

\usepackage{cite}

\usepackage{enumitem}
\usepackage[draft]{hyperref}

\usepackage{dsfont}
\newcommand{\Ber}{\mathop{\mathrm{Ber}}}

\newcommand{\field}[1]{\mathbb{#1}}

\newcommand{\indicator}{\mathds{1}} 

\newcommand{\Exp}[2][]{\ensuremath{\mathbb{E}_{#1}\left[#2 \right]}} 

\newcommand{\eqann}[2]{\overset{\mathclap{(\text{#2})}}{#1}} 
\newcommand{\eqannref}[1]{$(\text{#1})$}
\newcommand{\bv}[1]{\mathbf{#1}} 
\newcommand{\rv}[1]{\mathsf{#1}} 
\newcommand{\Prob}[1]{\ensuremath{\mathsf{P} \left( #1 \right)}} 

\newcommand{\mat}[1]{{#1}}

\newcommand{\minus}{\scalebox{0.75}[1.0]{\( - \)}}

\newcommand{\norm}[1]{\left\lVert #1 \right \rVert }

\newcommand{\pmf}[1]{\ensuremath{\mathsf{P}_{#1}}}

\newcommand{\simplex}[1]{\Delta_{#1}}

\newcommand{\set}[1]{\mathcal{#1}} 

 \DeclareMathOperator*{\argmax}{arg\,max} \DeclareMathOperator*{\argmin}{arg\,min}

\newtheorem{theorem}{Theorem}

\newtheorem{remark}{Remark}

\newtheorem{lemma}{Lemma}

\newenvironment{proof}[1][Proof]{\noindent\textbf{#1.} }{\ \rule{0.5em}{0.5em}}

\newtheorem{proposition}{Proposition}

\newacronym{awgn}{AWGN}{Additive White Gaussian Noise}

\newacronym{bac}{BAC}{Binary Asymmetric Channel}

\newacronym{bdsib}{BDSIB}{Binary Double-Sided Information-Bottleneck}
\newacronym{bec}{BEC}{\textit{binary erasure channel}}

\newacronym{bms}{BMS}{Binary Memoryless Symmetric}

\newacronym{bs}{BS}{base station}

\newacronym{bsc}{BSC}{\textit{binary symmetric channel}}
\newacronym{bscs}{BSCs}{\textit{binary symmetric channels}}

\newacronym{ceb}{CEB}{\textit{conditional entropy bound}}

\newacronym{comib}{COMIB}{\textit{compound information bottleneck}}

\newacronym{cr}{CR}{\textit{common reconstruction}}

\newacronym{cran}{C-RAN}{cloud radio access network}

\newacronym{cp}{CP}{Central Proccesor}

\newacronym{dmc}{DMC}{\textit{discrete memoryless channel}}

\newacronym{dmmac}{DM-MAC}{Discrete Memoryless Multiple Access Channel}

\newacronym{dms}{DMS}{Discrete Memoryless Source}

\newacronym{dpi}{DPI}{Data Proccesing Inequality}

\newacronym{dsbs}{DSBS}{\textit{doubly symmetric binary source}}

\newacronym{dsib}{DSIB}{Double-Sided Information-Bottleneck}

\newacronym{epi}{EPI}{Entropy Power Inequality}

\newacronym{gdsib}{GDSIB}{Gaussian Double-Sided Information Bottleneck}

\newacronym{gp}{GP}{Gelf'and-Pinsker}

\newacronym{ib}{IB}{\textit{information bottleneck}}
\newacronym{iid}{i.i.d.}{independent and identically distributed}

\newacronym{infcomb}{IC}{\textit{information combining}}

\newacronym{kld}{KLD}{\textit{Kullback–Leibler divergence}}

\newacronym{lhs}{LHS}{Left Hand Side}

\newacronym{sgdsib}{SGDSIB}{Scalar Gaussian Double-Sided Information-Bottleneck}

\newacronym{mac}{MAC}{multiple access channel}

\newacronym{mgl}{MGL}{Mrs. Gerber's lemma}

\newacronym{mi}{MI}{mutual information}

\newacronym{mmse}{MMSE}{minimum mean squared error}

\newacronym{mu}{MU}{Mobile User}

\newacronym{oblib}{OBLIB}{Oblivious Information Bottleneck}

\newacronym{pf}{PF}{\textit{Privacy Funnel}}

\newacronym{pmf}{PMF}{probability mass function}

\newacronym{ssib}{SSIB}{Single-Sided Information Bottleneck}
\newacronym{sgssib}{SGSSIB}{Scalar Gaussian Single-Sided Information Bottleneck function}
\newacronym{sawgnssib}{SAWGNSSIB}{Scalar AWGN Single-Sided Information Bottleneck function}

\newacronym{snr}{SNR}{Signal to Noise Ratio}

\newacronym{svd}{SVD}{Singular Value Decomposition}

\newacronym{tibo}{TIBO}{Ternary-Input Binary-Output }

\newacronym{tito}{TITO}{Ternary-Input Ternary-Output }

\newacronym{qiqo}{QIQO}{Quaternary-Input Quaternary-Output }

\newacronym{rhs}{RHS}{right hand side}

\newacronym{rv}{RV}{Random Variable}

\newacronym{rssib}{RSSIB}{Reversed Single-Sided Information Bottleneck function}
\newacronym{rsawgnssib}{RSAWGNSSIB}{Reversed Scalar AWGN Single-Sided Information Bottleneck function}

\newacronym{tv}{TV}{\textit{Total Variation}}

\newacronym{wlog}{WLOG}{long}

\newacronym{wtc}{WTC}{Wiretap Channel}
\usepackage{setspace}
\PassOptionsToPackage{normalem}{ulem}
\usepackage{ulem}

\title{The Compound Information Bottleneck Outlook}

\author{
	\IEEEauthorblockN{Michael Dikshtein, Nir Weinberger,  and Shlomo Shamai (Shitz)}
	
	\IEEEauthorblockA{Department of Electrical and Computer Engineering, Technion, Haifa 3200003, Israel}
	
	\IEEEauthorblockA{ 		
		Email: \{michaeldic@campus., nirwein@, sshlomo@ee.\}technion.ac.il
}
}	
\begin{document}
	\maketitle
	\begin{abstract}
     We formulate and analyze the \acrlong{comib} \emph{programming}. In this problem, a Markov chain $ \rv{X} \rightarrow \rv{Y} \rightarrow \rv{Z} $ is assumed with fixed marginal distributions  $\mathsf{P}_{\rv{X}}$ and  $\mathsf{P}_{\rv{Y}}$, and the mutual information between $ \rv{X} $ and $ \rv{Z} $ is sought to be maximized over the choice of conditional probability of $\rv{Z}$ given $\rv{Y}$ from a given class, under the \textit{worst choice} of the joint probability of the pair $(\rv{X},\rv{Y})$ from a different class. We consider  several classes based on extremes of: mutual information; minimal correlation; total variation; and the relative entropy class. We  provide values, bounds, and various characterizations for specific instances of this problem: the binary symmetric case, the scalar Gaussian case, the vector Gaussian case and the symmetric modulo-additive case. Finally, for the general case, we propose a Blahut-Arimoto type of alternating iterations algorithm to find a consistent solution to this problem.
	\end{abstract}
	\section{Introduction and Problem Formulation}
	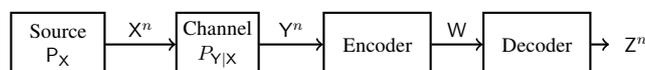
\begin{figure}[b]
	\centering
	\begin{tikzpicture}[thick,scale=0.7, every node/.style={scale=0.7}]
		\node[rectangle, draw = black, minimum width = 1.5cm, minimum height = 1.2cm, text width = 1.5cm, align = center]  (source) at (0,0) {Source $\pmf{\rv{X}}$};
		\node[rectangle, draw = black, minimum width = 1.5cm, minimum height = 1.2cm, align = center]  (channel) at (3,0) {Channel \\ $ P_{\rv{Y} | \rv{X}} $};
		\node[rectangle, draw = black, minimum width = 1cm, minimum height = 1.2cm, text width = 1.75cm, align = center]  (enc) at (6,0) {Encoder};
		\node[rectangle, draw = black, minimum width = 1cm, minimum height = 1.2cm, text width = 1.75cm, align = center]  (dec) at (9,0) {Decoder};
		\node (out) at (10.5,0) {};

		\draw[->] (source)  -- node [above]  {$ \rv{X}^n $}  (channel);
		\draw[->] (channel) -- node[above] {$ \rv{Y}^n $}  (enc);
		\draw[->] (enc) --  node[above] {$ \rv{W}  $} (dec);
		\draw[->] (dec) -- (out) node[right] {$ \rv{Z}^n $};
		
		\end{tikzpicture}
	\caption{Block diagram of Remote Source Coding.}
	\label{figure:remote_source_coding_diagram}
\end{figure}


The \acrfull{ib} methodology \cite{Tishby1999} plays a central role in data compression problems such as remote source coding and compression in oblivious relays, and more recently, it has found application in serving as a theoretical analysis tool to machine-learning algorithms, e.g. \cite{Tishby2015} (see \autoref{section:related_work} for a detailed overview). Another important aspect of the \acrshort{ib} methodology is that it provides a universal distortion measure for data compression when the desired distortion measure is either unavailable or cannot be defined. Nonetheless, in most practical cases, the distribution of the source involved in the \acrshort{ib} problem is also not known with perfect accuracy (e.g., when it is estimated from a finite sample). In this paper, this aspect motivates us to introduce a \textit{compound} version of the \acrshort{ib} problem, in which the source distribution is only known to belong to a given class, and the representation chosen by the \acrshort{ib} method is chosen to be the best possible under the worst-case choice within the class. We next exemplify this in two different compression scenarios -- remote source coding and oblivious relays. 

First, consider the compound remote source coding system  \cite{fontana1981universal,Dembo2003, Weissman2004} illustrated in  \autoref{figure:remote_source_coding_diagram}. Let $\pmf{\rv{X}}$ be a source of information generating the sequence $\rv{X}^n$. The encoder observes $\rv{Y}^n$ which is a noisy version of $\rv{X}^n$. Then, the encoder produces a compressed representation $\rv{W}$, which is later on mapped by the decoder to the reconstructed sequence $\rv{Z}^n$. The distortion is evaluated between $\rv{X}^n$ and $\rv{Z}^n$, while the rate is the relative number of bits required to represent $\rv{W}$. The encoder's goal is to find a compression strategy that extracts from $\rv{Y}^n$ the relevant information regarding $\rv{X}^n$, when the distribution of the channel $\pmf{\rv{Y}|\rv{X}}$ is not known in advance and cannot be accurately learned. This compound setting generalizes the classical remote source coding model studied by Dobrushin and Wolf \cite{Dobrushin1962,Wolf1970}. A different, yet related, problem of  compound rate-distortion is in terms of distortion measure mismatch \cite{Lapidoth1997}. In particular, consider a setting where the lossy compression codebook is generated for the purpose of minimizing the distortion under the distortion measure $d_0(\cdot,\cdot)$, but the average distortion of the reconstructed sequence is evaluated via a different distortion measure, $d_1(\cdot,\cdot)$. Furthermore, $d_1(\cdot,\cdot)$ can be a member of a certain class of distortion measures deviated from the nominal distortion $d_0(\cdot,\cdot)$. The compound \acrshort{ib} problem studied in this paper, can be interpreted as a remote source coding, in which the (logarithmic) distortion measure is determined by one member from the class of possible $\pmf{\rv{X}\rv{Y}}$, and thus is not completely specified.

\begin{figure}[b]
	\centering
	\begin{tikzpicture}[thick,scale=0.7, every node/.style={scale=0.7}]
		\coordinate (in) at (0,0);
		\node[rectangle, draw = black, minimum width = 1.5cm, minimum height = 1.2cm, text width = 2cm, align = center]  (transmitter) at (1.5,0) {Transmitter};
		\node[rectangle, draw = black, minimum width = 1.5cm, minimum height = 1.2cm, align = center]  (channel) at (4.5,0) {Channel \\ $ P_{\rv{Y} | \rv{X}} $};
		\node[rectangle, draw = black, minimum width = 1cm, minimum height = 1.2cm, text width = 1.75cm, align = center]  (relay) at (7.5,0) {Relay };
		\node[rectangle, draw = black, minimum width = 1cm, minimum height = 1.2cm, text width = 1.75cm, align = center]  (user) at (10.5,0) {User};
		\node (out) at (12,0) {};

		\draw[->] (in) node [left] {$ \rv{M}  $} --  (transmitter) ;
		\draw[->] (transmitter) --  node[above] {$ \rv{X}^n $}   (channel);
		\draw[->] (channel) -- node[above] {$ \rv{Y}^n $}  (relay);
		\draw[->] (relay) --  node[above] {$ \rv{W}  $} (user);
		\draw[->] (user) -- (out) node[right] {$ \hat{\rv{M}}  $};
		
		\end{tikzpicture}
	\caption{Block diagram of the Oblivious Relay Network.}
	\label{figure:oblivious_relay_diagram}
\end{figure}
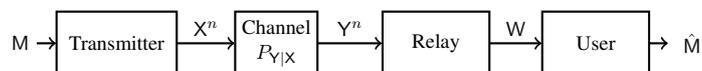

Second, consider the oblivious communication system illustrated in \autoref{figure:oblivious_relay_diagram}, which is a simplified model for cloud communication with oblivious processing \cite{Aguerri2019}. The network consists of a transmitter, a relay and a user. The channel from the transmitter to the relay is modeled as a \acrfull{dmc} $ \pmf{\rv{Y}|\rv{X}} $. The relay communicates messages to the receiver through a noiseless backhaul link of finite capacity. 
The transmitter maps the message $ \rv{M} $ to a codeword $ \rv{X}^n $ and transmits it through a \acrshort{dmc} $\mathsf{P}_{\rv{Y}|\rv{X}}$ to the relay.  The relay is unaware of communicating parties' codebook, 
but is capable of learning the marginal distribution of $ \rv{Y} $ from the received sequence $ \rv{Y}^n $. The relay represents $\rv{Y}^n$ with an index $ \rv{W} $ and sends it to the user via the noiseless finite capacity backhaul link. The receiver then decodes $\hat{\rv{M}}$. The system designer's goal is to construct a reliable communication scheme with the highest rate possible robust to the model constraints. 

In practice, the relay is usually oblivious regarding the statistical characteristics of the channel $\pmf{\rv{Y}|\rv{X}}$, but may assume that it belongs to a member of some defined class of channels. In fact, even if large number of samples had been available to learn the channel, and sophisticated learning algorithms are deployed, $\pmf{\rv{Y}|\rv{X}}$ can not be learned since the codebook is random, and typically changes per message (motivated, e.g., by cryptographic requirements). In this event, the codebook-oblivious relay performs a remote source coding with, loosely speaking, a compound distortion measure that gauges the ability of the receiver to decode the transmitted message.  Therefore, this scenario too falls into the framework of compound \acrshort{ib}. 


Formally, we define the \acrfull{comib} problem as follows. Let $(\rv{X},\rv{Y}) $ be a pair of random variables, and fix their  marginals to $ \mathsf{P}_{\rv{X}} $ and $ \mathsf{P}_{\rv{Y}} $, respectively. Consider all random variables $ \rv{Z} $ satisfying the Markov chain $ \rv{X} \rightarrow \rv{Y} \rightarrow \rv{Z} $. Unlike  the standard \acrshort{ib} problem, in which the joint distribution of $ \mathsf{P}_{\rv{X} \rv{Y}} $ is fixed, here we consider an uncertainty set for this joint distribution, and aim to solve
\begin{equation} \label{eq:definition_compound_ib}
	R^{\text{com}}_{\scaleto{\mathsf{P}_{\rv{X}}\mathsf{P}_{\rv{Y}}}{4pt}}(\set{P}_{\set{X}\set{Y}},\set{D}_{\set{Z}|\set{Y}}) 
	= \max_{\mathsf{P}_{\rv{Z}| \rv{Y}} \in \set{D}_{\set{Z}|\set{Y}}} \min_{\mathsf{P}_{\rv{X} \rv{Y}} \in \set{P}_{\set{X}\set{Y}}} I(\rv{X};\rv{Z}),
\end{equation}
where $I(\rv{X};\rv{Z})$ is the mutual information between $\rv{X}$ and $\rv{Z}$.
Thus, the set $\set{D}_{\set{Z}|\set{Y}} $ is the set of possible representations, and the set $\set{P}_{\set{X}\set{Y}}$ is the uncertainty set of the joint distribution. The class $\set{D}_{\set{Z}|\set{Y}} $ will be the usual \acrshort{ib} class, i.e.,
$
    \set{D}_{\set{Z}|\set{Y}} = \left\{
    \mathsf{P}_{\rv{Z}|\rv{Y}} \colon I(\rv{Y};\rv{Z}) \leq C_2
    \right\}
$,
or a restricted subset of this class, with an additional structure. The class $\set{P}_{\set{X}\set{Y}}$ will take one of the following variants:
\begin{itemize}
    \item \acrfull{pf} class:
    $
    \set{P}_{\set{X}\set{Y}} = \left\{
    \mathsf{P}_{\rv{X} \rv{Y}} \colon I(\rv{X};\rv{Y}) \geq C_1
    \right\}
    $. This class is motivated by trade-offs between privacy and utility, for example, of a health maintenance organization that wishes to share as much relevant information as possible  to a an external partner,  without disclosing the personal details of its patients. This setting can be modeled as a user that has two types of correlated data, a private data, represented by the random variable $\rv{Z}$, and a public data, represented by a random variable $\rv{Y}$, which he would like to share with an analyst. To diminish the inference capabilities of the analyst to extract private data from observing $\rv{Y}$, the user instead shares a distorted version of $\rv{Y}$ denoted by $\rv{X}$ \cite{Makhdoumi2014}. In the \acrshort{comib} problem studied here, the inference from $\rv{X}$ to $\rv{Z}$ is minimized over the representation $\mathsf{P}_{\rv{Z}|\rv{Y}}$, with the assumption that the disclosed information (in $\rv{X}$) will be as private as possible.
    \item \textit{Minimal Correlation}  class:
    $
    \set{P}_{\set{X}\set{Y}} = \left\{
    \mathsf{P}_{\rv{X} \rv{Y}} \colon \Exp{\rv{X} \rv{Y}} \geq \rho_1
    \right\}
    $.
    This class is motivated by the compressed representation canonical correlation analysis (CRCCA) \cite{Painsky2020}. The interpretation is similar to the privacy funnel case, only here the correlation replaces the mutual information as a measure of statistical dependence.
    \item \acrfull{tv}  class:
    $
    \set{P}_{\set{X}\set{Y}} = \left\{
    \mathsf{P}_{\rv{X} \rv{Y}} \colon d_{\text{TV}}(\mathsf{P}_{\rv{X}\rv{Y} }, \mathsf{P}_1 ) \leq D_1
    \right\}
    $, where the total variation distance between two probability vectors $\bv{p}$ and $\bv{q}$ is defined as
    $d_{\text{TV}} (\bv{p},\bv{q}) \triangleq \sum_{i=1}^n |p_i - q_i|$. This class is motivated by finite sample analysis for \acrshort{ib} setting \cite{Shamir2010}, where the true joint law $\pmf{1}$ of $(\rv{X},\rv{Y})$ is not known, but rather its empirical co-occurrence distribution, $\pmf{\rv{X}\rv{Y}}$, is used to calculate an estimate of the \acrshort{ib} functional. Thus, \acrshort{comib} method then provides bound on the extent in which the underlying distribution should be estimated in order to solve the \acrshort{ib} problem.
    \item \acrfull{kld}  class:
    $
    \set{P}_{\set{X}\set{Y}} = \left\{
    \mathsf{P}_{\rv{X} \rv{Y}} \colon D(\mathsf{P}_{\rv{X}\rv{Y} }|| \mathsf{P}_1 ) \leq \epsilon_1
    \right\}
    $. This class is commonly used by statisticians as a natural metric for model mismatch  \cite{McLachlan2008}, is considered as a natural geometric ``distance" between systems \cite{Amari2000}, and is utilized as a robustness measure for 
  arbitrary deviations of the prior from the nominal distribution in robust hypothesis testing problems \cite{Levy2009}. For the particular scenario of Gaussian nominal distribution, they were used to provide alternative bounds on MMSE \cite{Dytso2019}. Furthermore, they were applied to propose a reliable power distribution protocol in wireless communications \cite{Cao2020}.
\end{itemize}
For all the above classes, we will typically assume in the rest of the paper that the joint distribution is constrained to the given marginals, i.e., $ \sum_{x } \mathsf{P}_{\rv{X}\rv{Y}}(x,y) = \mathsf{P}_{\rv{Y}}(y) $ and $ \sum_{y } \mathsf{P}_{\rv{X}\rv{Y}}(x,y) =     \mathsf{P}_{\rv{X}}(x) $. 
Since the \acrshort{comib} problem generalizes the \acrshort{ib} problem, we next review the central results and approaches to the \acrshort{ib} problem,  before describing our results. As said, choosing the class $ \set{P}_{\set{X}\set{Y}} $ to a singleton,
i.e., $(\rv{X},\rv{Y})$ is a bivariate source characterized by a fixed joint probability law $  \mathsf{P}_{\rv{X}\rv{Y}} $, 
recovers the standard \acrshort{ib} problem \cite{Tishby1999}, namely,
\begin{equation} \label{eq:definition_ib}
	R^{\text{IB}}_{\scaleto{\mathsf{P}_{\rv{X}\rv{Y}}}{7pt}}(\set{D}_{\set{Z}|\set{Y}}) 
	= \max_{\mathsf{P}_{\rv{Z}| \rv{Y}} \in \set{D}_{\set{Z}|\set{Y}}}  I(\rv{X};\rv{Z}).
\end{equation}
For discrete alphabets, this problem was originally studied in \cite{Witsenhausen1975} as a method to characterize common information \cite{Gacs1973}. The \acrshort{ib} method is essentially a remote source coding problem \cite{Dobrushin1962, Wolf1970}, choosing the distortion measure as the logarithmic loss, and thus recovers remote source coding by taking $\set{D}_{\set{Z}|\set{Y}}$ as a maximal distortion constraint set.

In addition, \acrfull{pf},  a dual problem to the \acrshort{ib} framework  \cite{Makhdoumi2014,PinCalmon2017}, can also be recovered from \eqref{eq:definition_compound_ib} by setting $\set{P}_{\set{X}\set{Y}}$ as \acrshort{pf} family and $\set{D}_{\set{Z}|\set{Y}} $ to contain a singleton, that is,
\begin{equation} \label{eq:definition_pf}
	R^{\text{PF}}_{\scaleto{\mathsf{P}_{\rv{Z}|\rv{Y}}}{7pt}}(C_1) 
	= \min_{\mathsf{P}_{\rv{X}  \rv{Y}} \colon I(\rv{X};\rv{Y}) \geq C_1} I(\rv{X};\rv{Z}),
\end{equation} 
Therefore, under a \acrshort{pf} constraint, the problem introduced in \eqref{eq:definition_compound_ib} is actually a composition of the \acrshort{ib} 
and \acrshort{pf} problems.
This makes the problem in  \eqref{eq:definition_compound_ib} rather delicate -- e.g., if $ (\rv{Y},\rv{Z}) $ are jointly Gaussian, even the  standard \acrshort{pf} rate is zero, since one can use the channel from $ \rv{Y} $ to $ \rv{X} $ to describe the less significant bits of $ \rv{Y} $ \cite{Shamai2021}. 


The \acrshort{ib} problem is a non-convex optimization problem and a general closed form solution does not exist except for some particular settings. It was approached via several strategies. When $ (\rv{X},\rv{Y}) $ is a \acrlong{dsbs} (\acrshort{dsbs}) \cite{Wyner1976} with transition probability $ p $, it was shown in \cite{Zaidi2020} that binary symmetric channels are optimal via \acrlong{mgl} \cite{Wyner1973} (see also the examples in \cite{Witsenhausen1975} and \cite{Sutskover2005}).  
When $ (\rv{X},\rv{Y}) $ are jointly multivariate Gaussians,
it was shown in  \cite{Chechik2005} that the optimal distribution of $ (\rv{X},\rv{Y},\rv{Z}) $ is also jointly Gaussian. The optimality of  the Gaussian test channel can be  proved using the \acrfull{epi} \cite{Dembo1991}, or by utilizing 
the I-MMSE relation and Single Crossing Property \cite{Guo2013,Bustin2013}. 
Moreover, under the I-MMSE framework, the proof can be readily extended to Jointly Gaussian Random Vectors $ (\bv{X},\bv{Y}) $  \cite{Bustin2013}.
In a different, and more general case, when $ (\rv{X},\rv{Y},\rv{Z}) $ are discrete random variables, a locally optimal $ \mathsf{P}_{\rv{Z}|\rv{Y}} $ can be found by iteratively solving a set of self-consistent equations. A generalized Blahuto-Arimoto algorithm was proposed to solve those equations in  \cite{Blahut1972,Arimoto1972,Tishby1999,Hassanpour2017,Aguerri2021}. Finally, a particular case of deterministic mappings from $ \rv{X} $ to $ \rv{Y} $ was considered in \cite{Slonim2002}. \looseness=-1

In this work, we address the compound setting for the \acrshort{ib} problem, with the goal of providing similar results. First, we address the \acrshort{dsbs} and Gaussian (scalar and vector) settings. Second, we analyze the \acrshort{kld} class for $\set{P}_{\set{X}\set{Y}}$ for the particular choice of jointly Gaussian random variables. 
Then, we consider general modulo additive channels, with modulo additive representations, and provide various bounds on the \acrshort{comib} function with \acrshort{pf}-based compound set, and then with \acrshort{tv}-based compound set. Finally, we return to the general discrete alphabet case with \acrshort{pf} based compound set, and propose an alternating algorithm, which essentially iterates between the maximization over $\mathsf{P}_{\rv{Z}| \rv{Y}}$ (an \acrshort{ib} problem) and  minimization over $\mathsf{P}_{\rv{X} \rv{Y}}$ (a \acrshort{pf} problem). 
We further specialize this algorithm to the modulo-additive setting, obtaining an elegant and efficient computational method.

\subsection{Related work} \label{section:related_work}
In many problems in learning, there is an interest to represent data $\rv{Y}$, with a compressed version  $\rv{Z}$ that captures as much relevant information as possible with a fixed number of bits. One possible approach to handle such problem is via rate distortion theory for lossy source coding. However, the utilization of rate distortion theory, requires specifying a distortion function first, and it is usually intractable finding such function for real data scenarios. A pioneering work by Tishby et al. \cite{Tishby1999}, suggested the \acrshort{ib} framework, where additional variable $\rv{X}$ determines relevance (for example, it can be the labeling of the data). The quality of distortion is measured by the mutual information between $\rv{X}$ and $\rv{Z}$, thus revealing a more natural distortion measure.
This framework is closely related to a variety of problems in information theory, such as \textit{remote source coding} \cite{Wolf1970},  \acrfull{ceb} \cite{Witsenhausen1975}, \acrlong{cr} \cite{Steinberg2009}, and \acrfull{infcomb} \cite{Sutskover2005,Land2006}. See an overview in a recent comprehensive tutorial on the
IB method and related problems \cite{Zaidi2020}.
Applications of the \acrshort{ib} problem in machine-learning are detailed in \cite{Chechik2005,Tishby2015,Farajiparvar2018,Goldfeld2020}. 

In the coding-theoretic context, it has been recently shown that the \acrshort{ib} method can be used to reduce the data transfer rate and computational complexity in 5G low-density parity check (LDPC) decoders \cite{Lewandowsky2016,Stark2020}. Furthermore, it is also related to construction of good polar codes \cite{Bhatt2021}.  In this problem, the value of the capacity of the polarized channels is required in order to identify the location of "frozen bits" in the codeword. However, the  output-alphabet size of the polarized channels increases exponentially, and so quantization is employed in order to reduce the computational complexity. 
The quality of the quantization scheme is then assessed via  mutual information preservation. It can be shown that the corresponding \acrshort{ib} problem upper bounds the mutual-information after quantization technique. 

General quantization algorithms based upon the \acrshort{ib} method were considered in \cite{Stark2018,Shah2019, Shah2019a}. 
Furthermore, a relationship between the KL means algorithm, and the \acrshort{ib} method has been discovered in \cite{Kurkoski2017}.  
In \cite{Pensia2020} a robust \acrshort{ib} program was proposed, with the goal of extracting features that are simultaneously relevant and robust. Unlike in this paper, therein the channel from $ \rv{Y} $ to $ \rv{Z} $ is made robust, as measured in terms of Fisher Information.

With more generality, the \acrshort{ib} problem connects to many other timely aspects. These include game theory and Nash equilibrium \cite{Kazikli2020}, \emph{capital investment} \cite{Erkip1998}, \emph{distributed learning} \cite{Farajiparvar2018}, \emph{deep learning} \cite{Tishby2015,Alemi2017,ShwartzZiv2017,Gabrie2019,Goldfeld2018} and \emph{convolutional neural networks} \cite{Cheng2018,Yu2021}.


\subsection{Notations and Preliminaries}
Throughout the paper, random variables are denoted using a sans-serif font, e.g., $ \rv{X} $, their realizations are denoted by the respective lower-case letters, e.g., $ x $, and their alphabets are denoted by the respective calligraphic letters, e.g., $ \mathcal{X} $. 
The cardinality of a finite set, say $ \mathcal{X}$, is denoted by $ |\mathcal{X}|$. The probability distribution function of $ \rv{X} $, the joint distribution function of $ \rv{X} $ and $ \rv{Y} $, and the conditional distribution of $ \rv{X} $ given $ \rv{Y} $ are denoted by $ \pmf{\rv{X}} $, $ \pmf{\rv{X} \rv{Y}} $ and $ \pmf{\rv{X}|\rv{Y}} $ respectively. The expectation of $ \rv{X} $ is denoted by $ \Exp{\rv{X}} $. Random vectors and their realizations are denoted in the respective bold font, say $\bv{X}$ and $\bv{x}$.


Marginal probability vector is denoted by a lowercase boldface letter, i.e.,
	$\bv{q} \triangleq \left\{ \Prob{\rv{X}=x}\right\}_{x\in \set{X}}$.
The $ n-1 $ dimensional simplex, i.e, the set of all $ n $-ary probability vectors,  is denoted by $\simplex{n}$. For an integer $n$, the set of indices from $ 1 $ to $ n $ is denoted by $[n]\triangleq \{1,\ldots,n\}$. The standard  $ k $-th basis vector of $ \field{R}^n $ is denoted by $ \bv{e}^{(n)}_k $, i.e.,
	$[\bv{e}^{(n)}_k]_j \triangleq \delta_{jk}$,
where $\delta_{jk}$ is Kronecker's delta.
Furthermore, the all ones vector is denoted by $\bv{e}_n$, the uniform distributed probability vector is denoted by $ \bv{u}_n = \bv{e}_n/n $, and the all zeros vector is denoted by $ \bv{0}_{n}=[0,\ldots,0]^T$. Subscript $_n$ and superscript $^{(n)}$ are omitted when the dimension is clear from context.
The transition matrix $T$ from $ \rv{X} $ to $ \rv{Y} $ is denoted by
	$T_{ij} \triangleq \Prob{\rv{Y}=i|\rv{X}=j}$, $ i \in \set{Y}, j \in \set{X}$.

All logarithms are taken to the natural base. 
The entropy function in $ \field{R}^n $ is the function $ h: \field{R}^n_+ \mapsto \field{R}$, given by
$ h(\bv{x})  \triangleq -\sum_{i=1}^n x_i \log x_i$. When $\bv{q}$ is the probability vector of random variable $\rv{X}$, then $h(\bv{q})$ is the entropy of $\rv{X}$.

The indicator function $\indicator_{\set{S}} (x)$  of a set $\set{S}$ is denoted by 
\begin{equation}
    \indicator_{\set{S}}(x) \triangleq
    \begin{cases}
    1, & x \in \set{S} \\
    0, & x \notin \set{S}.
    \end{cases}
\end{equation}


The \acrfull{dsbs} $(\rv{X},\rv{Y})$ with parameter $\alpha$ is defined via the following joint \acrshort{pmf}
    	$\mathsf{P}_{\rv{X}\rv{Y}}(x,y) = \frac{1}{2} (\alpha \cdot \indicator(x \neq y) + (1-\alpha)
    	\indicator(x=y))$.
One's complement is denoted with a bar, i.e., $ \bar{x} = 1-x $. The binary convolution of $ x,y \in [0,1] $ is defined as $ x * y \triangleq x \bar{y} + \bar{x} y $.
The binary entropy function is defined as
$h_b(p) \triangleq -p \log p - (1-p) \log (1-p)$.
The inverse of the binary entropy function restricted to the domain $[0,1/2] $ is denoted by $ h_b^{-1}(\cdot) : [0,\log 2] \rightarrow [0,1/2] $. The maximum of $x$ and $1$ is denoted by $[x]^+ \triangleq \max \{x,1\}$. Similarly, the minimum of $x$ and $1$ is denoted by $[x]^- \triangleq \min \{x,1\}$.

A simple way to obtain solutions to \eqref{eq:definition_compound_ib} is by establishing a saddle point property. We briefly remind the reader this property as it will be used multiple times in the proofs. 
\begin{lemma}[Optimality of Saddle Point {\cite[Sec. 5.4.2]{Boyd2014}}] \label{lemma:optimality_saddle_point}
    Suppose there exists a saddle point $(\Tilde{w},\Tilde{z})$, satisfying
    $
        f(\Tilde{w},\Tilde{z}) = \inf_{w \in \set{W}} f(w,\Tilde{z}) 
    $ and $
        f(\Tilde{w},\Tilde{z}) = \sup_{z \in \set{Z}} f(\Tilde{w},z) 
    $,
    then
    \begin{equation}
        f(\Tilde{w},\Tilde{z}) = \sup_{z \in \set{Z}} \inf_{w \in \set{W}} f(w,z) .
    \end{equation}
\end{lemma}

	\section{Results - An Overview}
	\subsection{Binary $\rv{Y}$}
	In this section, we begin with a simple, yet canonical, example of binary random variables, for which full  characterization of  \eqref{eq:definition_compound_ib}  can be found.


Suppose $ \rv{Y} $ is a $ \Ber(0.5) $ random variable with \acrshort{pf} type of $\set{P}_{\set{X}\set{Y}}$ (no cardinality constraint on $\rv{Z}$). Let $R^{\text{bin}}(C_1,C_2)$ denote the \acrshort{comib} with a \acrshort{pf} constraint for this setting. The optimal solution here depends on the cardinality of $\rv{X}$, and possibly of $\rv{Z}$.
\begin{proposition} \label{proposition:comib_binaryXY}
    Assume that $\rv{X}$ is binary. Then, the optimal $ (\rv{X},\rv{Y}) $ are distributed as a \acrfull{dsbs} with parameter $ \alpha $, 
    where $ \alpha =  h_b^{-1} (1-C_1) $.
    Furthermore, the optimal $ \mathsf{P}_{\rv{Z}|\rv{Y}} $ in this case is a \acrshort{bsc} with parameter $ \beta =  h_b^{-1} (1-C_2) $. The compound rate is thus
    	$ R^{\text{bin}}(C_1,C_2) = 1- h_b(\alpha * \beta) $.
\end{proposition} 
The proof of \autoref{proposition:comib_binaryXY} appears in  \autoref{appendix:proof_of_proposition_comib_binaryXY}.

Next, assume that $\rv{Y}$ is $\Ber(0.5)$, but there are no constraints on neither $\rv{X}$ nor $\rv{Z}$.
\begin{proposition} \label{proposition:comib_binaryY_X_general}
     The optimal $\mathsf{P}_{\rv{Z}|\rv{Y}}$ is a \acrshort{bsc} with parameter $\delta = h_b^{-1} (1-C_2)$, while the optimal $\mathsf{P}_{\rv{X}|\rv{Y}}$ is a \acrshort{bec} with parameter $\epsilon = 1-C_1$. The optimal rate in such case is $R^{\text{bin}}(C_1,C_2) = C_1 \cdot C_2  $. 
\end{proposition}

Prop. \ref{proposition:comib_binaryY_X_general}  can be established by combining \cite[IV.C]{Witsenhausen1975} with \cite[Thm. 1]{Sutskover2005} and \autoref{lemma:optimality_saddle_point}, and its proof is omitted. We also note in passing that it appears to be challenging to find a closed-form analytical solution for the asymmetric binary setting, i.e., when $\rv{Y}$ is not uniform. 

\begin{remark}
    Note that in this section there is no explicit constraint on the marginal probability of $\rv{X}$, rather its cardinality. Making such assumption might make this problem  trivial. For example, assuming that $\rv{X} \sim \Ber(1/2)$ restricts $\pmf{\rv{X}\rv{Y}}$ to a \acrshort{dsbs}.
\end{remark}

\subsection{Scalar Gaussian $ (\rv{X},\rv{Y}) $}
    We begin with a fundamental scenario where the marginal distributions of $ \rv{X} $ and $ \rv{Y} $ are both Gaussian. Note that in contrast to the symmetric uniform Bernoulli setting, which restricts the  channel from $ \rv{X} $ to $ \rv{Y} $ being a \acrshort{bsc}, here, Gaussianity of the marginals does not imply the joint distribution of $ (\rv{X},\rv{Y}) $ being Gaussian \cite[Ch. 4.7]{Bertsekas2002}. Thus, the result of the following theorem is more complicated than that of \autoref{proposition:comib_binaryXY}. Let $R^{\text{sc-G}}(\rho,C)$ denote the value of \eqref{eq:definition_compound_ib} with $ \set{P}_{\set{X}\set{Y}} $ being the minimum correlation class with parameter $\rho > 0 $ and $\set{Q}_{\set{Z}|\set{Y}}$ being the \acrshort{ib} bottleneck class with parameter $C$.\looseness=-1

\begin{theorem} \label{theorem:oblib_gaussianXY}
   It holds that $R^{\text{sc-G}}(\rho,C) = \minus \frac{1}{2} \log( 1 \minus \rho^2 \rho_C^2) $,
	with $ \rho_C^2 = 1-2^{-2 C} $, and jointly Gaussian $ (\rv{X},\rv{Y},\rv{Z}) $ is the unique optimizer of \eqref{eq:definition_compound_ib}.
	
\end{theorem}

The proof of this theorem appears in \autoref{section:oblib_proof_of_theorem_gaussianXY}.

    \subsection{Vector Gaussian $ (\bv{X},\bv{Y}) $}
Now, suppose that $\bv{X}$ and $\bv{Y}$ are jointly Gaussian random vectors of dimension $n$.
Let $R^{\text{vec-G}}(C_1,C_2)$ denote the value of \eqref{eq:definition_compound_ib}, with $ \set{P}_{\set{X}\set{Y}} $ being the \acrshort{pf} constraint with capacity $C_1$, and $\set{Q}_{\set{Z}|\set{Y}}$ is the \acrshort{ib} bottleneck class with capacity $C_2$.
\begin{theorem} \label{theorem:oblib_VectorGaussianXY} It holds that
        $R^{\text{vec-G}}(C_1,C_2) = -\frac{n}{2} \log (1-\rho_1^2 \rho_2^2)$,
where $\rho_k^2 = 1-2^{-2C_k /n}$ for $ k \in \{1,2\}$. The optimal triplet $(\bv{X},\bv{Y},\bv{Z})$ is jointly Gaussian with independent components. 
\end{theorem}

In particular, this result establishes that the worst case channel $\mathsf{P}_{\rv{Y}|\rv{X}}$ is an \acrlong{awgn}, and its optimal representation $\mathsf{P}_{\rv{Z}|\rv{Y}}$ is also white.

The proof of this theorem is given in \autoref{section:oblib_proof_of_theorem_VectorgaussianXY}.

\subsection{Additive Channels with KL-divergence Constraint}
Suppose $\rv{Y} \sim \mathcal{N}(0,1)$ and the channel from $\rv{Y}$ to $\rv{X}$ is an additive 
Gaussian
noise channel,
namely, there exist a random variable $\rv{W} \sim \mathcal{N}(0,\sigma^2) $  such that
    $\rv{X} = \rv{Y} + \rv{W}$.

Let $R^{\text{KL-G}}(\epsilon_1,C_2)$ denote the value of \eqref{eq:definition_compound_ib} with $ \set{P}_{\set{X}\set{Y}} $ being the \acrshort{kld} constraint with ``distance" $\epsilon_1$ and $\set{Q}_{\set{Z}|\set{Y}}$ is the \acrshort{ib} bottleneck class with capacity $C_2$.
where $\rv{N}_0 \sim \mathcal{N}(0,\sigma^2_0)$.
We have the following result.
\begin{theorem} \label{theorem:comib_kld_gaussian}
Let $\sigma_*$ be the solution to
    $\frac{1}{2} \log \frac{\sigma^2}{\sigma_0^2} + \frac{\sigma^2}{2 \sigma_0^2} - \frac{1}{2} = \epsilon_1$.
The \acrshort{comib} rate with \acrshort{kld} constraint  is given by:
\begin{equation}
     R(\epsilon_1,C_2)^{\text{KL-G}} = \frac{1}{2} \log \left(
    \frac{1}{1-(1-2^{-2C_2}) \frac{1}{1+\sigma^2_*}}
    \right).
\end{equation}
\end{theorem}
    
The proof of \autoref{theorem:comib_kld_gaussian} is given in 
Supplementary Material.

	
	


	\subsection{Modulo Additive Channels with \acrshort{pf} Constraint}
	\label{section:comib_modulo_additive_pf}
	In this section, we return to the (general) discrete alphabet case, yet we 
restrict our attention to a symmetric setting with the following assumptions:
\begin{align}
    \label{eq:comib_moduloXY_assumption_Pxy}
    \set{P}_{\set{X}\set{Y}}
    & \triangleq
    \left\{
    \mathsf{P}_{\rv{X}\rv{Y}} \colon \rv{X} \sim \text{unif}[n], \rv{Y} = \rv{X} \oplus \rv{W}, H(\rv{W}) \leq \eta_1)
    \right\}, \\
    \label{eq:comib_moduloXY_assumption_Qzy}
    \set{Q}_{\set{Z}|\set{Y}}
    & \triangleq
    \left\{
    \mathsf{P}_{\rv{Z}|\rv{Y}} \colon \rv{Z} = \rv{Y} \oplus \rv{V}, H(\rv{V}) \geq \eta_2)
    \right\}.
\end{align}
%
This setting implies $ |\set{X}| = |\set{Y}| = |\set{Z}| = n $.   Moreover, it also holds that $ \rv{Z} = \rv{X} \oplus \rv{W} \oplus \rv{V}$, where $ \oplus $ is a modulo-$ n $ additive operator, so that $ \rv{X} \rightarrow \rv{Y} \rightarrow \rv{Z} $ holds.  
Using $ H(\rv{W}) \equiv H(\mathsf{P}_{\rv{W}})$ and $ H(\rv{V}) \equiv H(\mathsf{P}_{\rv{V}})$,
we observe that 
$I(\rv{X};\rv{Z})=\log n-H(\mathsf{P}_{\rv{W}}*\mathsf{P}_{\rv{V}})$, where $*$ is the $n$-ary convolution operator. Thus, 
the solution to \eqref{eq:definition_compound_ib} is equivalent to the solution of 
\begin{equation}\label{eq:definition_compound_ib_modulo}
	R^{\text{mod}}(\eta_{1},\eta_{2}) \triangleq 
	\min_{ \scaleto{\mathsf{P}_{\rv{V}} \colon  H(\mathsf{P}_{\rv{V}}) \geq \eta_{2}}{7pt}}
	\max_{
	\scaleto{
	\mathsf{P}_{\rv{W}}\colon  H(\mathsf{P}_{\rv{W}})\leq\eta_{1}}{7pt}
	}
		H(\mathsf{P}_{\rv{W}}*\mathsf{P}_{\rv{V}}).
\end{equation} 
In \eqref{eq:comib_moduloXY_assumption_Qzy} we have confined the channel $\mathsf{P}_{\rv{Z}|\rv{Y}}$ to be modulo additive, which may be too restrictive in general. Nonetheless, when the \acrshort{ib} function is \textit{strictly} convex, the modulo additive channel assumption for $\set{Q}_{\set{Z}|\set{Y}}$ can be relaxed. Indeed:
\begin{proposition} \label{proposition:modulo_additiveness_Pzgy}
    Fix a joint \acrshort{pmf} $\mathsf{P}_{\rv{X}\rv{Y}}\in\set{P}_{\set{X}\set{Y}}$, where $\set{P}_{\set{X}\set{Y}}$ is as defined in \eqref{eq:comib_moduloXY_assumption_Pxy}. Denote by $T$ the transition probability matrix from $\rv{Y}$ to $\rv{X}$. Assume that function $R^{\text{CEB}}_{\scaleto{T}{4pt}}(\eta)$ defined by
    \begin{equation} \label{eq:ssib_problem_definition}
	     R^{\text{CEB}}_{\scaleto{T}{4pt}}(\eta) \triangleq 
		   \min_{\mathsf{P}_{\rv{Z}|\rv{Y}} \colon H(\rv{Y}|\rv{\rv{Z}})  \geq   \eta }
		    H(\rv{X}|\rv{Z})
    \end{equation}
    is a strictly convex function of $\eta$, then it is equivalent to the following problem:
    \begin{equation} \label{eq:ssib_problem_definition2}
	    g_{\scaleto{T}{4pt}}(\eta) \triangleq 
		\min_{\bv{p} \in \simplex{n} \colon h_n(\bv{p}) \geq \eta}
		    h_n(T \bv{p}),
    \end{equation}
    where $\simplex{n}$ is the $n$-dimensional simplex, and the optimal channel from $\rv{Y}$ to $\rv{Z}$ is also a modulo additive channel.
\end{proposition}

Thus, if the strict convexity holds then modulo additive channels form a saddle point in  \eqref{eq:definition_compound_ib_modulo}, and are thus optimal via \autoref{lemma:optimality_saddle_point} (in the restricted class of modulo additive $ \mathsf{P}_{\rv{X}\rv{Y}}$). We postpone the proof of Prop. 3 to Supplementary Material.
\begin{remark}
    \autoref{proposition:modulo_additiveness_Pzgy} establishes equivalence between the problems addressed in \cite{Witsenhausen1974} and \cite{Witsenhausen1975}.
    However, as was shown in \cite{Witsenhausen1974}, the function $g_T(\eta)$ is not convex in general, and therefore  we cannot universally utilize \autoref{proposition:modulo_additiveness_Pzgy}. We may use it  only for regions of $\eta$ where $g_T(\eta)$ is convex. Nonetheless, it was shown in \cite{Witsenhausen1974} that  $g_T(\eta)$ is convex for all binary channels and noiseless channels. 
\end{remark}

We will next show that in the low-SNR regime, specifically, when $ \eta_{1} \geq \log (n-1) $, the optimal distribution achieving \eqref{eq:definition_compound_ib_modulo} has a unique structure, characterized by generalized Hamming channels. We first give a proper definition of such channels.
A \acrshort{pmf}  $\bv{p} \in \simplex{n}$ is called $(\alpha,n)$-Hamming \cite{Witsenhausen1974}, if for some $\alpha\in[0,1]$, it is of the form 
\begin{equation}
	\bv{p} = \alpha\cdot \bv{e}_n +\bar \alpha \cdot\bv{u}_n=\left(\alpha+\frac{\bar \alpha}{n},\frac{\bar \alpha}{n},\ldots\frac{\bar \alpha}{n}\right)\label{eq: Hamming channel}.
\end{equation}
That is, $\bv{p}$ is an $\alpha$-mixture between
the deterministic \acrshort{pmf} $\boldsymbol{e}_{n}$ and the completely noisy \acrshort{pmf} $\bv{u}_n$ (a uniform distribution over $ n $).
Also note that as $\alpha>0$ then $\bv{p}$ is ordered where the first
probability is the largest and all the other $n - 1$ probabilities
are smaller and equal to each other. 
For negative values of $\alpha$, the vector on the RHS of (\ref{eq: Hamming channel})
is a \acrshort{pmf} only if $\alpha\in[-\frac{1}{n-1},0)$. In that case it has
a full support, the first probability is the smallest, and all
the other $n-1$ probabilities are the largest and equal to  each other.
Note also that $ \bv{p} =  \bv{u}_n $ for $\alpha=0$
and then $h(\bv{p})=\log n$, while 
$\bv{p}= (0, \bv{u}_{n-1}^T)^T
$
for $\alpha=-\frac{1}{n-1}$ and then $h(\bv{p})=\log(n-1)$. We thus generalize
the Hamming \acrshort{pmf} for all $\alpha\in[-1,0]$ as follows. A \acrshort{pmf} $\bv{p}$
is $(\alpha,n,k)$ negative-Hamming if 
\begin{equation} \label{eq:negative_hamming_definition}
	 \bv{p} = [\alpha\cdot \bv{e}_{k}+ \bar \alpha \bv{u}_{k},\boldsymbol{0}_{n-k}] ,
\end{equation}
where 
$k\in[n]$ is such that $\alpha\in(-\frac{1}{k-1},0]$.
Strictly speaking, an $(\alpha,n,k)$  negative-Hamming probability vector has a support $k$, with first $k-1$ equal elements and the $k$-th element is smaller than the first $k-1$ ones.

\begin{theorem}\label{theorem:oblib_main_result}
	Consider the optimization problem defined in \eqref{eq:definition_compound_ib_modulo}, and assume that $ \eta_1 \geq \log (n-1) $.
	Then, the optimal $ \mathsf{P}_{\rv{V}} $ and $ \mathsf{P}_{\rv{W}} $ are a regular Hamming channel with parameters $ (\alpha,n) $ and a negative Hamming channel with parameters $ (\beta,n,n) $, respectively, where $ \alpha $ is the positive root of 
	\begin{equation} \label{eq:comib_alpha_eta2}
		\eta_2 + \left(\alpha + \frac{\bar{\alpha}}{n}\right) \log \left(\alpha + \frac{\bar{\alpha}}{n}\right) + \frac{(n-1)\bar{\alpha}}{n} \log 	\frac{\bar\alpha}{n} = 0,
	\end{equation}
	and $ \beta $ is the negative root of
	\begin{equation} \label{eq:comib_beta_eta1}
		\eta_1  + \left(\beta + \frac{\bar{\beta}}{n}\right) \log \left(\beta + \frac{\bar{\beta}}{n}\right) + \frac{(n-1)\bar{\beta}}{n} \log \frac{\bar\beta}{n} = 0.
	\end{equation}
	Furthermore,
	\begin{equation}
		R^{\scaleto{\text{mod}}{4pt}}(\eta_{1},\eta_{2})  =  \minus 
		\left( \alpha \beta + \frac{\overline{\alpha \beta}}{n}  \right) 
		\log  \left( \alpha \beta  +  \frac{\overline{\alpha \beta}}{n} \right)  \minus  \frac{(n \minus 1)\overline{\alpha \beta}}{n} \log 	\frac{\overline{\alpha \beta}}{n}.
	\end{equation}
\end{theorem}
We postpone the proof of this theorem to  \autoref{section:oblib_proof_of_theorem_low_snr}.
\begin{remark}
    This elegant result does not extends to the regime $ \eta_1 \in (0, \log(n-1)) $, as the following counterexample demonstrates. Suppose $ \mathsf{P}_{\rv{W}} = \bv{p} $ is a negative Hamming channel with parameters $ (0.46,3,2) $, and take $ \eta_2 = 0.7 $. In this case, the positive Hamming point is given by
	$ \bv{q}^{+} = (0.866 , 0.067 , 0.067)^T $,
which achieves an output entropy of
$ 	h(\bv{p} * \bv{q}^{+}) = 1.179 $ (bits). However, taking
	$\bv{q}^* = (0.857 , 0.031 , 0.112)^T $
gives us
$ 	h(\bv{p} *\bv{q}^*) = 1.165 < h(\bv{p} *\bv{q}^{+}) $ (bits).
\end{remark}


%

We next provide bounds on \eqref{eq:definition_compound_ib_modulo} which complement the result of \autoref{theorem:oblib_main_result}.

\begin{theorem}\label{theorem:oblib_main_result_bounds}
	If $ \eta_1 \in (0,\log(n\minus 1)) $, then
	\begin{multline}
	R^{\text{mod}}(\eta_{1},\eta_{2}) \leq
		\minus \left(\alpha \beta + \frac{\alpha \bar{\beta}}{k} + \frac{\bar{\alpha}}{n} \right) \log \left(\alpha \beta + \frac{\alpha \bar{\beta}}{k} + \frac{\bar{\alpha}}{n} \right) \\
	\minus  (k\minus 1) \left( \frac{\alpha \bar{\beta}}{k} + \frac{\bar{\alpha}}{n} \right) \log \left( \frac{\alpha \bar{\beta}}{k} + \frac{\bar{\alpha}}{n} \right) \minus
	  (n\minus k) \left(  \frac{\bar{\alpha}}{n} \right) \log \left(  \frac{\bar{\alpha}}{n} \right)
		   ,
	\end{multline}
	where $\alpha$ is the positive root of \eqref{eq:comib_alpha_eta2} and $\beta$ is the parameter of the negative Hamming \acrshort{pmf} \eqref{eq:negative_hamming_definition} with entropy $\eta_1$.
	If $ n=3 $, then
	\begin{equation}
		R^{\scaleto{\text{mod}}{4pt}} (\eta_{1},\eta_{2}) \geq (1+\beta)   h_b \left(  \frac{1\minus \alpha}{3}    \right) +(1+\beta)  \left(   1\minus \frac{1\minus \alpha}{3}  \right) \minus \beta \eta_2,
	\end{equation}
	where $\alpha$ is the positive root of \eqref{eq:comib_alpha_eta2} and $\beta$ is the parameter of the negative Hamming \acrshort{pmf} \eqref{eq:negative_hamming_definition} with entropy $\eta_1$.
	If $n > 3$, then
	\begin{equation}
		R^{\scaleto{\text{mod}}{4pt}} (\eta_{1},\eta_{2})  \geq \minus  \left( \alpha \beta + \frac{\overline{\alpha \beta}}{n}  \right)  \log  \left( \alpha \beta  + \frac{\overline{\alpha \beta}}{n}  \right) \minus  \frac{(n \minus 1)\overline{\alpha \beta}}{n} \log 	\frac{\overline{\alpha \beta}}{n},
    \end{equation}
    with $\alpha$ and $\beta$ being the positive roots of \eqref{eq:comib_alpha_eta2}, and  \eqref{eq:comib_beta_eta1}.
\end{theorem}
The proof of this theorem is relegated to \autoref{section:oblib_proof_of_main_result_bounds}.




    

Finally, we consider the high-SNR regime, namely the scenario where $\eta_1 $ is small. In such case we have the following characterization of the optimal distributions and rate.
\begin{theorem} \label{theorem:comib_modulo_high_snr}
    Suppose $\eta \ll 1$,
    then
    \begin{equation}
        	R^{\scaleto{\text{mod}}{4pt}} (\eta_{1},\eta_{2}) -\eta_2 = \alpha \beta \log \left(1+\frac{\alpha n}{1-\alpha}\right)\cdot (1+o(1)),
    \end{equation}
    with $\alpha$ and $\beta$ being the positive roots of \eqref{eq:comib_alpha_eta2}, and  \eqref{eq:comib_beta_eta1}, and $o(1)$ vanishes when $\eta_1\downarrow 0$. Asymptotically, optimal $\mathsf{P}_{\rv{W}}$ and $\mathsf{P}_{\rv{V}}$ are both positive Hamming distributions satisfying the constraints with equality.
\end{theorem}
The proof of this theorem is relegated to \autoref{section:oblib_proof_of_high_modulo}.
	
	\section{Modulo Additive Channels with \acrshort{tv} Constraint}
	\label{section:tv_constraint}

Let $\delta\in(0,2)$ be given, and a nominal  modulo additive
channel represented by $P_{\rv{W}}^{(0)}$. In this section, the constraint $H(\rv{W}) \leq \eta_1$ in $\set{P}_{\set{X}\set{Y}}$ from the previous section is replaced with the constraint 
$d_{\text{TV}} (\mathsf{P}_{\rv{W}},\mathsf{P}_{\rv{W}}^{(0)})\leq\delta$ (the set $\set{Q}_{\set{Z}|\set{Y}}$ remains the same). We denote the resulting \acrshort{comib} value as $R^{\text{TV}}(\delta,\eta_{2})$. 

A natural approach is to relate $R^{\text{TV}}(\delta,\eta_{2})$ to the standard
bottleneck problem $R(0,\eta_{2})\equiv R_{T}^{\text{CEB}}(\eta_{2})$
via the continuity of entropy in the total variation metric. This
idea was used, e.g., in \cite{Shamir2010}, to establish
generalization bounds for the bottleneck problem, that is, in the
regime of vanishing $\delta$. Here, we present a tighter
result, valid for any $\delta\in(0,1)$. To this end, recall that
the entropy difference of two \acrshort{pmf}s in $\Delta_{n}$ of total variation
$\delta$ is bounded by 
$
\omega(\delta,n) \triangleq \frac{1}{2}\delta\log(n-1)+h_{b}\left(\frac{\delta}{2}\right)
$ \cite{Audenaert2006,Zhang2007}.\looseness=-1
\begin{proposition}
\label{prop: entropy continuity bound}For any $\delta\in(0,1)$
\begin{equation}
\left|R^{\text{TV}}(\delta,\eta_{2})-R_{T}^{\text{CEB}}(\eta_{2}))\right|\leq\omega(\delta,n),
\end{equation}
where $R_{T}^{\text{CEB}}(\eta_{2})$ from \eqref{eq:ssib_problem_definition} is computed at $\mathsf{P}_{\rv{W}}^{(0)}$. 
\end{proposition}

Proposition \ref{prop: entropy continuity bound} relates the compound
IB to the standard IB problem, however, the latter is, in general,
difficult to compute (and requires, for example, an alternating
minimization algorithm, as in \autoref{section:alternating_algorithm}). In what
follows, we will state computable upper and lower bounds to $R^{\text{TV}}(\delta,\eta_{2})$. To this end, let $T$ be a channel
transition matrix, and let $\theta(T)\in[0,1]$ be the Dobrushin contraction
coefficient of $T$ \cite{dobrushin1956central} 
\begin{align}
\theta(T) &\triangleq \max_{\bv{p},\bv{q}\in\simplex{n}\colon \bv{p}\neq \bv{q}}\frac{d_{\text{TV}} (T\bv{p},T\bv{q})}{d_{\text{TV}} (\bv{p},\bv{q})} \\
&= \frac{1}{2}\max_{i,i'\in[n]\colon i\neq i'}d_{\text{TV}} (T_{i},T_{i'}),
\end{align}
where $T_{i}$ is the $i$th row of $T$ (the second inequality is a "two-point characterization"). 
Thus, at worst case, the computation of $\theta(T)$
requires $n^{2}-n$ total variation distance calculations.
Furthermore, if $T\in[0,1]^{n\times n}$ is obtained by $n$ permutations
of a \acrshort{pmf}, then only $n-1$ total variation distance calculations are
required. Second, let 
$
\Gamma(\delta)\triangleq\min_{\bv q\in\Delta_{n}\colon d_{\text{TV}} (\bv q,\bv{u}_n)\leq\delta}H(\bv q)
$
be the minimal entropy over a total variation ball centered at $\bv{u}_n$.
This problem has a closed-form solution \cite[Thm. 3]{ho2010interplay} as follows: If $1-1/n\leq\delta/2$
then the optimal solution is $\bv{q}=(1,0,\ldots,0)$ and $\Gamma(\delta)=0$.
Otherwise, let $n_{0}(\delta)\triangleq\lfloor n+1-n\delta/2\rfloor$,
and then the optimal solution is 
$\bv q^{*}=\left(1/n+\delta/2,1/n\ldots,1/n,(n-n_{0}(\delta)+1)/n-\delta/2,0,\ldots,0\right)$
(there are $n_0-2$ terms of $1/n$ so the support size of this solution is $n_{0}$). Therefore, for $\delta\in[0,2-2/n]$
the function $\Gamma(\delta)$ is strictly positive and strictly decreasing
with extreme values of $\Gamma(0)=\log n$ and $\Gamma(2-2/n)=0$.
So, there exists an inverse function to $\Gamma(\delta)$, which we
denote by $D(\eta):[0,\log n]\to[0,2-2/n]$. Third, for a given $\bv{p}^{(0)}\in \simplex{n}$, let
$\Phi(\delta;\bv p^{(0)})\triangleq\max_{\bv q\in\Delta_{n}\colon d_{\text{TV}} (\bv q,\bv p^{(0)})\leq\delta}H(\bv q)$
be the maximal entropy over a total variation ball centered at
$\bv{p}^{(0)}$. This problem also has a closed-form solution \cite[Thm. 2]{ho2010interplay}  as follows: Let $\mu$ and $\nu$ be
such that $\sum_{i=1}^{n}(p_{i}^{(0)}-\mu)_{+}=\sum_{i=1}^{n}(\nu-p_{i}^{(0)})_{+}=\delta/2$.
If $\nu\geq\mu$ then $\Phi(\delta;\bv p^{(0)})=\log n$ and the maximizing
distribution $\bv q^{*}=\bv{u}_n$ is uniform. Otherwise, $\bv q^{*}$
is such that $\bv q_{i}^{*}=\min\{\max(p_{i}^{(0)},\mu),\nu\}$, and
its entropy is the maximum. 
\begin{theorem} \label{theorem:comib_total_variation_bounds}
Let $T(\mathsf{P}_{\rv{W}})$ be the channel transition matrix which corresponds
to $n$ cyclic permutations of $\mathsf{P}_{\rv{W}}$. Then,
\begin{equation}
R^{\text{TV}}(\delta,\eta_{2})\geq\max_{\mathsf{P}_{\rv{W}}\colon d_{\text{TV}} (P_{\rv{W}},P_{\rv{W}}^{(0)})\leq\delta}\Gamma\left(\theta(T(\mathsf{P}_{\rv{W}}))\cdot D(\eta)\right),
\end{equation}
and that 
\begin{equation}
R^{\text{TV}}(\delta,\eta_{2})\leq\min_{P_{\rv{V}}\colon H(P_{\rv{V}})=\eta_{2}}\Phi\left(\theta(T(\mathsf{P}_{\rv{V}}))\cdot\delta;T(\mathsf{P}_{\rv{V}})\bv p^{(0)}\right).
\end{equation}
\end{theorem}
Since $\Gamma(\delta)$, its inverse $D(\eta)$, as well as $\Phi(\delta;\bv p^{(0)})$
are all efficiently computable, the expressions in the lower bound can be computed for any given $T(\mathsf{P}_{\rv{W}})$.
In general, the optimization over $\mathsf{P}_{\rv{W}}$ 
in the lower bound is computationally
difficult. However, \emph{any} arbitrary choice of $\mathsf{P}_{\rv{W}}$ which
satisfies the constraint leads to a valid lower bound, and any global optimization algorithm can be used. Analogous statements hold  
for $\mathsf{P}_{\rv{V}}$ in the upper bound. It should be noted that the optimization
of the lower bound requires finding the minimal $\theta(T(\mathsf{P}_{\rv{W}}))$,
whereas $\mathsf{P}_{\rv{V}}$ in the upper bound affects both the contraction coefficient
$\theta(T(\mathsf{P}_{\rv{V}}))$ and the transformed nominal \acrshort{pmf} $T(\mathsf{P}_{\rv{V}})\bv p^{(0)}$. 

Note that as $g_{T}(\eta)\geq\eta$ always holds \cite[Lemma 5 (c)]{Witsenhausen1974}, the lower bound of \autoref{theorem:comib_total_variation_bounds} requires optimizing over $\mathsf{P}_{\rv{W}}$
for which $\theta(T(\mathsf{P}_{\rv{W}}))<1$. In general $\theta(T)<1$ only if
no two rows of $T$ are orthogonal. Here, since the rows of $T(\mathsf{P}_{\rv{W}})$
are circular permutations of $\mathsf{P}_{\rv{W}}$, it holds that $\theta(T)<1$
if and only if the support of $\mathsf{P}_{\rv{W}}$ is strictly larger than $n/2$. 
\begin{remark}
The proof of \autoref{theorem:comib_total_variation_bounds}, given at  \autoref{appendix:oblib_proof_tv_bounds},  provides a lower bound on Witsenhausen's function $g_{T}(\eta)$ from \cite{Witsenhausen1974}, which may be of independent interest. 
\end{remark}

	
	\section{An Alternating Optimization Algorithm} \label{section:alternating_algorithm}
	We return in this section to the general $(C_1,C_2)$ \acrshort{pf} compound set. Applying a two-phase Lagrangian methodology, we obtain a set of self-consistent equations for $ \mathsf{P}_{\rv{X} \rv{Y}} $ and $ \mathsf{P}_{\rv{Z}| \rv{Y}} $. We then propose a Blahuto-Arimoto type iterative algorithm that solves those equations. The proofs are given in Supplementary Material.

\subsection{The Inner Lagrangian}
Fix $ \pmf{\rv{Z}|\rv{Y}} $ that satisfies $ I(\rv{Y};\rv{Z}) \leq C_2 $ and consider the inner minimization problem from \eqref{eq:definition_compound_ib}, given by \eqref{eq:definition_pf},
where the joint \acrshort{pmf} is constrained to have some fixed marginal distributions, namely, there exist $\pmf{\rv{X}}$ and $\pmf{\rv{Y}}$ such that $\sum_{y \in \set{Y}} \pmf{\rv{X} \rv{Y} } (x,y) = \pmf{\rv{X}}(x) $ and $\sum_{x \in \set{X}} \pmf{\rv{X}\rv{Y}} (x,y) = \pmf{\rv{Y}}(y) $.
For $ \lambda_1 \geq 0 $, the respective Lagrangian of the \acrshort{pf} problem \eqref{eq:definition_pf} is given by,
\begin{equation} \label{eq:definition_oblivious_ib_minimization_lagrangian}
	\mathcal{L}_{\min} (\pmf{\rv{X}\rv{Y}} , \lambda_1, \boldsymbol{\mu},\boldsymbol{\nu}) =  I(\rv{X};\rv{Z}) - \lambda_1  I(\rv{X};\rv{Y})
	+ \sum_{x \in \set{X}}  \mu_x  \sum_{y \in \set{Y}} \pmf{\rv{X}\rv{Y}} (x,y)  
	+ \sum_{y \in \set{Y}}  \nu_y  \sum_{x \in \set{X}} \pmf{\rv{X}\rv{Y}} (x,y) 
	.
\end{equation}


\begin{proposition} \label{proposition:oblib_stationarity_inner_minimization}
	Any stationary point $ \pmf{\rv{X}\rv{Y}}^* $ of \eqref{eq:definition_oblivious_ib_minimization_lagrangian} satisfies
	\begin{equation} \label{eq:oblib_stationarity_pXY}
		\mathsf{P}^*_{\rv{X}\rv{Y}} (x,y) =  \frac{\pmf{\rv{X}}(x) \pmf{\rv{Y}}(y) e^{- \beta_1   D\left( \pmf{\rv{Z}|\rv{Y}} (\cdot|y) || \pmf{\rv{Z}|\rv{X}} (\cdot|x)   \right)}}{Z_1 (x,y,\beta_1)},
	\end{equation}
	where $ \beta_1 \triangleq 1/ \lambda_1 $ and $ Z_1(x,y,\beta_1) $ is the proper marginalization function, which verifies that $\pmf{\rv{X}\rv{Y}}$ has the desired marginals $\pmf{\rv{X}}$ and $\pmf{\rv{Y}}$.  Furthermore,  the optimal $ \pmf{\rv{Z}|\rv{X}} (z|x) $ is given by
	\begin{equation} \label{eq:oblib_stationarity_pZgX}
		\pmf{\rv{Z}|\rv{X}} (z|x) = \frac{1}{\pmf{\rv{X}}(x)} \sum_{y \in \set{Y}} \pmf{\rv{Z}|\rv{Y}} (z|y) \mathsf{P}^*_{\rv{X} \rv{Y}} (x,y).
	\end{equation}
\end{proposition}

\begin{remark}
Note that the problem of computing $\rv{Z}_1(x,y,\beta_1)$ is of independent interest. We propose an alternating algorithm which is summarized in \autoref{algorithm:oblib_minimization_marginalization}.
\end{remark}

\begin{algorithm}[t]
	\SetAlgoLined
	\SetKwInput{Init}{Initialize}
	\KwIn{$ \pmf{\rv{X}} $, $ \pmf{\rv{Y}} $, $ \mathsf{Q}_{\rv{X}\rv{Y}} $ and $NumIter$}
	\Init{$ \mathsf{Q}_{\rv{X}\rv{Y}}^{(0)} = \mathsf{Q}_{\rv{X}\rv{Y}}  $.}
	\For{ $ t= 1$ \KwTo $IterNum$}{
		$\mathsf{Q}_{\rv{X}} (x) = \sum_{y \in \set{Y}} \mathsf{Q}_{\rv{X}\rv{Y}}^{(t-1)} (x,y)$ \;
		$\mathsf{Q}_{\rv{Y}} (y) = \sum_{x \in \set{X}} \mathsf{Q}_{\rv{X}\rv{Y}}^{(t-1)} (x,y)$ \;
		$\mathsf{Q}_{\rv{X}\rv{Y}}^{un} (x,y) = \frac{ \pmf{\rv{X}}(x) \cdot \pmf{\rv{Y}}(y) \mathsf{Q}_{\rv{X}\rv{Y}}^{(t-1)} (x,y)}{\mathsf{Q}_{\rv{X}} (x)  \cdot \mathsf{Q}_{\rv{Y}} (y) }$ \;
		$\mathsf{Q}_{\rv{X}\rv{Y}}^{(t)} (x,y) = \frac{\mathsf{Q}_{\rv{X}\rv{Y}}^{un} (x,y)}{\sum_{x,y} \mathsf{Q}_{\rv{X}\rv{Y}}^{un} (x,y)}$
	}
	\KwOut{$\mathsf{Q}_{\rv{X}\rv{Y}}^{(IterNum)} (x,y) $}
	\caption{marginalization(args)}
	\label{algorithm:oblib_minimization_marginalization}
\end{algorithm}

The system of equations characterizing the stationary points in \eqref{eq:oblib_stationarity_pXY} and \eqref{eq:oblib_stationarity_pZgX} must hold simultaneously for consistency. An alternating iteration algorithm is a common approach to solve these equations.

\begin{proposition}
	Equations \eqref{eq:oblib_stationarity_pXY} and \eqref{eq:oblib_stationarity_pZgX} are satisfied simultaneously at the minimum of the Lagrangian \eqref{eq:definition_oblivious_ib_minimization_lagrangian}
	where the minimization is performed independently over the convex sets of $ \{\pmf{\rv{X}\rv{Y}} (x,y)\} $ and $ \{\pmf{\rv{Z}|\rv{X}} (z|x) \} $,
	\begin{equation}
		\min_{\pmf{\rv{Z}|\rv{X}} (z|x) } \min_{\pmf{\rv{X}\rv{Y}} (x,y)} \mathcal{L}_{\min} (\pmf{\rv{X}\rv{Y}} , \lambda_1, \mu,\nu) .
	\end{equation}
	These independent conditions correspond precisely to alternating interactions of \eqref{eq:oblib_stationarity_pXY} and \eqref{eq:oblib_stationarity_pZgX}. Denoting by $ t $ the iteration step, we obtain \autoref{algorithm:oblib_minimization}.
\end{proposition}

\begin{algorithm}[b]
	\SetAlgoLined
	\SetKwInput{Init}{Initialize}
	\KwIn{$ \pmf{\rv{X}} $, $ \pmf{\rv{Y}} $, $ \pmf{\rv{Z}|\rv{Y}} $ and $ \beta_1 $}
	\Init{Arbitrary $ \pmf{\rv{X}\rv{Y}}^{(0)}  $ with valid marginals, $t=1$.}
	\While{Variation in $ I(\rv{X};\rv{Z}) $ is greater then $ \epsilon $}{
		Compute $ \pmf{\rv{Z}|\rv{X}}^{(t)} (z|x) = \frac{\sum_{y \in \set{Y}} \pmf{\rv{Z}|\rv{Y}} (z|y) \pmf{\rv{X} \rv{Y}} ^{(t-1)} (x,y)}{\pmf{\rv{X}}(x)}  $ \;
		Set $ \pmf{\rv{X}\rv{Y}}^{(t)} (x,y) =  \frac{ \pmf{\rv{X}} (x) \pmf{\rv{Y}} (y) e^{- \beta_1    D\left( \pmf{\rv{Z}|\rv{Y}} (\cdot|y) || \pmf{\rv{Z}|\rv{X}}^{(t)} (\cdot|x) \right) }}{Z_1(x,y,\beta_1)} $\;
		Find $ Z_1(x,y,\beta_1) $ s.t. $  \pmf{\rv{X}\rv{Y}}^{(t)}  $ has valid marginals (see \autoref{algorithm:oblib_minimization_marginalization}) \; 
		$t = t+1$;
	}
	\KwOut{$\mathsf{P}^*_{\rv{X}\rv{Y}} $}
	\caption{pf\_iterator(args)}
	\label{algorithm:oblib_minimization}
\end{algorithm}
\subsection{The Outer Lagrangian}
Note that maximization of $ I(\rv{X};\rv{Z}) $ for a fixed $ \mathsf{P}_{\rv{X}\rv{Y}} $ that satisfies $ I(\rv{X};\rv{Y}) \geq C_1 $ is just the standard \acrlong{ib}, the proposed here technique is identical to the one suggested in \cite{Tishby1999}. For completeness, the respective algorithm from \cite[Thm. 5]{Tishby1999} is summarized in \autoref{algorithm:oblib_maximization}.
	\begin{algorithm} [t]
		\SetAlgoLined
		\SetKwInput{Init}{Initialize}
		\SetKwInput{Cmp}{Compute}
		\SetKwInput{Set}{Set}
		\KwIn{$ \mathsf{P}_{\rv{X} \rv{Y}} $,  and $ \beta_2 $}
		\Init{Arbitrary $ \mathsf{P}_{\rv{Z}|\rv{Y}}^{(0)}  $, $ s=1 $.}
		\While{Variation in $ I(\rv{X};\rv{Z}) $ is greater then $ \epsilon $}{
			$ \mathsf{P}_{\rv{Z}|\rv{Y}}^{(s)} (z|y) 
			= \frac{\mathsf{P}_{\rv{Z}}^{(s - 1)} (z)}{Z(y,\beta_2)}  \cdot 
			e^{ \minus \beta_2   D\left( \mathsf{P}_{\rv{X}|\rv{Y}} (\cdot|y) || \mathsf{P}_{\rv{X}|\rv{Z}}^{(s-1)} (\cdot|z) \right) }$\;
			 $ \mathsf{P}_{\rv{Z}}^{(s)} (z) =  \sum_{y \in \set{Y}} \mathsf{P}_{\rv{Y}} (y) \mathsf{P}_{\rv{Z}|\rv{Y}}^{(s-1)} (z|y) $\;
			 $ \mathsf{P}_{\rv{X}|\rv{Z}}^{(s)} (x|z) =  \sum_{y \in \set{Y}} \mathsf{P}_{\rv{X}|\rv{Y}} (x|y) \mathsf{P}_{\rv{Y}|\rv{Z}}^{(s)} (y|z) $\;
			$ s = s+1 $ \;
		}
		\KwOut{$\mathsf{P}^*_{\rv{Z}|\rv{Y}} $}
		\caption{ib\_iterator(args)}
		\label{algorithm:oblib_maximization}
	\end{algorithm}

\subsection{The Compound Algorithm}
We have proposed two algorithms that aim to solve the underlying maximum and minimum optimization problems in a isolated manner. The algorithm we propose for the \acrshort{comib} problem intervenes them together with an objective to find the solution simultaneously. There are two natural approaches to handle this problem. The first one is to alternate between the steps of each algorithm until convergence. The second one is to run the first algorithm until convergence and then the other one, and so on. We have found the second type of algorithms to be more effective, and this is summarized in  \autoref{algorithm:hopping_optimization}.
\begin{algorithm} 
	\SetAlgoLined
	\SetKwInput{Init}{Initialize}
	\SetKwInput{Set}{Set}
	\KwIn{$ \mathsf{P}_{\rv{X}} $, $ \mathsf{P}_{\rv{Y}} $, $ C_1 $ and $ C_2 $}
	\Init{$ \mathsf{P}_{\rv{Z}|\rv{Y}}^{(0)}  $ and $ \mathsf{P}_{\rv{X}\rv{Y}}^{(0)}  $ with valid marginals .}
	\While{Variation in $ I(\rv{X};\rv{Z}) $ is greater then $ \epsilon $}{
		\For{$ \beta_1 \in \field{R}_+ $}{
			$\mathsf{P}^*_{\rv{X}\rv{Y}}(\beta_1^*) = pf\_iterator(\mathsf{P}_{\rv{X}},\mathsf{P}_{\rv{Y}},\mathsf{P}_{\rv{Z}|\rv{Y}}^{(0)},\beta_1) $;
		}
		Find $\mathsf{P}^*_{\rv{X}\rv{Y}}(\beta_1^*) $ s.t. $ I(\rv{X};\rv{Y}) = C_1 $ ;\\
		\Set{$ \mathsf{P}^*_{\rv{X}\rv{Y}} (\beta_1^*) \mapsto \mathsf{P}_{\rv{X}\rv{Y}}^{(0)}$;} 
		\For{$ \beta_2 \in \field{R}_+ $}{
			$\mathsf{P}^*_{\rv{Z}|\rv{Y}}(\beta_2) = ib\_iterator(\mathsf{P}^{(0)}_{\rv{X}\rv{Y}},\beta_2) $;
		}
		Find $\beta_2^* $ s.t. $ I(\mathsf{P}^*_{\rv{Z}|\rv{Y}} (\beta_2^*)) = C_2 $;\\
		\Set{$ \mathsf{P}^*_{\rv{Z}|\rv{Y}} (\beta_2^*) \mapsto \mathsf{P}_{\rv{Z}|\rv{Y}}^{(0)}$ ;}
	}
	\KwOut{$ P^*_{\rv{X}\rv{Y}} $,$ P^*_{\rv{Z}|\rv{Y}} $}
	\caption{\acrshort{comib} Programming}
	\label{algorithm:hopping_optimization}
\end{algorithm}
	
	\section{Alternating Optimization Algorithm for Modulo Additive Channels}
	\label{section:alternating_algorithm_modulo}
	In this section we specialize the alternating algorithm developed in \autoref{section:alternating_algorithm} for the modulo additive channel introduced in \autoref{section:comib_modulo_additive_pf}. In particular, we propose here a method to solve \eqref{eq:definition_compound_ib_modulo}. The proofs are given in Supplementary Material.

\subsection{Maximization Algorithm}
For some fixed $\pmf{\rv{V}}$ that satisfies $H(\pmf{\rv{V}}) \geq \eta_2 $, consider the following maximization problem:
\begin{equation}
    \varphi_T(\eta) \triangleq \max_{
	\scaleto{
	\mathsf{P}_{\rv{W}}\colon  H(\mathsf{P}_{\rv{W}})\leq\eta_{1}}{7pt}
	}
		H(\mathsf{P}_{\rv{W}}*\mathsf{P}_{\rv{V}}).
\end{equation}
The respective Lagrangian is given by
\begin{equation} \label{eq:comib_modulo_max_lagrangian}
    L^{\varphi}(\bv{p},\lambda_1) = -h(T_v \bv{p}) + \lambda_1(h(\bv{p})-\eta_1),
\end{equation}
where $T_v$ is a transition matrix with columns being the cyclic permutations of $\bv{p}_v$. Maximizing $L^{\varphi}(\bv{p},\lambda_1)$ can be given an exact formal solution.
\begin{proposition}
The maximizer of \eqref{eq:comib_modulo_max_lagrangian}, $\bv{p}_w^*$, satisfies
\begin{equation}
    \bv{p}_{w}^* = \frac{e^{\beta_1 T_v^T \log \bv{q}_w}}{Z_1(\beta_1)},
\end{equation}
where $Z_1(\beta_1)$ is the partition function,
and $\bv{q}_w $ is given by
    $\bv{q}_w = T_v \bv{p}_w^*$.
\end{proposition}

The self-consistent 
equations can be turned into converging,
alternating iterations as given in the following proposition.

\begin{proposition}
The set of self-consistent equations is satisfied simultaneously at the maxima of \eqref{eq:comib_modulo_max_lagrangian}, where the maximization is done independently over the convex set of the normalized distributions, $\bv{p}, \bv{q} \in \simplex{n} $. Namely,
\begin{equation}
    \max_{\bv{q} \in \simplex{n}} \max_{\bv{p} \in \simplex{n}} \Phi [\bv{p},\bv{q}] = \max_{\bv{q} \in \simplex{n}} \max_{\bv{p} \in \simplex{n}} h(\bv{q}) - \lambda_1 h(\bv{p}) .
\end{equation}
This maximization is performed by the converging alternating iterations. Denoting by $t$ the iterations step, we obtain \autoref{algorithm:oblib_modulo_maximization}.
\end{proposition}

\begin{algorithm}[b]
		\SetAlgoLined
		\SetKwInput{Init}{Initialize}
		\SetKwInput{Set}{Set}
		\KwIn{$ \bv{p}_v $,  and $ \beta_1 $}
		\Set{$T_v = $ cyclic permutations of $\bv{p}_v$}
		\Init{Arbitrary valid $ \bv{p}^{(0)} \in \simplex{n} $, $\bv{q}^{(0)} = T_v \bv{p}^{(0)}$ $t=1$.}
		\While{Variation in $ h(T_v \bv{p}) $ is greater then $ \epsilon $}{
			Compute $ \bv{p}^{(t)} = \frac{e^{\beta_1 T_v^T \log \bv{q}^{(t-1)}}}{Z_1(\beta_1)} $ \;
			Set $ \bv{q}^{(t)} = T_v \bv{p}^{(t)} $\;
			$t = t+1$;
		}
		\KwOut{$\bv{p}_w^* = \bv{p}^{(t-1)} $}
		\caption{pf\_modulo\_iterator(args)}
		\label{algorithm:oblib_modulo_maximization}
\end{algorithm}

\subsection{Minimization Algorithm}
In a very similar manner, fix $\pmf{\rv{W}}$ that satisfies $H(\pmf{\rv{W}}) \leq \eta_1 $, and consider the respective minimization problem, namely, \acrshort{ib} for modulo additive channels:
\begin{equation}
    g_T(\eta) \triangleq \min_{
	\scaleto{
	\mathsf{P}_{\rv{V}}\colon  H(\mathsf{P}_{\rv{V}})\geq\eta_{2}}{7pt}
	}
		H(\mathsf{P}_{\rv{W}}*\mathsf{P}_{\rv{V}}).
\end{equation}
The respective Lagrangian is given by
\begin{equation} \label{eq:comib_modulo_min_lagrangian}
    L^{g}(\bv{p},\lambda_2) = h(T_w \bv{p}) + \lambda_2(\eta_2- h(\bv{p})),
\end{equation}
where $T_w$ is a transition matrix with columns being the cyclic permutations of $\bv{p}_w$. Minimizing $L^{g}(\bv{p},\lambda_2)$ can be given an exact formal solution.
\begin{proposition}
The minimizer of \eqref{eq:comib_modulo_min_lagrangian},  $\bv{p}_{v}^*$,  satisfies
\begin{equation}
    \bv{p}_{v}^* = \frac{e^{\beta_2 T_w^T \log \bv{q}_v}}{Z_2(\beta_2)},
\end{equation}
where $Z_2(\beta_2)$ is the partition function, and $\bv{q}_v $ is given by
    $\bv{q}_v = T_w \bv{p}_v^*$.
\end{proposition}

The self-consistent equations can be turned into converging, alternating iterations as given in the following proposition.

\begin{proposition}
The set of self-consistent equations are satisfied simultaneously at the minima of \eqref{eq:comib_modulo_min_lagrangian}, where the minimization is done independently over the convex set of the normalized distributions, $\bv{p}, \bv{q} \in \simplex{n} $. Namely,
\begin{equation}
    \min_{\bv{q} \in \simplex{n}} \min_{\bv{p} \in \simplex{n}} \mathcal{G} [\bv{p},\bv{q}] = \min_{\bv{q} \in \simplex{n}} \min_{\bv{p} \in \simplex{n}} h(\bv{q}) - \lambda_2 h(\bv{p}) .
\end{equation}
This minimization is performed by the converging alternating iterations. Denoting by $s$ the iterations step, we obtain \autoref{algorithm:oblib_modulo_minimization}.
\end{proposition}

\begin{algorithm}[t]
		\SetAlgoLined
		\SetKwInput{Init}{Initialize}
		\SetKwInput{Set}{Set}
		\KwIn{$ \bv{p}_w $,  and $ \beta_2 $}
		\Set{$T_w = $ cyclic permutations of $\bv{p}_w$}
		\Init{Arbitrary valid $ \bv{p}^{(0)} \in \simplex{n} $, $\bv{q}^{(0)} = T_w \bv{p}^{(0)}$ $t=1$.}
		\While{Variation in $ h(T_w \bv{p}) $ is greater then $ \epsilon $}{
			Compute $ \bv{p}^{(s)} = \frac{e^{\beta_2 T_w^T \log \bv{q}^{(s-1)}}}{Z_2(\beta_2)} $ \;
			Set $ \bv{q}^{(s)} = T_w \bv{p}^{(s)} $\;
			$s = s+1$;
		}
		\KwOut{$\bv{p}_v^* = \bv{p}^{(s-1)} $}
		\caption{ib\_modulo\_iterator(args)}
		\label{algorithm:oblib_modulo_minimization}
\end{algorithm}
 
\subsection{Compound Algorithm}
We combine the maximization and minimization methods into alternating procedure in order to solve \eqref{eq:definition_compound_ib_modulo}, as described in \autoref{algorithm:modulo_hopping_optimization}.

\begin{algorithm} 
	\SetAlgoLined
	\SetKwInput{Init}{Initialize}
	\SetKwInput{Set}{Set}
	\KwIn{$ n $, $ \eta_1 $ and $ \eta_2 $}
	\Init{$ \bv{p}_w^{(0)}  $ and $ \bv{p}_w^{(0)}  $  .}
	\While{Variation in $ h(\bv{p}_w^{(0)}*\bv{p}_v^{(0)}) $ is greater then $ \epsilon $}{
		\For{$ \beta_1 \in \field{R}_+ $}{
			$\bv{p}_w^*(\beta_1) = pf\_modulo\_iterator(\bv{p}_v^{(0)},\beta_1) $;
		}
		Find $\beta_1^* $ s.t. $ h(\bv{p}^*_{w}(\beta_1^*)) = \eta_1 $ ;\\
		\Set{$ \bv{p}^*_{w} (\beta_1^*) \mapsto \bv{p}_{w}^{(0)}$;} 
		\For{$ \beta_2 \in \field{R}_+ $}{
			$\bv{p}^*_{v}(\beta_2) = ib\_modulo\_iterator(\bv{p}^{(0)}_{w},\beta_2) $;
		}
		Find $\beta_2^* $ s.t. $ h(\bv{p}_v (\beta_2^*)) = \eta_2 $;\\
		\Set{$ \bv{p}^*_{v} (\beta_2^*) \mapsto \bv{p}_{v}^{(0)}$;}
	}
	\KwOut{$ \bv{p}_w^*, \bv{p}_v^* $}
	\caption{\acrshort{comib} Modulo Programming}
	\label{algorithm:modulo_hopping_optimization}
\end{algorithm}
	
 	\section{Numerical Simulations}
\begin{figure}[t]
    \centering
    \begin{subfigure}[b]{0.49\textwidth}
        \centering
        \input{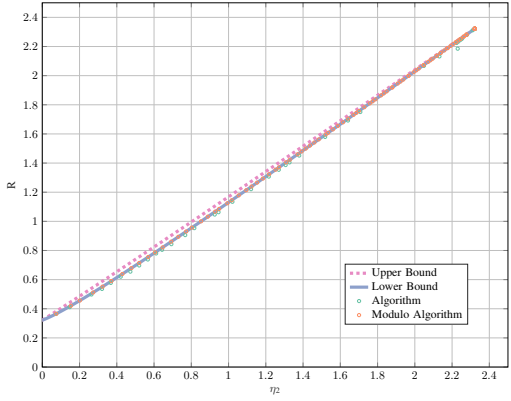}
        \caption{ $\eta_1=0.32$.}
         \label{fig:oblib_bounds_N5_eta1_0p32}
    \end{subfigure}
    \begin{subfigure}[b]{0.49\textwidth}
        \centering
        \input{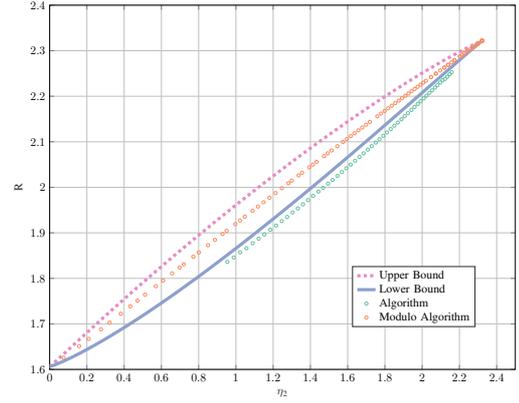}
        \caption{ $\eta_1=1.61$.}
        \label{fig:oblib_bounds_N5_eta1_1p61}
    \end{subfigure}
    \caption{Bounds on COMIB function with \acrshort{pf} constraint for $n=5$.}
    \label{fig:oblib_bounds_N5}
\end{figure}

 	We evaluate both the analytical bounds derived in \autoref{theorem:oblib_main_result_bounds} and the algorithms developed in Sec. \ref{section:alternating_algorithm} and Sec.  \ref{section:alternating_algorithm_modulo} by comparing their results on a common example. A representative examples of $n=5, 10 $ and various rate constraints is shown in Figs. \ref{fig:oblib_bounds_N5},  and \ref{fig:oblib_bounds_N10}. 
As expected, the algorithm's output lies between the upper and lower bounds. It is also  somewhat closer to the lower bound, which hints that lower bound is tighter than the upper bound, and it is the latter that should be improved. In addition, we have  evaluated the algorithm from \autoref{section:alternating_algorithm}, which is not constrained to modulo-additive channels. As expected, better rates are obtained when the constraint is relaxed, but they are only slightly smaller. Furthermore, it is evident that the unconstrained setting has better performance as $n$ grows, indicating that the test-channel can better align its structure in order to approach the bottleneck constraints. The analytical bounds are tight in the extreme points of $\eta_2$ and also for large alphabets.

\begin{figure}
    \centering
%
%
\definecolor{mycolor1}{rgb}{0.30196,0.68627,0.29020}%
\definecolor{mycolor2}{rgb}{0.92900,0.69400,0.12500}%
\definecolor{mycolor3}{rgb}{0.59608,0.30588,0.63922}%
\definecolor{mycolor4}{rgb}{0.30100,0.74500,0.93300}%

\begin{tikzpicture}[scale=0.4, every node/.style={scale=1.5}]

\begin{axis}[%
width=6.028in,
height=4.754in,
at={(1.011in,0.642in)},
scale only axis,
xmin=0,
xmax=4,
xlabel style={font=\color{white!15!black}},
xlabel={$\eta{}_\text{2}$},
ymin=0.5,
ymax=4,
ylabel style={font=\color{white!15!black}},
ylabel={R},
label style={font=\Large},
axis background/.style={fill=white},
axis x line*=bottom,
axis y line*=left,
xtick={0,0.5,...,4},
xmajorgrids,
ymajorgrids,
legend style={at={(0.6,0.2)},anchor=west,legend cell align=left, align=left, draw=white!15!black}
]
\addplot [color=mycolor3,line width=3.0pt]
  table[row sep=crcr]{%
0	0.948413864620117\\
0.434098955067613	1.26384290311001\\
0.868197910135226	1.59073420336227\\
1.30229686520284	1.92898103983228\\
1.73639582027045	2.2573634346068\\
2.17049477533807	2.56986839446496\\
2.60459373040568	2.93862347277558\\
3.03869268547329	3.25566031827402\\
3.47279164054091	3.56614122457228\\
3.90689059560852	3.90689059560852\\
};
\addlegendentry{Lower Bound}

\addplot [color=mycolor2,line width=3.0pt]
  table[row sep=crcr]{%
0	1.76363485638578\\
0.434098955067613	1.93356028546423\\
0.868197910135226	2.14731834985547\\
1.30229686520284	2.38062950110581\\
1.73639582027045	2.62584497009253\\
2.17049477533807	2.87976096302891\\
2.60459373040568	3.12462606167211\\
3.03869268547329	3.38450628870519\\
3.47279164054091	3.64754410050753\\
3.90689059560852	3.90689030255709\\
};
\addlegendentry{Upper Bound}

\end{axis}

\begin{axis}[%
width=7.778in,
height=5.833in,
at={(0in,0in)},
scale only axis,
xmin=0,
xmax=1,
ymin=0,
ymax=1,
axis line style={draw=none},
ticks=none,
axis x line*=bottom,
axis y line*=left,
legend style={legend cell align=left, align=left, draw=white!15!black}
]
\end{axis}
\end{tikzpicture}%
    \caption{Bounds on COMIB function with \acrshort{tv} constraint for $n=15$ and $\delta=0.3$.}
    \label{fig:comib_tv_bounds_N15_delta_0p3}
\end{figure}
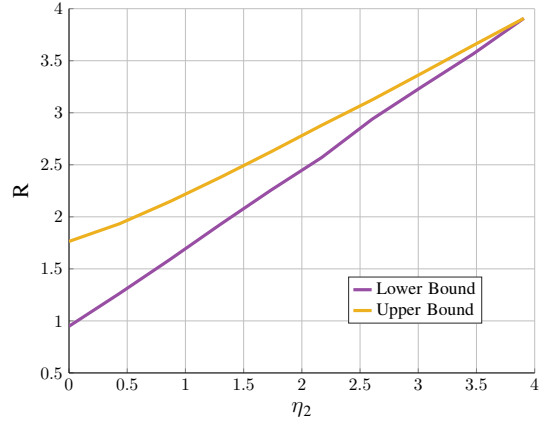

We also evaluate the bounds derived for the \acrshort{tv} class setting in \autoref{section:tv_constraint}. An example for $n=15$, and $\delta = 0.3$, and $\mathsf{P}_{\rv{W}}^{(0)}\propto \exp(2i)$ for $i\in[15]$ (and $0$ otherwise) is illustrated in \autoref{fig:comib_tv_bounds_N15_delta_0p3}. The bounds are fairly close and  tighten for large values of $\eta_2$, but should be tightened for lower values.

\begin{figure}[t]
    \centering
    \begin{subfigure}[b]{0.49\textwidth}
        \centering
        \input{figures/comib_N10_eta1_0p32}
        \caption{ $\eta_1=0.32$.}
         \label{fig:comib_N10_eta1_0p32}
    \end{subfigure}
    \begin{subfigure}[b]{0.49\textwidth}
        \centering
        \input{figures/comib_N10_eta1_2p32}
        \caption{ $\eta_1=2.32$.}
        \label{fig:comib_N10_eta1_2p32}
    \end{subfigure}
    \caption{Bounds on COMIB function with \acrshort{pf} constraint for $n=10$.}
    \label{fig:oblib_bounds_N10}
\end{figure}







\section{Selected Proofs}

	\subsection{Proof of \autoref{proposition:comib_binaryXY}}
	\label{appendix:proof_of_proposition_comib_binaryXY}
	We utilize \autoref{lemma:optimality_saddle_point}. First direction - assuming a particular $\pmf{\rv{X}\rv{Y}}$ and solving the respective maximization problem over $\pmf{\rv{Z}|\rv{Y}}$. Suppose $(\rv{X},\rv{Y})$ is a \acrshort{dsbs} with parameter $\alpha$, then it is known  \cite{Witsenhausen1975} that the optimal $ \mathsf{P}_{\rv{Z}|\rv{Y}} $ in this case is a \acrshort{bsc} with parameter $ \beta =  h_b^{-1} (1-C_2) $.

Opposite direction - fixing a specific $\pmf{\rv{Z}|\rv{Y}}$ and solving the respective minimization problem over $\pmf{\rv{X}\rv{Y}}$. Suppose $(\rv{Y},\rv{Z})$ is a \acrshort{dsbs} with parameter $\beta$. Denoting
    $\alpha_x \triangleq \Prob{\rv{Y}=1|\rv{X}=x}$,
we obtain
    $I(\rv{X};\rv{Z}) = 1- \sum_{x\in \{0,1\}} h_b(\alpha_x*\beta) \pmf{\rv{X}}(x)$,
and
    $I(\rv{X};\rv{Y}) = 1- \sum_{x\in \{0,1\}} h_b(\alpha_x) \pmf{\rv{X}}(x)$.
Thus, this problem is equivalent to the following maximization problem:
\begin{equation}
    \begin{aligned}
    &R(\eta_1) =
    && \underset{\{\alpha_x\}}{\text{maximize}}
    && \Exp{h_b(\alpha_\rv{X}*\beta)} \\
    &&& \text{subject to}
    && \Exp{h_b(\alpha_{\rv{X}})} \leq \eta_1.
    \end{aligned}
\end{equation}
The respective Lagrangian is given by
\begin{equation}
    L(\alpha_0,\alpha_1,\pmf{\rv{X}},\lambda)
    = \Exp{h_b(\alpha_\rv{X}*\beta)}-\lambda\left[ 
    \Exp{h_b(\alpha_{\rv{X}})} - \eta_1
    \right] = \Exp{h_b(\alpha_{\rv{X}}*\beta) - \lambda h_b(\alpha_{\rv{X}}) } + \lambda \eta_1.
\end{equation}
Define
    $f(\alpha) \triangleq h_b(\alpha *\beta) - \lambda h_b(\alpha)$,
and let 
    $\alpha^* \in \argmax_{\alpha \in [0,1]} f(\alpha)$.
Note that
\begin{equation}
    f(\Bar{\alpha}) = h_b(\Bar{\alpha}*\beta) - \lambda h_b(\Bar{\alpha}) =  h_b(\alpha*\beta) - \lambda h_b(\alpha) = f(\alpha).
\end{equation}
Thus, $\Bar{\alpha}^*$ also maximizes $f(\alpha)$. Therefore,
\begin{equation}
    L(\alpha_0,\alpha_1,\pmf{\rv{X}},\lambda)
    = \Exp{f(\alpha_{\rv{X}})} + \lambda \eta_1
    \leq f(\alpha^*) + \lambda \eta_1,
\end{equation}
with equality when $\alpha_0 = 1-\alpha_1 = \alpha^*$ and $\rv{X} \sim \Ber (0.5)$. Finally note that $\rv{X} \sim \Ber(0.5) $ and $\rv{Y} \sim \Ber(0.5) $ restrict $(\rv{X},\rv{Y})$  to a \acrshort{dsbs} with parameter $\alpha$, thus completing the proof.


	\subsection{Proof of \autoref{theorem:oblib_gaussianXY}} \label{section:oblib_proof_of_theorem_gaussianXY}
	We utilize the saddle point property of \autoref{lemma:optimality_saddle_point}. Assume that $ (\rv{X},\rv{Y}) $ are jointly Gaussian with covariance matrix
\begin{equation}
	K_{\rv{X}\rv{Y}} = 
	\begin{pmatrix}
		1 & \rho_1 \\
		\rho_1 & 1
	\end{pmatrix}.
\end{equation}

Due to \cite[Thm. 7.1]{Guo2013} and \cite{Chechik2005}, the solution to \eqref{eq:definition_ib}
 is a Gaussian $ f_{\rv{Z}|\rv{Y}} $.
Thus, $ (\rv{Y},\rv{Z}) $ are also jointly Gaussian 
with covariance matrix
\begin{equation}
	K_{\rv{Y}\rv{Z}} = 
	\begin{pmatrix}
		1 & \rho_2 \\
		\rho_2 & 1
	\end{pmatrix},
\end{equation}
where
	$\rho_2^2 = 1-2^{-2 C_2}$.
This further implies that $ (\rv{X},\rv{Y},\rv{Z}) $ are jointly Gaussian. 
Thus
\begin{equation}
	\max_{f_{\rv{Z}|\rv{Y}} \colon I(\rv{Y};\rv{Z}) \leq C } I(\rv{X};\rv{Z})= \frac{1}{2} \log \frac{1}{1-\rho_1^2 \cdot \rho_2^2}.
\end{equation}

Now consider the opposite direction of the saddle point property. Suppose that the optimal channel from $ \rv{Y} $ to $ \rv{Z} $ is a Gaussian channel, i.e., there exists $ \rv{V} \sim \mathcal{N}(0,1) $, independent of $\rv{Y}$, such that
	$\rv{Z} =  \rho_2 \rv{Y} + \sqrt{1-\rho_2^2} \rv{V}$,
where
	$\rho_2^2 = 1-2^{-2 C_2}$.
We aim to solve the following minimization problem:
\begin{equation}
    \begin{aligned}
    &\underset{f_{\rv{X}\rv{Y}}}{\text{minimize}} &&I(\rv{X};\rv{Z}) \\
    &\text{subject to} 
    && \Exp{\rv{X} \rv{Y}} \geq \rho_1. 
    \end{aligned}
\end{equation}
We proceed to lower bound $I(\rv{X};\rv{Z})$ from below:
\begin{align}
    I(\rv{X};\rv{Z}) 
    &= h(\rv{Z}) - h(\rv{Z}|\rv{X}) \\
    &\geq \frac{1}{2} \log 2\pi e - \frac{1}{2} \log 2\pi e \Exp{(\rv{Z} - \Exp{\rv{Z}|\rv{X}})^2} \\
    &\eqann{\geq}{a} \frac{1}{2} \log 2\pi e - \frac{1}{2} \log 2\pi e \Exp{(\rv{Z} - \Exp{\rv{Z}\rv{X}} \cdot \rv{X})^2} \\
    &= \frac{1}{2} \log \frac{1}{\Exp{(\rho_2\rv{Y} + \sqrt{1-\rho_2^2}\rv{V} -\rho_2 \Exp{\rv{X}\rv{Y}} \rv{X})^2}} \\
    &= \frac{1}{2} \log \frac{1}{1 - \rho_2^2 (\Exp{\rv{X}\rv{Y}})^2} \\
    &\geq \frac{1}{2} \log \frac{1}{1-\rho_1^2 \rho_2^2},
\end{align}
where \eqannref{a}  follows since the optimal MMSE estimator of $\rv{Z}$ given $\rv{X}$ has lower error than the linear estimator.
This lower bound can attained by taking 
 $(\rv{X},\rv{Y})$ jointly Gaussian with correlation $\rho_1$. 
 
Summarizing the above, we have shown that if $(\rv{X},\rv{Y})$ are jointly Gaussian with correlation $\rho_1$ then the maximum of $I(\rv{X};\rv{Z})$ is attained with jointly Gaussian $(\rv{Y},\rv{Z})$ satisfying $I(\rv{X};\rv{Z}) = C_2$. We have also shown that assuming that $(\rv{Y},\rv{Z})$ are jointly Gaussian satisfying $I(\rv{Y};\rv{Z}) = C_2$, then jointly Gaussian $(\rv{X},\rv{Y})$ with correlation $\rho_1$ minimize $I(\rv{X};\rv{Z})$. Hence, by the saddle point property, they are the optimal choice for the problem. This completes the proof of the theorem.

 	\subsection{Privacy Funnel for Jointly Gaussian Vectors}
 	\label{appendix:pf_vector_gaussian}
 	\begin{theorem} \label{theorem:comib_pf_vector_gaussian}
    Suppose $\bv{X} \rightarrow \bv{Y} \rightarrow \bv{Z}$ constitute a jointly Gaussian vector Markov chain with positive definite marginal covariance matrices $\Sigma_{\bv{X}}$, $\Sigma_{\bv{Y}}$, and $\Sigma_{\bv{Z}}$ respectively, and that the cross-covariance matrix of $\bv{Z}$ and $\bv{Y}$ is given by $\Sigma_{\bv{Z}\bv{Y}}$.
    Denote by $\Sigma_{\bv{Y}\bv{X}} $ the cross-covariance matrix of the optimal solution to the \acrshort{pf} problem \eqref{eq:definition_pf}.
    Further, let $U_1^T \Lambda V_1$ be the \acrfull{svd} of $\Sigma_{\bv{Y}}^{-1/2} \Sigma_{\bv{Y}\bv{X}} \Sigma_{\bv{X}}^{-1/2}$ and $U_2^T \Gamma V_2$ be the \acrshort{svd} 
    of $\Sigma_{\bv{Z}}^{-1/2} \Sigma_{\bv{Z}\bv{Y}} \Sigma_{\bv{Y}}^{-1/2}$.
    
    The underlying Gaussian \acrshort{pf} problem can be relaxed to the following optimization problem:
    \begin{equation}
        \begin{aligned}
            &R^{\mathsf{PF}-G}(C_1) =
            && \underset{U_1 \in \set{U}(N), \{\lambda_{i}\}}{ \text{minimize}}
            && - \frac{1}{2} \log \det( I
     - V_2^T \Gamma^2 V_2 U_1^T \Lambda^2 U_1 
    ) \\
            &&& \text{subject to} 
            &&  -\sum_{i=1}^N  \frac{1}{2} \log(1-\lambda_i^2) = C_1,
        \end{aligned}
    \end{equation}
    where $\set{U}(N)$ is the set of all $N \times N$ unitary matrices, called the unitary group, and $\{\lambda_i\}$ are the entries of the diagonal matrix $\Lambda$.
\end{theorem}

\begin{proof}
Suppose that $\bv{Y}$ and $\bv{Z}$ are jointly Gaussian vectors with covariance matrix $\Sigma_{\bv{Z}\bv{Y}}$, then there exists $\bv{V} \sim \mathcal{N}(0,\Sigma_{\bv{V}})$ with $\Sigma_{\bv{V}} = \Sigma_{\bv{Z}} - \Sigma_{\bv{Z}\bv{Y}} \Sigma_{\bv{Y}}^{-1} \Sigma_{\bv{Z}\bv{Y}}^T $ such that
	$\bv{Z} = \Sigma_{\bv{Z}\bv{Y}} \Sigma_{\bv{Y}}^{-1} \bv{Y} + \bv{V}$.
Consider the \acrshort{svd}
of $\Sigma_{\bv{Z}} ^{-1/2} \Sigma_{\bv{Z} \bv{Y}} \Sigma_{\bv{Y}} ^{-1/2} = U_2^T \Gamma V_2$ where $U_2$ and $V_2$ are two orthogonal matrices and $\Gamma$ is a diagonal matrix with singular values on the diagonal.

We further define the following transformations $\tilde{\bv{Z}} = \tilde{T}_z \bv{Z}$ and $\tilde{\bv{Y}} = \tilde{T}_y \bv{Y}$, where $\tilde{T}_z = U_2 \Sigma_{\bv{Z}} ^{-1/2}$ and $\tilde{T}_y = V_2 \Sigma_{\bv{Y}} ^{-1/2}$. Note that
\begin{align}
    \Sigma_{\tilde{\bv{Z}}} &= \tilde{T}_z \Sigma_{\bv{Z}} \tilde{T}_z^T = I_{n_z}, \\
    \Sigma_{\tilde{\bv{Y}}} &= \tilde{T}_y \Sigma_{\bv{Y}} \tilde{T}_y^T = I_{n_y}, \\ 
    \Sigma_{\tilde{\bv{Z}} \tilde{\bv{Y}}} &= \tilde{T}_z \Sigma_{\bv{Z} \bv{Y}} \tilde{T}_y^T = U_2 \Sigma_{\bv{Z}} ^{-1/2} \Sigma_{\bv{Z} \bv{Y}} \Sigma_{\bv{Y}} ^{-1/2} V_2^T = \Gamma.
\end{align}

We are interested in the PF optimization problem from \eqref{eq:definition_pf},
which is a minimization of convex function over the complement of an open convex set, therefore the minimum is obtained on the boundary of the set.

Since $\bv{Y}$ and $\bv{X}$ are jointly Gaussian, there exists $\bv{W} \sim \mathcal{N}(0,\Sigma_{\bv{Y}} - \Sigma_{\bv{Y}\bv{X}} \Sigma_{\bv{X}}^{-1} \Sigma_{\bv{Y}\bv{X}}^T)$ such that
    $\bv{Y} = \Sigma_{\bv{Y}\bv{X}} \Sigma_{\bv{X}}^{-1} \bv{X} + \bv{W}$.
Furthermore, considering the singular value decomposition of
$\Sigma_{\bv{Y}}^{-1/2}\Sigma_{\bv{Y}\bv{X}} \Sigma_{\bv{X}}^{-1/2} = U_1^T \Lambda V_1$,
the rate constraint obtains the following form:
\begin{align}
    I(\bv{Y};\bv{X}) 
    &= h(\bv{Y}) - h(\bv{W}) \\
    &= \frac{1}{2} \log \frac{\det \Sigma_{\bv{Y}}}{\det (\Sigma_{\bv{Y}} - \Sigma_{\bv{Y}\bv{X}} \Sigma_{\bv{X}}^{-1} \Sigma_{\bv{Y}\bv{X}}^T) } \\
    &=- \frac{1}{2} \log \det (I - \Lambda^2) \\
    &=-\sum_{i=1}^n \frac{1}{2} \log(1-\lambda_i^2),
\end{align}
where we identify $C_{1i} \triangleq - \frac{1}{2} \log(1-\lambda_i^2) $.
Next, consider the objective function. Note that
\begin{equation}
    \bv{Z} = \Sigma_{\bv{Z}\bv{Y}} \Sigma_{\bv{Y}}^{-1} \Sigma_{\bv{Y}\bv{X}} \Sigma_{\bv{X}}^{-1} \bv{X} + \Sigma_{\bv{Z}\bv{Y}}  \Sigma_{\bv{Y}}^{-1} \bv{W} + \bv{V},
\end{equation}
and so,
\begin{align}
    I(\bv{Z};\bv{X}) 
    &= h(\bv{Z}) - h( \Sigma_{\bv{Z}\bv{Y}} \Sigma_{\bv{Y}}^{-1} \bv{W} + \bv{V}) \\
    &=\frac{1}{2} \log \frac{\det \Sigma_{\bv{Z}}}{\det( 
    \Sigma_{\bv{Z}} -
    \Sigma_{\bv{Z}\bv{Y}}  \Sigma_{\bv{Y}}^{-1}  \Sigma_{\bv{Y}\bv{X}} \Sigma_{\bv{X}}^{-1} \Sigma_{\bv{Y}\bv{X}}^T
     \Sigma_{\bv{Y}}^{-1} \Sigma_{\bv{Z}\bv{Y}}^T
    )}\\
    &= - \frac{1}{2} \log \det( I
     - U_2^T \Gamma V_2 U_1^T \Lambda V_1 V_1^T \Lambda U_1 V_2^T \Gamma U_2^T
    ) \\
    &= - \frac{1}{2} \log \det( I
     - V_2^T \Gamma^2 V_2 U_1^T \Lambda^2 U_1 
    ).
\end{align}
This completes the proof of the theorem.
\end{proof}
	
	\subsection{Proof of \autoref{theorem:oblib_VectorGaussianXY}} \label{section:oblib_proof_of_theorem_VectorgaussianXY}

We extend here \autoref{theorem:oblib_gaussianXY} to a vector setting, by utilizing again the saddle point property from \autoref{lemma:optimality_saddle_point}. 
We begin with the first direction of the saddle point property. Assume that $\Sigma_{\bv{X}} = \Sigma_{\bv{Y}} = I$ and $\Sigma_{\bv{X}\bv{Y}} = \lambda I$. Then, by \cite{Chechik2005} 
we have
\begin{equation}
    C_2 = \sum_{i=1}^n \frac{1}{2} \log \left[ \frac{\lambda^2 (\nu - 1)}{1-\lambda^2} \right]
    = \frac{n}{2} \log \left[ \frac{\lambda^2 (\nu - 1)}{1-\lambda^2} \right].
\end{equation}
Therefore,
    $\nu^* - 1 = \frac{1-\lambda^2}{\lambda^2} 2^{\frac{2C_2}{n}}$,
and the respective \acrshort{ib} rate is given by
    $R(C_2) = -\frac{n}{2} \log \left(1-\lambda^2 (1-2^{-\frac{2C_2}{n}})\right)$.
Furthermore, since $I(\bv{X};\bv{Y}) = C_1$, then
    $\lambda^2 = 1-2^{-\frac{2C_1}{n}}$,
and thus,
\begin{equation}
    R(C_1,C_2) \leq -\frac{n}{2} \log \left( 1-(1-2^{-\frac{2C_1}{n}}) (1-2^{-\frac{2C_2}{n}}) \right).
\end{equation}

Now consider the opposite direction of the saddle point property. 
Suppose that 
$\Sigma_{\bv{Y}} = \Sigma_{\bv{Z}} =  I$
and 
$\Sigma_{\bv{Z}\bv{Y}} = \gamma I$. 
Thus, by \autoref{theorem:comib_pf_vector_gaussian}, the \acrshort{pf} problem is given by:
\begin{equation}
    \begin{aligned}
        &\underset{\{\lambda_i\}}{\text{minimize}} 
        && -\sum_{i=1}^n \frac{1}{2} \log (1-\gamma^2 \eta_i) \\
        & \text{subject to} 
        && -\sum_{i=1}^n \frac{1}{2} \log(1-\eta_i) \geq C_1,
    \end{aligned}
\end{equation}
where $\eta_i \triangleq \lambda_i^2$.
The respective Lagrangian has the form
\begin{equation}
    \tilde{L}(\{\eta_i\},\mu) 
    =
    -\sum_{i=1}^n \frac{1}{2} \log (1-\gamma^2 \eta_i) + \mu \left[
    C_1 + \sum_{i=1}^n \frac{1}{2} \log(1-\eta_i)
    \right].
\end{equation}
The KKT conditions are given by:
\begin{itemize}
    \item Stationarity:
    \begin{equation}
        \frac{\partial \tilde{L}}{\partial \eta_i}
        =
        \frac{\gamma^2 }{2(1-\gamma^2 \eta_i)} -  \frac{\mu}{2(1-\eta_i)} = 0,
    \end{equation}
    which implies
        $\eta_i = \frac{\gamma^2-\mu}{\gamma^2 (1-\mu)}$.
    \item Complementary Slackness:
        $\mu \left[
    C_1 + \sum_{i=1}^n \frac{1}{2} \log(1-\eta_i) \right] = 0$.
\end{itemize}
Note that $\mu =0$ gives us $\eta_i = 1 = \lambda_i^2$ which implies an infeasible rate. Therefore we assume $\mu^* > 0$ and we obtain that all $\eta_i = \lambda_i^2$ are equal, where $\mu $ is chosen to satisfy the constraint.

Therefore,
\begin{equation}
    R(C_1,C_2) \geq R^{\text{PF}-G}(C_1)
    = -\frac{n}{2} \log 
    \left[
    1-(1-2^{-\frac{2C_1}{n}}) (1-2^{-\frac{2C_2}{n}})
    \right].
\end{equation}
This completes the proof of the theorem.

	\subsection{Proof of \autoref{theorem:oblib_main_result}}
	\label{section:oblib_proof_of_theorem_low_snr}
	The main idea here is to show that the composition of positive and negative Hamming channels is a saddle point for \eqref{eq:definition_compound_ib_modulo} and then apply \autoref{lemma:optimality_saddle_point}. In particular, assuming that $ \mathsf{P}_{\rv{W}} $ is a negative Hamming channel satisfying $ H(\mathsf{P}_{\rv{W}}) \geq \log (n-1) $, we will first show that a positive Hamming channel $ \mathsf{P}_{\rv{V}} $, which satisfies $ H(\mathsf{P}_{\rv{V}}) = \eta_2 $, is the optimizer of
\begin{equation} \label{eq:oblib_minimize_problem}
	 \min_{\mathsf{P}_{\rv{V}}\colon H(\mathsf{P}_{\rv{V}}) \geq \eta_{2}}  H(\mathsf{P}_{\rv{W}}*\mathsf{P}_{\rv{V}}).
\end{equation}
Then, assuming that $ \mathsf{P}_{\rv{V}} $ is a positive Hamming channel, we will show that it implies the optimizer of
\begin{equation} \label{eq:oblib_maximize_problem}
	\max_{\mathsf{P}_{\rv{W}}\colon H(\mathsf{P}_{\rv{W}}) \leq \eta_{1}}  H(\mathsf{P}_{\rv{W}}*\mathsf{P}_{\rv{V}}),
\end{equation}
is a negative Hamming channel satisfying $ H(\mathsf{P}_{\rv{W}}) = \eta_1 $.
Our proof is based on an auxiliary lemma presented below.

We will begin our discussion by solving a simplified version of our problem, termed here as the optimization kernel. 
The proof follows and extends a basic idea from the proof of  \cite[Lemma 7]{Witsenhausen1974}.
\begin{lemma} \label{lemma:optimization_kernel}
	Let $ \bv{x} = (x_1,x_2,x_3)^T \in \field{R}^3_+ $, and consider the extreme points of the following function
	\begin{equation}
		f(\bv{x}) = -\sum_{i=1}^3 (a x_i + b) \log (a x_i + b),
	\end{equation}
	where $ 0 \neq a \in \field{R} $ and $ b > 0 $, over the set defined by the following system of equations
	\begin{align}
		x_1 \geq x_2 \geq x_3 &\geq 0, \\
		x_1+x_2+x_3 &= c_1 \label{eq:oblib_optimization_kernel_sumX},\\
		-\sum_{i=1}^3 x_i \log x_i &= c_2 \label{eq:oblib_optimization_kernel_entropyX}.
	\end{align}
	The function $ f(\bv{x}) $ decreases as $ x_1 $ increases along the arc in $ \field{R}^3 $ defined by \eqref{eq:oblib_optimization_kernel_sumX} and \eqref{eq:oblib_optimization_kernel_entropyX}. That is  if $ c_2 < \log 2 $, then the maximum is obtained for $ \bv{x} = (x_1^*,c_1-x_1^*,0) $ where $ x_1^* $ is the root of
	\begin{equation} \label{eq:kernel_root1}
		c_2+x_1\log x_1 + (c_1-x_1) \log (c_1-x_1) = 0.
	\end{equation}
	Otherwise, if $ c_2 \geq \log 2 $, then the maximum is obtained for $ \bv{x} = (x_1^*,x_1^*, c_1-2x_1^*) $ where $ x_1^* $ is the root of
	\begin{equation} \label{eq:kernel_root2}
		c_2+2x_1\log x_1 + (c_1-2x_1) \log (c_1-2x_1) = 0,
	\end{equation}
	and the minimum is obtained for $ \bv{x} = (\hat{x}_1,(c_1-\hat{x}_1)/2,(c_1-\hat{x}_1)/2) $ where $ \hat{x}_1 $ is the root of
	\begin{equation} \label{eq:kernel_root3}
		c_2+x_1\log x_1 + (c_1-x_1) \log (c_1-x_1)/2 = 0.
	\end{equation}
\end{lemma}

\begin{proof}
	The relations  \eqref{eq:oblib_optimization_kernel_sumX} and \eqref{eq:oblib_optimization_kernel_entropyX} imply
		$\mathrm{d} x_1 + \mathrm{d} x_2 + \mathrm{d} x_3 = 0$, and
		$\sum_{i=1}^3 (1+\log x_i) \mathrm{d} x_i = 0$,
	which further indicate
	\begin{equation}
		\mathrm{d} x_3 = \frac{\log x_1 - \log x_2}{\log x_2 - \log x_3} \mathrm{d}x_1.
	\end{equation}
	Thus
	\begin{align}
		&\mathrm{d} f(\bv{x}) 
		= -\sum_{i=1}^3 a(1+\log(ax_i+b)) \mathrm{d}x_i 
		= -a\sum_{i=1}^3 \log(ax_i+b) \mathrm{d}x_i \\
		&= -a \mathrm{d} x_1 \bigg[
		\log (ax_1+b) - \log (ax_2+b) 
		 - \frac{\log x_1 - \log x_2}{\log x_2 - \log x_3} \left( \log (ax_2+b) - \log (ax_3+b) \right)
		\bigg] \label{eq:oblib_optimization_kernel_df}.
	\end{align}
	Consider the function $ \phi(t) \triangleq a\log (ae^t+b) $. Note that
		$\frac{\mathrm{d} \phi}{\mathrm{d} t} = \frac{a^2 e^t}{a e^t+b}$,
	and
	\begin{equation}
		\frac{\mathrm{d}^2 \phi}{\mathrm{d} t^2} = \frac{a^2 e^t(a e^t+b) - a^3 e^{2t}}{(a e^t+b)^2} = \frac{a^2 b e^t }{(a e^t+b)^2} > 0,
	\end{equation}
	where the last inequality follows since $ b > 0 $. Thus $ \phi(t) $ is a convex function and by Chordal Slope Lemma \cite[Ch. 6.6]{Royden2010} we have with $ t_i = \log x_i $ that
	\begin{equation}
		\frac{a  \log  (  ax_1 +  b  )  -   a  \log ( ax_2  +  b  )}{\log x_1 - \log x_2} \geq  \frac{a  \log   (  ax_2  +  b  )  -   a \log  ( ax_3  +  b )}{\log x_2 - \log x_3}.
	\end{equation}
	Plugging this inequality in \eqref{eq:oblib_optimization_kernel_df} implies that the expression inside the brackets is strictly positive, and thus increasing $ x_1 $ results in decreasing of $ f(\bv{x}) $.
	
	If $ c_2 < \log 2 $ the endpoint that corresponds to the maximum value satisfies $ (x_1,c_1-x_1,0) $, where $ x_1 $ can be found by solving \eqref{eq:kernel_root1}.
	If $ c_2 \geq \log 2 $ the endpoint that corresponds to the maximum value satisfies $(x_1,x_1,c_1-2x_1) $ where $ x_1 $ can be found by solving  \eqref{eq:kernel_root2}.
	For any $ c_2 $, the minimum value is obtained for $ (x_1,(c_1-x_1)/2,(c_1-x_1)/2,) $ where $ x_1 $ is found by solving  \eqref{eq:kernel_root3}.
\end{proof}

We proceed to solve the first direction of the saddle point property, i.e., we will solve a maximization problem. The result is summarized in the following proposition.

\begin{proposition} \label{proposition:comib_modulo_maximization}
	Suppose that $ \mathsf{P}_{\rv{V}} $ is a regular $ (\alpha,n) $ Hamming. Then the solution of
	\begin{equation} \label{eq:oblib_maximize_problem2}
		\max_{\mathsf{P}_{\rv{W}}\colon H(\mathsf{P}_{\rv{W}}) \leq \eta_{1}}  H(\mathsf{P}_{\rv{W}}*\mathsf{P}_{\rv{V}}),
	\end{equation}
	is a negative Hamming channel $ (\beta,n,k) $.
\end{proposition}
\begin{remark}
	Note that here we do not have any constraint on $ k $, i.e., this result holds for any entropy constraint (i.e., value of $\eta_1$). 
\end{remark}

\begin{proof}
	The underlying optimization problem is a maximization of a concave function over the complement of a convex set, therefore, the optimal value lies on the boundary of the set, that is, we may restrict to all $ \mathsf{P}_{\rv{W}} $ satisfying $ H(\mathsf{P}_{\rv{W}}) = \eta_1 $.  Since $ h_n\left( \alpha \bv{p} + (1-\alpha) \bv{u}\right) $ and $ h_n(\bv{p}) $ are both invariant under permutations, one may assume $ p_1 \geq p_2 \geq \cdots \geq p_n $.
	Thus, \eqref{eq:oblib_maximize_problem2} may be reformulated in the following standard form:
	\begin{equation} \label{eq:oblib_maximize_problem3}
		\begin{aligned}
			\max_{\bv{p} \in  \field{R}^n} \quad & -\sum_{i=1}^n \left( \alpha p_i + \frac{(1-\alpha)}{n} \right) \log \left( \alpha p_i + \frac{(1-\alpha)}{n} \right) \\
			\textrm{s.t.} \quad & -\sum_{i=1}^n p_i \log p_i = \eta_1,
			\quad \sum_{i=1}^n p_i = 1\\
			& 1\geq p_1 \geq p_{2} \geq \cdots \geq p_n \geq 0  .
		\end{aligned}
	\end{equation}
	
	\begin{itemize}
		\item For $ n=3 $, \eqref{eq:oblib_maximize_problem3} is exactly the problem defined in \autoref{lemma:optimization_kernel} with $ a= \alpha  > 0$, $ b= \bar{\alpha} /n >0$, $ c_1=1 $ and $ c_2 = \eta_1 $, thus
		\begin{itemize}
			\item if $ \eta_1 < \log _2 $, the maximizer is $ \bv{p}^* = (p_1^*,1-p_1^*,0) $, where $ p_1^* = h_b^{-1}(\eta_1) $.
			\item otherwise, if $ \eta_1 \geq \log _2 $, the maximizer is $ \bv{p}^* = (p_1^*,p_1^*,1-2p_1^*) $, where $ p_1^* $ is the root of $ h_b( 2p_1^*) + 2p_1^* \log 2 = \eta_1 $ .
		\end{itemize}
		\item Suppose $ n>3 $ and assume by contradiction that $ \bv{p}^* $ is not a negative Hamming. Thus, there exist $ k_1,k_2,k_3 \in [n] $ with $ 1 > p^*_{k_1} > p^*_{k_2} > p^*_{k_3} >0 $. We will show that the output entropy can be further increased, thus contradicting the optimality of $ \bv{p}^* $. Assume that the remaining indices are kept fixed, thus they contribution to the output entropy is not changed. We are interested in the following problem:
		\begin{equation} \label{eq:oblib_maximize_problem_converse}
			\begin{aligned}
				&\max_{
					[p_{k_1}, p_{k_2}, p_{k_3}]^T
					 \in  \field{R}^3} 
				&& -  \sum_{i=1}^3    \left(  \alpha p_{k_i}   +  \frac{(1-\alpha)}{n}   \right)   \log   \left(   \alpha p_{k_i}   +  \frac{(1-\alpha)}{n}   \right) \\
				&\textrm{subject to} 
				&& -\sum_{i=1}^3 p_{k_i} \log p_{k_i} = \eta_1 + \sum_{i \notin \{k_1,k_2,k_3\} }    p_{i} \log p_{i} \\
				&&& 1\geq p_{k_1} \geq p_{k_2} \geq p_{k_3} \geq 0 , 
				\quad \sum_{i=1}^n p_{k_i} = 1 - \sum_{i \notin \{k_1,k_2,k_3\} } p_{i}.
			\end{aligned}
		\end{equation}
		The problem defined in \autoref{lemma:optimization_kernel} is identical to \eqref{eq:oblib_maximize_problem_converse} with $ a= \alpha  > 0$, $ b= \bar{\alpha} /n >0$, $ c_1=1 - \sum_{i \notin \{k_1,k_2,k_3\} } p_{i}$ and $ c_2 = \eta_1 + \sum_{i \notin \{k_1,k_2,k_3\} } p_{i} $, but it has a different maximizer, thus contradicting the optimality of $ \bv{p}^* $.
	\end{itemize}
	
\end{proof}

Now, we go ahead to solve the reverse direction of the saddle point property, i.e., we will solve a minimization problem. The result is summarized in the following proposition.

\begin{proposition} \label{proposition:comib_modulo_minimization}
	Suppose that $ \mathsf{P}_{\rv{W}} $ is a negative $(\beta,n,n) $ Hamming satisfying $ H(\mathsf{P}_{\rv{W}}) \geq \log (n-1) $. Then, the solution of
	\begin{equation} \label{eq:oblib_minimize_problem2}
	 \min_{\mathsf{P}_{\rv{V}}\colon H(\mathsf{P}_{\rv{V}}) \geq \eta_{2}}  H(\mathsf{P}_{\rv{W}}*\mathsf{P}_{\rv{V}}),
	\end{equation}
	is a regular Hamming $  \mathsf{P}_{\rv{V}} $ with parameter $\alpha$. 
\end{proposition}
The proof of \autoref{proposition:comib_modulo_minimization} is omitted due to space limitation and its resemblance to the proof of \autoref{proposition:comib_modulo_maximization}.

	\subsection{Proof of \autoref{theorem:oblib_main_result_bounds}} \label{section:oblib_proof_of_main_result_bounds}
	This theorem addresses the regime in which $ \rv{W} $ does not have a full support, which occurs when the entropy constraint on $ \rv{W} $ is below $ \log (n-1) $. The respective $(\beta,n,k)$ negative Hamming distribution is given by \eqref{eq:negative_hamming_definition} with $ k < n $.

Choosing $ \mathsf{P}_{\rv{V}} $ as a regular Hamming channel with parameter $ \alpha $ we obtain an upper bound. The resulting maximization problem, which is given by
\begin{equation}
	R(\eta_{1},\eta_{2}) \leq   \max_{\mathsf{P}_{\rv{W}}\colon H(\mathsf{P}_{\rv{W}})\leq\eta_{1}}H(\mathsf{P}_{\rv{W}}*\mathsf{P}_{\rv{V}}),
\end{equation}
yields a negative Hamming channel with parameters $(\beta,n,k)$. 
Plugging the definition of Hamming \acrshort{pmf} \eqref{eq:negative_hamming_definition},
we obtain the following upper bound
\begin{multline}
	R(\eta_{1},\eta_{2}) \leq  \minus \left(\alpha \beta + \frac{\alpha \bar{\beta}}{k} + \frac{\bar{\alpha}}{n} \right) \log \left(\alpha \beta + \frac{\alpha \bar{\beta}}{k} + \frac{\bar{\alpha}}{n} \right) 
	\\ +  (k\minus 1) \left( \frac{\alpha \bar{\beta}}{k} + \frac{\bar{\alpha}}{n} \right) \log \left( \frac{\alpha \bar{\beta}}{k} + \frac{\bar{\alpha}}{n} \right) \minus
	  (n\minus k) \left(  \frac{\bar{\alpha}}{n} \right) \log \left(  \frac{\bar{\alpha}}{n} \right).
\end{multline}

For the lower bound, we first address here the case with $ \log(n-2 ) < \eta_1 < \log (n-1) $.
The respective negative Hamming $ \mathsf{P}_{\rv{W}} $ is given by:
\begin{equation}
	\mathsf{P}_{\rv{W}} = \left[ \frac{\bar{\theta}}{n \minus 2},\frac{\bar{\theta}}{n\minus 2},\dots,\frac{\bar{\theta}}{n\minus 2}, \theta,0 \right] 
 =\left[ \frac{\bar{\theta}}{n\minus 2}\minus \theta + \theta,\frac{\bar{\theta}}{n\minus 2} \minus \theta +  \theta,\dots,\frac{\bar{\theta}}{n\minus 2}\minus \theta + \theta, \theta,0 \right],
\end{equation}
where $ \theta \leq (1-\theta) /(n-2) $ .
This gives rise to the following output probability vector
\begin{align}
	[\bv{q}]_j &= \theta \sum_{i \neq j} p_i + \frac{1-\theta}{n-2}\sum_{i \notin \{j, j+1\}}p_i 
	= \theta (1- p_j) + \left(\frac{1-\theta}{n-2} - \theta\right) ( 1- p_j - p_{j-1}).
\end{align}
Let $ \bv{p}' $ be the left shifted version of $ \bv{p} $, then the output \acrshort{pmf} takes the following form
\begin{align}
	\bv{q} &= \theta (1-\bv{p}) + \left(\frac{1-\theta}{n-2} - \theta\right) (1-\bv{p} - \bv{p}') 
	 = (n-1)\theta \frac{1-\bv{p}}{n-1} + (1-(n-1)\theta) \frac{1-\bv{p} - \bv{p}'}{n-2}.
\end{align}
Since entropy is a concave function, we have
\begin{align}
	h(\bv{q}) 
	&= h\left(  (n \minus 1)\theta \frac{1 \minus \bv{p}}{n \minus 1} + (1 \minus (n \minus 1)\theta) \frac{1 \minus \bv{p}  \minus  \bv{p}'}{n \minus 2} \right) \\
	&\geq  (n \minus 1)\theta h\left( \frac{1 \minus \bv{p}}{n \minus 1}  \right) + (1 \minus (n \minus 1)\theta) h\left(\frac{1 \minus \bv{p}  \minus  \bv{p}'}{n \minus 2} \right).
\end{align}
Thus,
\begin{equation} \label{eq:comib_pf_modulo_lb}
	h(\bv{q}) \geq     (n \minus 1)\theta h\left( \frac{1 \minus \bv{p}}{n \minus 1}  \right) + (1 \minus (n \minus 1)\theta) h\left(\frac{1 \minus \bv{p}  \minus  \bv{p}'}{n \minus 2} \right).
\end{equation}
We have already shown that the minimizer of the first term in RHS of \eqref{eq:comib_pf_modulo_lb} is a regular Hamming channel (See the proof for \autoref{theorem:oblib_main_result}). Therefore, 
\begin{equation} 
	\min_{ \scaleto{\bv{p} \colon h(\bv{p}) \geq \eta_{2}}{7pt}} \left\{  h\left( \frac{1 \minus \bv{p}}{n \minus 1}  \right)  \right\}  
	=   h_2 \left( \frac{1 \minus \alpha}{n} \right) + \left(  1 \minus \frac{1 \minus \alpha}{n} \right) \log (n \minus 1) . 
\end{equation}
As for the second term in the RHS of \eqref{eq:comib_pf_modulo_lb}, we can only solve it exactly for $ n=3 $, where $ 1- \bv{p} - \bv{p} ' = \bv{p}'' $, with $ \bv{p}'' $ being a different cyclic permutation of $ \bv{p} $. In such case we have
\begin{equation}
	\min_{ \bv{p} \colon h(\bv{p}) \geq \eta_{2}}   \left\{h(\bv{p}'')  \right\} = \eta_2.
\end{equation}
Thus for $ n=3 $ we obtain 
\begin{equation}
	\min_{ \scaleto{\bv{p} \colon h(\bv{p}) \geq \eta_{2}}{7pt}} h(\bv{q}) 
	\geq 
	 2\theta   h_2\left(\frac{\bar{\alpha}}{3}\right) + 2 \theta  \left(1\minus \frac{\bar{\alpha}}{3}\right) \log (2) + (1\minus 2\theta)\eta_2.
\end{equation}
By minimax inequality, this is also a lower bound on our problem, i.e.,
\begin{equation}
	R(\eta_1,\eta_2)  \geq    2  \theta   h_2\left(    \frac{\bar{\alpha}}{3}  \right)   +  2 \theta  \left(  1\minus \frac{\bar{\alpha}}{3}  \right) \log (2)   +  (1\minus 2\theta)\eta_2. 
\end{equation}
Using the standard notation given in \eqref{eq:negative_hamming_definition}, we have $ 1-\theta = \frac{1-\beta}{2} $, therefore $  \theta = \frac{1+\beta}{2} $, and thus establishing the lower bound.

For the case $n>3$ we can choose $P_{\rv{W}}$ to be a positive $(\beta,n)$-Hamming channel which has a full support. Plugging the respective Hamming \acrshort{pmf}s from \eqref{eq: Hamming channel}, we obtain the following lower bound:
\begin{equation}
		R(\eta_1,\eta_2)  \geq -  \left( \alpha \beta + \frac{\overline{\alpha \beta}}{n} \right)  \log  \left( \alpha \beta  + \frac{\overline{\alpha \beta}}{n}  \right)  +  \frac{(n \minus 1)\overline{\alpha \beta}}{n} \log 	\frac{\overline{\alpha \beta}}{n},
\end{equation}

where $ \alpha $ is the positive root of \eqref{eq:comib_alpha_eta2} 
and $ \beta $ is the positive root of \eqref{eq:comib_beta_eta1}.

	\subsection{Proof of \autoref{theorem:comib_modulo_high_snr}}
	\label{section:oblib_proof_of_high_modulo}
	Suppose $\eta_1 > 0 $ is small, and assume w.l.o.g. that $ p_1 $ is the largest element of $\bv{p}$. Consider the following bound on the entropy function:
    $\eta_1 \geq  h(\bv{p}) \geq h_b(1-p_1)$.
Thus, 
    $1- p_1 \leq h_b^{-1}(\eta_1) \rightarrow p_1 \geq 1- h_b^{-1}(\eta_1)$,
and therefore, every $\bv{p}_w$ satisfying $h_n(\bv{p}_w) = \eta_1 $ can be written as 
$\bv{p}_w = \bv{e}_1 + \boldsymbol{\epsilon}$, where $\bv{e}_1$ is an extreme point of $\simplex{n}$ and $\bv{e}_n^T \boldsymbol{\epsilon} = 0$, with $\bv{e}_n $ being the all ones vector, 
and the maximal absolute component of $\boldsymbol{\epsilon}$ tends to zero as $\eta_1\downarrow 0$ . Fix $\bv{p}_w$ as above and some $\bv{p}_{v} $, and consider the output distribution $\bv{q}$, given by
    $\bv{q} = \bv{p}_v*\bv{p}_w = \bv{p}_v + \bv{p}_v * \boldsymbol{\epsilon} = \bv{p}_v + T \boldsymbol{\epsilon}$,
where $T$ represents the transition probability matrix of a modulo channel defined by $\bv{p}_v$. Now, utilizing \emph{linear approximation theorem} \cite[Thm. 1.24]{Beck2014}, we obtain
\begin{equation}
    h(\bv{q}) = h(\bv{p}_v) + \nabla_{\boldsymbol{\epsilon}}^T h(\bv{p}_v) T \boldsymbol{\epsilon} + \frac{1}{2} \boldsymbol{\epsilon}^T T^T \nabla^2 h(\boldsymbol{\xi})  T\boldsymbol{\epsilon},
\end{equation}
for some $\boldsymbol{\xi} \in [\bv{p}_v, \bv{q}]$, where $[\bv{p}_v, \bv{q}]$ stands for the line connecting the points $\bv{p}_v$ and $\bv{q}$.

Consider the gradient
    $\nabla_{\bv{q}} h_n(\bv{q}) = -\log \bv{q} - \bv{e}_n$,
thus,
    $\nabla_{\boldsymbol{\epsilon}} h_n(\bv{q}) =  - T^T \left( \log \bv{q} + \bv{e}_n \right)$.
We have obtained the following first order approximation to the output entropy 
\begin{align}
    h_n(\bv{q}) 
    &= h_n(\bv{p}_v) - \left( \log \bv{p}_v + \bv{e}\right)^T T \boldsymbol{\epsilon} + \frac{1}{2} \boldsymbol{\epsilon}^T T^T \nabla^2 h(\boldsymbol{\xi})  T\boldsymbol{\epsilon}\\
    &= h_n(\bv{p}_v) - \left(T^T \log \bv{p}_v\right)^T \boldsymbol{\epsilon} - \bv{e}^T  \boldsymbol{\epsilon} + \frac{1}{2} \boldsymbol{\epsilon}^T T^T \nabla^2 h(\boldsymbol{\xi})  T\boldsymbol{\epsilon}\\
    &= \eta_2 - \left(T^T \log \bv{p}_v\right)^T \boldsymbol{\epsilon} + \frac{1}{2} \boldsymbol{\epsilon}^T T^T \nabla^2 h(\boldsymbol{\xi})  T\boldsymbol{\epsilon} .
\end{align}
We next validate that the matrix norm of the Hessian $\nabla^2 h(\boldsymbol{\xi})$ is bounded. Indeed, note that
\begin{equation}
    \nabla^2_{\boldsymbol{p}} h(\boldsymbol{p}) \bigg|_{\bv{p} = \boldsymbol{\xi}} = \text{diag}(\boldsymbol{\xi}) ^{-1},
\end{equation}
thus if $\boldsymbol{p}$ and $\boldsymbol{q}$ have both full support, then any point $\boldsymbol{\xi}$ on the line that connects them has full support. In the sequel, we will utilize \autoref{lemma:optimality_saddle_point} and take $\boldsymbol{p}_v$ with full support, therefore $\bv{q} = \boldsymbol{p}_v * \boldsymbol{p}_w$ also has full support (convolution of full support vector and nonnegative elements vector).
Hence, the first-order approximation to the original optimization problem at asymptotically low values of $\eta_1$ is 
\begin{equation} 
    f(\eta_1,\eta_2)
    = \max_{\substack{\bv{p}_v \colon\\ h(\bv{p}_v) = \eta_2 }}
    \min_{\substack{\boldsymbol{\epsilon} \colon \\ h(\bv{e}_1 + \boldsymbol{\epsilon}) = \eta_1 }}  
    \left(T^T \log \bv{p}_v\right)^T \boldsymbol{\epsilon} + O(\norm{\boldsymbol{\epsilon}}^2).
\end{equation}
Note that the rows of the matrix $T$ are cyclic permutations of $\bv{p}_v$, thus
\begin{align}
    -[T^T \log \bv{p}_v]_i
    &= -(\Pi_i \bv{p}_v)^T \log \bv{p}_v \\
    &= - \bv{p}_v^{(i)^T}\log \bv{p}_v \\
    &= -\sum_{j=1}^n p_{v,i+j} \log p_{v,j} \\
   k=i+j \rightarrow &= -\sum_{k=1}^n p_{v,k} \log p_{v,k-i} \\
    &= D(\bv{p}_v || \Pi_i \bv{p}_v) + h(\bv{p}_v).
\end{align}
Therefore,
\begin{align}
    \left(T^T \log \bv{p}_v\right)^T \boldsymbol{\epsilon} 
    &= -\sum_{i=1}^n \epsilon_i \left[ D(\bv{p}_v || \Pi_i \bv{p}_v) + h(\bv{p}_v)\right] \\
    &= -\sum_{i=1}^n \epsilon_i  D(\bv{p}_v || \Pi_i \bv{p}_v),
\end{align}
and we have obtained the following relaxed optimization problem:
\begin{equation} \label{eq:comib_high_snr_approximation}
    g(\eta_1,\eta_2)
    = \min_{\bv{p}_v \colon h(\bv{p}_v) = \eta_2 }
    \max_{\boldsymbol{\epsilon} \colon h(\bv{e}_1 + \boldsymbol{\epsilon}) = \eta_1}
    \sum_{i=1}^n \epsilon_i  D(\bv{p}_v || \Pi_i \bv{p}_v).
\end{equation}

Suppose that $\bv{p}_v$ is $(\alpha,n)$-Hamming, then
\begin{equation}
    D(\bv{p}_v || \Pi_i \bv{p}_v) =
    \begin{cases}
     \alpha  \log \left( 1 + \frac{\alpha n}{1-\alpha} \right), & i\neq 0 \\
     0, & i=0.
    \end{cases}
\end{equation}
In such case, since the objective function is symmetric and convex, the optimal $\boldsymbol{\epsilon}$ will have the following form:
\begin{equation} \label{eq:comib_modulo_high_snr_proof_epsilonI}
    \epsilon_i = 
    \begin{cases}
    \theta, & i \neq 0 \\
    -(n-1) \theta, & i= 0.
    \end{cases}
\end{equation}
On the other hand, suppose that $\boldsymbol{\epsilon}$ satisfies \eqref{eq:comib_modulo_high_snr_proof_epsilonI} 
then, the objective function is given by
    $\theta \sum_{i=1}^n   D(\bv{p}_v || \Pi_i \bv{p}_v)$,
which is also minimized by a positive Hamming choice of $\bv{p}_v$. Since the point $(\bv{p}_{w},\bv{p}_v)$, simultaneously minimizes the minimum problem and maximizes the maximum problem, then by \autoref{lemma:optimality_saddle_point}, it is the solution to the minimax 
problem in \eqref{eq:comib_high_snr_approximation}.
	
	\subsection{Proof of \autoref{theorem:comib_total_variation_bounds}}
	\label{appendix:oblib_proof_tv_bounds}
	To prove the lower bound, note that by the minimax inequality  
\begin{equation}
    R(\delta,\eta_{2})\geq\max_{\pmf{\rv{W}}\colon d_{TV}(\pmf{\rv{W}},\pmf{\rv{W}}^{(0)})\leq\delta}\min_{\pmf{\rv{V}} \colon H(\pmf{\rv{V}})\geq\eta_{2}}H(\pmf{\rv{W}}*\pmf{\rv{V}}),
\end{equation}
and so any arbitrary choice of $\pmf{\rv{W}}$ with $d_{TV}(\pmf{\rv{W}},\pmf{\rv{W}}^{(0)})\leq\delta$
leads to a valid lower bound. Fix $\pmf{\rv{W}}$ and denote for brevity
$T\equiv T(\pmf{\rv{W}})$. Then, the lower bound is given by $R(\delta,\eta_{2})\geq g_{T}(\eta_{2})$,
where 
\begin{equation}
g_{T}(\eta)=\min_{\bv p\in\Delta_{n}\colon H(\bv p)\geq\eta}H(T\bv p)
\end{equation}
is Witsenhausen's function from {[}44{]}. Let $\bv{p}^{*}$ the minimizer
of $g_{T}(\eta)$. It then holds
\begin{align}
g_{T}(\eta) & \overset{(a)}{\geq}\min_{\bv p\in\Delta_{n}\colon d_{TV}(\bv p,\bv u)\leq d_{TV}(\bv p^{*},\bv u)}H(T\bv p)\\
 & \overset{(b)}{\geq}\min_{\bv p\in\Delta_{n}\colon d_{TV}(T\bv p,T\bv u)\leq\theta(T)\cdot d_{TV}(\bv p^{*},\bv u)}H(T\bv p)\\
 & \overset{(c)}{\geq}\min_{\bv q\in\Delta_{n}\colon d_{TV}(\bv q,\bv u)\leq\theta(T)\cdot d_{TV}(\bv p^{*},\bv u)}H(\bv q)\\
 & \overset{(d)}{\geq}\min_{\bv q\in\Delta_{n}\colon d_{TV}(\bv q,\bv u)\leq\theta(T)\cdot D(\eta)}H(\bv q)\\
 & =\Gamma\left(\theta(T)\cdot D(\eta)\right),
\end{align}
where $(a)$ follows since $\bv{p}^{*}$ is optimal for $g_{T}(\eta)$,
$(b)$ follows from the definition of $\theta(T)$, $(c)$ follows
by setting $\bv q=T\bv p$ and relaxing the constraint that $\bv q=T\bv p$,
and $(d)$ follows since $H(\bv{p}^{*})\geq\eta$, so that 
\begin{equation}
d_{TV}(\bv p^{*},\bv u)\leq\max_{\bv p\in\Delta_{n}\colon H(\bv p)\geq\eta}d_{TV}(\bv p,\bv u)=D(\eta),
\end{equation}
where the last equality holds for $\eta\in[0,\log n]$ by the definition
of $D(\eta)$ is the inverse of $\Gamma(\delta)$. This lower bound
on $g_{T}(\eta)$ then completes the proof of the lower bound for
$R(\eta_{2})$. 

To prove the upper bound, choose an arbitrary $\pmf{\rv{V}}$ with $H(\pmf{\rv{V}})=\eta_{2}$,
and let $T\equiv T(\pmf{\rv{V}})$ be the corresponding channel matrix. Then,
\begin{align}
R(\delta,\eta_{2}) & \leq\max_{\bv p\in\Delta_{n}\colon d_{TV}(\bv p,\bv p^{(0)})\leq\delta}H(T\bv p)\\
 & \overset{(a)}{\leq}\max_{\bv p\in\Delta_{n}\colon d_{TV}(T\bv p,T\bv p^{(0)})\leq\theta(T)\cdot\delta}H(T\bv p)\\
 & \overset{(b)}{\leq}\max_{\bv q\in\Delta_{n}\colon d_{TV}(\bv q,T\bv p^{(0)})\leq\theta(T)\cdot\delta}H(q)\\
 & =\Phi\left(\theta(T)\cdot\delta;T\bv p^{(0)}\right),
\end{align}
where $(a)$ follows from the definition of $\theta(T)$ and $(b)$
follows by setting $\bv q=T\bv p$ and relaxing the constraint that
$\bv q=T\bv p$.

	

\section{Concluding Remarks}
	In this paper we have introduced the \acrshort{comib} programming problem. As for the  regular, non-compound \acrshort{ib} setting, the underlying optimization problem is non-convex, and no general closed form solution exists. We have thus obtained various characterizations for the binary setting, the Gaussian settings, and derived upper and lower bounds for modulo additive channels with \acrshort{pf} constraints, and with \acrshort{tv} constraints. Under some qualifying conditions, Gaussian distributions and Hamming channels were shown to be extermal. Finally, we have proposed an alternating optimization algorithm that finds a locally optimal solution. 

Future research directly related to the results of this paper, calls for further tightening these bounds, and establishing additional settings in which the optimal channels and representations can be analytically characterized. 
In addition, convergence rates of \acrshort{ib}, \acrshort{pf}, and \acrshort{comib} algorithms remains an open problem.

A plausible different research direction might be to extend the compound setting analysed here to a compound version of the distributed bottleneck problem \cite{Zaidi2020}, and in particular, one that examines a robust oblivious \acrshort{cran} with many users and many relays (where robustness is measured with respect to channel uncertainty). An open problem is whether the white noise channel is still optimal for the compound multiterminal Gaussian setting.
In addition, the compound setting may be combined with the broadcast approach for the  \acrshort{ib} problem \cite{Steiner2021}. 
In this setting, the encoder's goal is to maximize the average serviceable rate, leveraging multilayer coding strategy, that is, to achieve differential communication  rates -- the better the channel is, the higher the rate to the specific user. This encoding strategy can be combined with the worst-case choice of representation studied here. 

Finally, the compound setting discussed here may play a role in  finite-sample analysis of deep-learning algorithms. As said, in real world applications, the true \acrshort{pmf} $\pmf{\rv{X}\rv{Y}}$ is not known, but rather it is to be estimated from finite sample data \cite{Shamir2010}. Although the amount of data required to obtain a good estimation of $\pmf{\rv{X}\rv{Y}}$ is possibly enormous (due to the curse of dimensionality), it is possible that under some settings it is much smaller if only the solution to the  \acrshort{ib} problem is of interest. Compound methodology presented here might be beneficial in providing non-vacuous bounds and robust compression strategies for finite sample scenarios, in the setting where the total variation between the true joint distribution and the estimated distribution is non-vanishing. 
	
	\section*{Acknowledgment}
	The work has been supported by the 
	European Union's Horizon 2020 Research And Innovation Programme,
	grant agreement no. 694630, by the ISF under Grant 1791/17, and by the WIN consortium via the
	Israel minister of economy and science.

	\appendix
	This section contains supplementary material that supports the paper "The Compound Information Bottleneck Outlook", and was less important to be included in the main body of the paper due to space limitations. We provide here a list of proofs and definitions that complement the results in the main article.

	\subsection{Auxiliary Results for the Gaussian setting}
	\label{appendix:ib_scalar_gaussian}
    The following auxiliary result is well known and mentioned here for self-sustainability. For alternative variant of this result, the interested reader is referred to \cite{Guo2013}. The uniqueness of the result and the proof provided here is the specific application of the \acrshort{epi} on the \acrshort{ib} problem (rather than I-MMSE on \acrshort{infcomb} as in \cite{Guo2013}).
\begin{lemma} \label{lemma:comib_scalar_gaussian_ib}
    Suppose that $\rv{X} \rightarrow \rv{Y} \rightarrow \rv{Z}$ constitute a Markov chain, where $\rv{X}$ and $\rv{Y}$ are unit variance jointly Gaussian random variables with correlation $\rho_1$. 
    Then, it holds that the value of the \acrshort{ib} program is \cite{Tishby1999}
    \begin{align}
        R^{IB} (C_2) &=
         \max_{\pmf{\bv{Z}|\bv{Y}} \colon I(\rv{Y};\rv{Z}) = C_2}
         I(\rv{X};\rv{Z}) \\
        &=\frac{1}{2} \log \frac{1}{1-\rho_1^2 \rho_2^2},
    \end{align}
    where $\rho_2^2 = 1-2^{-2C_2}$,
    and the optimizing distribution is a jointly Gaussian triplet $(\rv{X},\rv{Y},\rv{Z})$ with covariance matrix
    \begin{equation}
        \begin{pmatrix}
            1 & \rho_1 & \rho_1 \rho_2 \\
            \rho_1 & 1 & \rho_2 \\
            \rho_1 \rho_2 & \rho_2 & 1
        \end{pmatrix}.
    \end{equation}
\end{lemma}
\begin{proof}
The main tools used in  proof of the lemma are the \acrfull{epi} \cite{Dembo1991}, and the scaling property of the differential entropy function \cite{Cover2006}. The objective function has the following form in the Gaussian case:
\begin{align}
    I(\rv{X};\rv{Z})
    &= h(\rv{X}) - h(\rv{X}|\rv{Z}) \\
    &= \frac{1}{2} \log 2\pi e - h \left( \rho_1 \rv{Y} + \sqrt{1-\rho_1^2} \cdot  \rv{W} \bigg| \rv{Z} \right).
\end{align}
The last term can be bounded from below using \acrshort{epi},
\begin{align}
    h \left( \rho_1 \rv{Y} + \sqrt{1-\rho_1^2} \cdot \rv{W} \bigg| \rv{Z} \right)
    &\geq \frac{1}{2} \log \left( 2^{2 h(\rho_1 \rv{Y}  | \rv{Z})} + 2^{2 h\left(\sqrt{1-\rho_1^2} \rv{W}  | \rv{Z}\right)} \right) \\
    &= \frac{1}{2} \log \left( \rho_1^2 \cdot 2^{2 h( \rv{Y}  | \rv{Z})} +(1-\rho_1^2) \cdot 2^{2 h( \rv{W}  )} \right) \\
    &= \frac{1}{2} \log \left( \rho_1^2 \cdot 2^{2(h(\rv{Y}) -  I( \rv{Y}  ; \rv{Z}))} +(1-\rho_1^2) 2\pi e \right) \\
    &\eqann{=}{a} \frac{1}{2} \log 2\pi e \left( \rho_1^2 2^{- 2C_2} + 1-\rho_1^2 \right),
\end{align}
where \eqannref{a} follows since $h(\rv{Y}) = \frac{1}{2} \log 2\pi e$.
Thus,
\begin{equation} \label{eq:ib_gaussian_ub}
    I(\rv{X};\rv{Z}) \leq \frac{1}{2} \log \frac{1}{1- \rho_1^2(1-2^{-2C_2})}.
\end{equation}
Since the inequality in \eqref{eq:ib_gaussian_ub} follows from \acrshort{epi}, it can be attained with equality if we choose $\rv{Z} \sim \mathcal{N}(0,1)$, $\rv{V} \sim \mathcal{N}(0,1)$ and
\begin{equation}
    \rv{Y} = \rho_2 \rv{Z} + \sqrt{1-\rho_2^2} \rv{V},
\end{equation}
such that
\begin{equation}
    \rv{X} = \rho_1 \rho_2 \rv{Z} + \rho_1 \sqrt{1-\rho_2^2} \rv{V} + \sqrt{1-\rho_1^2} \rv{W},
\end{equation}
with
    $\rho_2^2 = 1-2^{-2C_2}$.
\end{proof}


	\subsection{Information Bottleneck for Jointly Gaussian Vectors}
	\label{appendix:ib_vector_gaussian}
 	\begin{theorem} \label{theorem:ib_jointly_gaussian_vectors}
    Suppose that $\bv{X}$ and $\bv{Y}$ are jointly Gaussian vectors, with positive-definite covariance matrices $\Sigma_{\bv{X}} $ and, respectively, $\Sigma_{\bv{Y}} $, and a cross-covariance matrix $\Sigma_{\bv{X}\bv{Y}}$. Let the $i$th eigenvalue of $\Sigma_{\bv{X}}^{-\frac{1}{2}}  \Sigma_{\bv{X}\bv{Y}} \Sigma_{\bv{Y}}^{-\frac{1}{2}}$ be $d_i$. Assume that $\bv{X} \rightarrow \bv{Y} \rightarrow \rv{Z}$ constitute a Markov chain and consider the following optimization problem 
    \begin{equation}
        \begin{aligned}
        & R^{IB}(C_2) =
        && \underset{f_{\rv{Z}|\bv{Y}}}{\text{maximize}}
        && I(\bv{X};\rv{Z}) \\
        &&& \text{subject to}
        && I(\bv{Y};\rv{Z}) \leq C_2.
        \end{aligned}
    \end{equation}
    Then the maximum is achieved by a jointly Gaussian triple $(\bv{X},\bv{Y},\rv{Z})$ and
    \begin{equation}
        R(C_2) 
        =
        \frac{1}{2} \sum_{i=1}^N
        \log
        \left[
        \frac{\nu-1}{\nu (1-d_i^2)}
        \right]^+,
    \end{equation}
    where the water-filling level $\nu$ is chosen such that
    \begin{equation}
    \sum_{i=1}^N  \frac{1}{2}\log\left[ \frac{d_i^2 (\nu -1)}{1-d_i^2} \right]^+ = C_2.
    \end{equation}
\end{theorem}
This result recovers the \acrshort{ib} curve from \cite[Sec. 2]{Goldfeld2020} and \cite{Meidlinger2014}. The proof we provide here has more information-theoretic flavor, which utilizes information measures, \acrshort{epi}, sufficient statistics and diaganolization. Further, it is more easier generalizable to discrete and continuous time models. The proof is mainly based on ideas from \cite{Tian2009} and we find it more rigorous opposite to \cite{Chechik2005}.

\begin{proof}
Suppose $\bv{X}$ and $\bv{Y}$ are jointly Gaussian  random vectors with covariance matrix $\Sigma_{\bv{X}\bv{Y}}$. It is easy to verify that without loss of generality, we can write
\begin{equation}
    \bv{X} = K_{xy} \bv{Y} + \bv{W},
\end{equation}
where $K_{xy} = \Sigma_{\bv{X}\bv{Y}} \Sigma_{\bv{Y}}^{-1} $  and $\Sigma_{\bv{W}} = \Sigma_{\bv{X}} - \Sigma_{\bv{X}\bv{Y}} \Sigma_{\bv{Y}}^{-1} \Sigma_{\bv{X}\bv{Y}}^T$.

By the \acrfull{svd} theorem \cite{Horn2012}, there exist unitary matrices $U$ and $V$ and a diagonal matrix $D$ such that 
\begin{equation}
    \Sigma_{\bv{X}}^{-\frac{1}{2}}  \Sigma_{\bv{X}\bv{Y}} \Sigma_{\bv{Y}}^{-\frac{1}{2}}  = U^T D V.
\end{equation}
Denote $T_x \triangleq U \Sigma_{\bv{X}}^{-\frac{1}{2}}$, $T_y \triangleq V \Sigma_{\bv{Y}}^{-\frac{1}{2}}$ and let $\tilde{\bv{X}} \triangleq T_x \bv{X}$. Thus, the mapping from $\tilde{\bv{X}}$ to $\bv{X}$ is a bijection. Furthermore $ \Sigma_{\tilde{\bv{X}}}  = I$,
implying $\tilde{\bv{X}}$ is a random Gaussian vector with independent unit variance entries. Similarly defining $\tilde{\bv{Y}} = T_y \bv{Y}$ where $T_y = V \Sigma_{\bv{Y}}^{-\frac{1}{2}}$, we obtain $\tilde{\bv{Y}}$ a random Gaussian vector with unit-variance independent entries, i.e, $\Sigma_{\tilde{\bv{Y}}}  = I$. 

Further note that,
\begin{align}
    \tilde{\bv{X}} 
    &= T_x \bv{X} \\
    &= U \Sigma_{\bv{X}}^{-\frac{1}{2}} \left( K_{xy} \bv{Y} + \bv{W} \right) \\
    &= U \Sigma_{\bv{X}}^{-\frac{1}{2}} \left( \Sigma_{\bv{X}\bv{Y}} \Sigma_{\bv{Y}}^{-1} \bv{Y} + \bv{W} \right) \\
    &= U \Sigma_{\bv{X}}^{-\frac{1}{2}} \Sigma_{\bv{X}\bv{Y}} \Sigma_{\bv{Y}}^{-\frac{1}{2}} V^T V \Sigma_{\bv{Y}}^{-\frac{1}{2}} \bv{Y} + U \Sigma_{\bv{X}}^{-\frac{1}{2}} \bv{W}  \\
    &= U U^T D V V^T \tilde{\bv{Y}} + U \Sigma_{\bv{X}}^{-\frac{1}{2}} \bv{W}  \\
    &= D \tilde{\bv{Y}} + U \Sigma_{\bv{X}}^{-\frac{1}{2}} \bv{W}.
\end{align}
Defining $\tilde{\bv{W}} \triangleq U \Sigma_{\bv{X}}^{-\frac{1}{2}} \bv{W}$, we obtain
\begin{align}
    \Sigma_{\tilde{\bv{W}}}
    &= \Exp{\tilde{\bv{W}}\tilde{\bv{W}}^T}\\
    &= U \Sigma_{\bv{X}}^{-\frac{1}{2}} \Sigma_{\bv{W}} \Sigma_{\bv{X}}^{-\frac{1}{2}} U^T \\
    &= U \Sigma_{\bv{X}}^{-\frac{1}{2}} \left( \Sigma_{\bv{X}} - \Sigma_{\bv{X}\bv{Y}} \Sigma_{\bv{Y}}^{-1} \Sigma_{\bv{X}\bv{Y}}^T
    \right)
    \Sigma_{\bv{X}}^{-\frac{1}{2}} U^T \\
    &= I - D^2.
\end{align}
Thus,
\begin{equation}
    \tilde{\rv{X}}_i = d_i \tilde{\rv{Y}}_i +\sqrt{1-d_i^2} \tilde{\rv{W}}_i.
\end{equation}

Consider the mutual information constraint on the pair $(\bv{Y},\rv{Z})$. Since the transform $T_y$ is full rank, there is no loss of information, i.e.,
\begin{align}
    I(\bv{Y};\rv{Z})
    &= I(\tilde{\bv{Y}};\rv{Z}) \\
    &= h(\tilde{\bv{Y}}) - h(\tilde{\bv{Y}}|\rv{Z}) \\
    &= \sum_{i=1}^N h(\tilde{\rv{Y}}_i) - h(\tilde{\rv{Y}}_i|\tilde{\rv{Y}}^{i-1}, \rv{Z}).
\end{align}
Identifying $\rv{Z}_i \triangleq (\rv{Z},\tilde{\rv{Y}}^{i-1})$ and denoting
$C_{2,i} = I(\tilde{\rv{Y}}_i; \rv{Z}_i) $ we obtain the following representation of the bottleneck constraint
\begin{equation}
    \sum_{i=1}^{n_y} C_{2,i}  \leq C_2, \qquad C_{2,i} = I(\tilde{\rv{Y}}_i; \rv{Z}_i).
\end{equation}
Now consider the objective function. Similarly, since $T_x$ has full rank, there is no loss of information,
\begin{align}
    I(\bv{X};\rv{Z})
    &= I(\tilde{\bv{X}};\rv{Z}) \\
    &= h(\tilde{\bv{X}}) - h(\tilde{\bv{X}}|\rv{Z}) \\
    &= \sum_{i=1}^N h(\tilde{\rv{X}}_i) - h(\tilde{\rv{X}}_i|\tilde{\rv{X}}^{i-1}, \rv{Z})\\
    &\eqann{\leq}{a} \sum_{i=1}^N h(\tilde{\rv{X}}_i) - h(\tilde{\rv{X}}_i|\tilde{\rv{X}}^{i-1},\tilde{\rv{Y}}^{i-1}, \rv{Z}) \\
    &\eqann{=}{b} \sum_{i=1}^N h(\tilde{\rv{X}}_i) - h(\tilde{\rv{X}}_i|\tilde{\rv{Y}}^{i-1}, \rv{Z}) \\
    &= \sum_{i=1}^N h(\tilde{\rv{X}}_i) - h(\tilde{\rv{X}}_i| \rv{Z}_i) \\
    &= \sum_{i=1}^N I(\tilde{\rv{X}}_i;\rv{Z}_i),
\end{align}
where \eqannref{a} follows since conditioning reduces differential entropy, and equality in \eqannref{b} is due to Markov chain $\tilde{\rv{X}}_i \rightarrow (\tilde{\rv{Y}}^{i-1},\rv{Z}) \rightarrow \tilde{\rv{X}}^{i-1}$.
Further, by \autoref{lemma:comib_scalar_gaussian_ib},
\begin{equation}
     I(\tilde{\rv{X}}_i;\rv{Z}_i)
     \leq \frac{1}{2} \log \frac{1}{1-d_i^2 (1-2^{-2C_{2i}})},
\end{equation}
and equality is achieved for $\rv{Z}_i \sim \mathcal{N}(0,1) $, $\rv{V}_i \sim \mathcal{N}(0,1)$ and
\begin{equation}
    \tilde{\rv{Y}}_i = \rho_2 \rv{Z}_i + \sqrt{1-\rho_2^2} \rv{V}_i,
\end{equation}
where $\rho_2^2 = 1-2^{-2C_{2i}}$.
Thus, we have relaxed our original optimization problem to the following one:
\begin{equation}
    \begin{aligned}
    &R(C_2) =
    && \underset{\{C_{2,i} \}}{\text{maximize}}
    && \sum_{i=1}^N \frac{1}{2} \log \frac{1}{1-d_i^2 (1-2^{-2C_{2i}})} \\
    &&& \text{subject to}
    && \sum_{i=1}^N C_{2i} \leq C_2.
    \end{aligned}
\end{equation}
We apply KKT conditions to solve the underlying optimization problem. The respective Lagrangian is given by
\begin{equation}
    L(\{C_{2i}\},\lambda) 
    = \sum_{i=1}^N \frac{1}{2} \log \frac{1}{1-d_i^2 (1-e^{-2C_{2i}})} - \lambda \left(
    \sum_{i=1}^N C_{2i} - C_2
    \right).
\end{equation}
The KKT conditions are given by:
\begin{itemize}
    \item Stationarity:
    \begin{equation}
        \frac{\partial L}{\partial C_{2i}} 
        = \frac{d_i^2 e^{-2C_{2i}}}
        {1-d_i^2(1-e^{-2C_{2i}})} -\lambda = 0.
    \end{equation}
    Thus,
    \begin{equation}
        e^{2C_{2i}} (1-d_i^2)+d_i^2 = \frac{d_i^2}{\lambda} \rightarrow C_{2i} = \frac{1}{2} \ln \frac{d_i^2 (1 -\lambda) }{ \lambda(1-d_i^2)}.
    \end{equation}
    \item Complementary Slackness:
    \begin{equation}
        \lambda \left(
    \sum_{i=1}^N C_{2i} - C_2
    \right) = 0.
    \end{equation}
    \item Constraints:
    \begin{align}
        C_{2i} &\geq 0, \\
        \sum_{i=1}^{N} C_{2,i} &\leq C_2.
    \end{align}
\end{itemize}
Since $\lambda = 0$ is infeasible solution, therefore $\lambda^* > 0 $ and the last constraint must be satisfied with equality, then $\lambda$ is chosen as the solution to
\begin{equation}
    \sum_{i=1}^N C_{2i} = C_2.
\end{equation}
Further denote $\nu = \frac{1}{\lambda}$, the optimal solution has the following water-filling form:
\begin{equation}
    C_{2i} =  \frac{1}{2}\log\left[ \frac{d_i^2 (\nu -1)}{1-d_i^2} \right]^+ .
\end{equation}
Thus,
\begin{equation}
    R(C_2) =
    \sum_{i=1}^N
    \frac{1}{2} 
    \log 
    \frac{1}{1-d_i^2\left(1 -
    \left[ \frac{1-d_i^2}{d_i^2 (\nu -1)} \right]^-
    \right)
    },
\end{equation}
where $\nu$ is chosen to satisfy the rate constraint with equality.
\end{proof}

	\subsection{Proof of Thm. 3}
	\label{section:oblib_proof_of_theorem_gaussianXY_kld}
	We begin with an upper bound. Due to maximin
inequality \cite[Sec. 5.4.1]{Boyd2014}, we have the following bound:
\begin{equation}
    R(\epsilon_1,C_2) = \max_{f_{\rv{Z}|\rv{Y}} \colon I(\rv{Y};\rv{Z}) \leq C_2  }
    \min_{\rv{W} \colon D(\rv{W}||\rv{N}_0) \leq \epsilon_1}
     I(\rv{X};\rv{Z})
     \leq
     \min_{\rv{W} \colon D(\rv{W}||\rv{N}_0) \leq \epsilon_1}
     \max_{f_{\rv{Z}|\rv{Y}} \colon I(\rv{Y};\rv{Z}) \leq C_2  }
     I(\rv{X};\rv{Z}).
\end{equation}
Thus, choosing a specific $\rv{W}$ that satisfies the constraint will also provide an upper bound. We choose $\rv{W}$ as a Gaussian random variable, namely, $\rv{W} \sim \mathcal{N}(0,\sigma^2)$. It follows from the standard scalar Gaussian \acrshort{ib}, that it is optimal to choose $(\rv{X},\rv{Y},\rv{Z})$ jointly Gaussian. In particular, there exist $\rv{Z} \sim \mathcal{N}(0,1)$, $\rv{V} \sim \mathcal{N}(0,1)$, such that
    $\rv{X} = \rv{Y} + \rv{W} = \rho_2 \rv{Z} + \sqrt{1-\rho_2^2 }\rv{V} + \rv{W}$,
where $\rho_2 = 1-2^{-2C_2}$ .
The upper bound in such case is given by
\begin{equation} \label{eq:comib_KL_gaussian_upper_bound_priorOpt}
    R(\epsilon_1,C_2)
    \leq \min_{\sigma^2 \colon D(\rv{W}||\rv{N}_0) \leq \epsilon_1} \frac{1}{2} \log \left(
    \frac{1}{1-(1-2^{-2C_2}) \frac{1}{1+\sigma^2}}
    \right).
\end{equation}
Note that the expression on the RHS of \eqref{eq:comib_KL_gaussian_upper_bound_priorOpt} is decreasing in $\sigma^2$, and so it remains to minimize over the choice of $\sigma^2$. Consider the relative entropy constraint with $\rv{W} \sim \mathcal{N}(0,\sigma_{\rv{W}})^2 $. We have
\begin{equation} \label{eq:comib_KL_gaussian_constraint}
    \epsilon_1
    \geq D(\rv{W}||\rv{N}_0) 
    = \frac{1}{2} \log \frac{\sigma^2}{\sigma_0^2} + \frac{\sigma^2}{2 \sigma_0^2} - \frac{1}{2}.
\end{equation}
Letting $\sigma_*^2$ be the solution of \eqref{eq:comib_KL_gaussian_constraint} with equality, it then follows that
\begin{equation}
    R(\epsilon_1,C_2)
    \leq  \frac{1}{2} \log \left(
    \frac{1}{1-(1-2^{-2C_2}) \frac{1}{1+\sigma^2_*}}
    \right).
\end{equation}

We proceed to develop a lower bound. First note that
\begin{equation}
    R(\epsilon_1,C_2) \geq
    \min_{\rv{W} \colon D(\rv{W}||\rv{N}_0) \leq \epsilon_1}
     I(\rv{X};\rv{Z}),
\end{equation}
for some $f_{\rv{Z}|\rv{Y}}$ that satisfies the \acrshort{ib} constraint. We further choose 
    $\rv{Y} = \rho_2 \rv{Z} + \sqrt{1-\rho_2^2} \rv{V}$, 
where $\rv{Z} \sim \mathcal{N}(0,1)$, $\rv{V} \sim \mathcal{N}(0,1)$, and
    $\rho_2^2 = 1-2^{-2C_2}$.
Consider the objective function,
\begin{align}
    I(\rv{X};\rv{Z})
    &= h(\rv{Y} + \rv{W}) - h(\sqrt{1-\rho_2^2}\rv{V}+\rv{W}) \\
    &= \frac{1}{2} \log \frac{1+\sigma^2}{1-\rho_2^2 + \sigma^2} \\
    &\geq \frac{1}{2} \log \frac{1}{1 - \frac{\rho_2^2}{1+\sigma_*^2} }.
\end{align}
Thus,
    $R(\epsilon_1,C_2) \geq 
    -\frac{1}{2} \log \left(1 - \frac{\rho_2^2}{1+\sigma_*^2} \right)$.

	\subsection{Proof of Prop. 3}
	\label{appendix:modulo_additiveness_Pzgy}
	Let $T$ be the transition probability matrix from $ \rv{Y} $ to $ \rv{X} $ and consider the following optimization problem
\begin{equation}
    \begin{aligned}
    &R^{\text{CEB}}_{\scaleto{T}{4pt}}(\eta) =
    &&  \underset{\bv{p}_{z},q_z}{\text{minimize}} 
    && \sum_{z \in \set{Z}} h_n(T \bv{p}_z) q_z \\
    &&& \text{subject to} 
    && \sum_{z \in \set{Z}} h_n(\bv{p}_z) q_z \geq \eta, \\
    &&&&& \sum_{z \in \set{Z}}  \bv{p}_z q_z = \bv{u}_n.
    \end{aligned}
\end{equation}
By \cite{Witsenhausen1975}, $R(\eta)$ is convex and it suffices to consider $ |\set{Z}| \leq n+1$.

The Lagrangian of the respective problem is given by
\begin{align}
    L\left(\{\bv{p}_z\}, \bv{q}, \lambda \right)
    &= \sum_{z \in \set{Z}} h_n(T \bv{p}_z) q_z + \lambda \left(\eta - \sum_{z \in \set{Z}} h_n(\bv{p}_z) q_z \right)   \\
    &= \sum_{z \in \set{Z}} \left[ h_n(T \bv{p}_z) -  \lambda h_n(\bv{p}_z)  \right] q_z + \lambda \eta,
\end{align}
where $\lambda \geq 0 $ and the set of the Lagrangian parameters is defined over:
\begin{equation}
     \set{F} \triangleq \left\{  \{\bv{p}_z \}_{z\in \set{Z}} \in \simplex{n}, \bv{q} \in \simplex{n}  \colon  \sum_{z \in \set{Z}}  \bv{p}_z q_z = \bv{u}_n \right\}.
\end{equation}
The respective dual objective function is given by
\begin{align}
    q(\lambda) &= \min_{(\bv{p}_z,\bv{q}) \in \set{F}} \left\{ L\left(\{\bv{p}_z\}, \bv{q}, \lambda \right) \right\} \\
    &= \min_{(\bv{p}_z,\bv{q}) \in \set{F}} \left\{ \sum_{z \in \set{Z}} \left[ h_n(T \bv{p}_z) -  \lambda h_n(\bv{p}_z)   \right] q_z \right\} + \lambda \eta.
\end{align}

\begin{proposition}
The solution of the minimization problem defining $q(\lambda)$ is a modulo additive channel from $\rv{Z}$ to $\rv{Y}$.
\end{proposition}

\begin{IEEEproof}
    Let $ \{\hat{\bv{p}}_z \}_{z \in \set{Z} } $ be the solution of the minimization problem above, and assume on the contrary that it does not represents a modulo additive channel.
    Consider the following function:
    \begin{equation}
        \phi_{\lambda}(\bv{p}) \triangleq h_n(\bv{T} \bv{p}) - \lambda h_n(\bv{p}).
    \end{equation}
    Suppose that $ \bv{p}^* $ minimizes $\phi_{\lambda}(\bv{p})$ over the set $ \{\hat{\bv{p}}_z \}_{z \in \set{Z} } $, namely,
    \begin{equation}
        \bv{p}^* \triangleq \argmin_{ \bv{p} \colon \{\hat{\bv{p}}_z \}_{z \in \set{Z} } } \phi_\lambda (\bv{p}).
    \end{equation}
    Since $ T $ is the transition matrix of modulo additive channel it has a symmetry group of size $n$ that consists of cyclic permutation matrices $\{\Pi_k\}_{k=1}^n$. We construct the following set:
    \begin{equation}
        \bv{p}^*_k \triangleq \Pi_k \bv{p}^* \quad \forall k \in [n].
    \end{equation}
    Note that since $ T \Pi_k = \Pi_k T$, we obtain
    \begin{equation}
        h_n(T \bv{p}^*_k) = h_n(T \Pi_k \bv{p}^*) = h_n(\Pi_k T   \bv{p}^*) = h_n(T \bv{p}^*),
    \end{equation}
    and
    \begin{equation}
        h_n( \bv{p}^*_k) = h_n( \Pi_k \bv{p}^*) = h_n(\bv{p}^*).
    \end{equation}
    Therefore
    \begin{equation}
        \phi_{\lambda}(\bv{p}^*_k)  = \phi_{\lambda}(\bv{p}^* ), \quad \forall k \in [n].
    \end{equation}
    Furthermore, since
    \begin{equation}
        \sum_{k=1}^n \frac{1}{n} \bv{p}_k^* = \bv{u}_n,
    \end{equation}
    then $(\{ \bv{p}_k^*\}_{k=1}^n, \bv{u}) \in \set{F}$, and also satisfy
    \begin{equation}
        \sum_{k \in [n]} \frac{1}{n} \left[ h_n(T \bv{p}_k^*) -  \lambda h_n(\bv{p}_k^*)   \right]
        = \phi_{\lambda} (\bv{p}^*) 
        \leq \sum_{z \in \set{Z}} \left[ h_n(T \hat{\bv{p}}_z) -  \lambda h_n(\hat{\bv{p}}_z)   \right] q_z,
    \end{equation}
    that is, achieve the minimal objective.
    This contradicts our initial assumption, therefore implying optimality of the modulo additive channels.
\end{IEEEproof}

Returning to the proof of Prop. 3, we have the following equivalent dual objective function:
\begin{equation}
    q(\lambda) = \min_{\bv{p} \in \simplex{n}} \left\{ \phi_{\lambda} (\bv{p}) \right\} + \lambda \eta \\
    = \min_{\bv{p} \in \simplex{n}} \left\{ h_n(\bv{T} \bv{p}) - \lambda h_n(\bv{p}) \right\} + \lambda \eta.
\end{equation}

Denote $\psi(\lambda) \triangleq \min_{\bv{p} \in \simplex{n}} \{ \phi_\lambda (\bv{p}) \}$
and consider the dual problem, given by
\begin{equation} \label{eq:comib_modulo_optimality_tildeR}
        \tilde{R}(\eta) = 
        \max_{\lambda \geq 0} \psi(\lambda) + \lambda \eta .
\end{equation}

Note that by definition $\tilde{R}(\eta)$ is the conjugate function of $\psi(\lambda)$, and therefore convex in $\eta$ \cite[Ch. 3.3]{Boyd2014}. Furthermore, as was shown in \cite{Witsenhausen1975}, strong duality holds for the general $T$ and in particular for modulo additive $T$. Thus, 
    $R(\eta) = \tilde{R} (\eta)$.

The next question is whether for every $ \lambda \in [0,1] $ we have a unique $\eta \in [0,\log n]$. Since $\lambda$ is the slope of the tangent to $\tilde{R}(\eta)$ at $\eta$, this is equivalent to $\tilde{R}(\eta) $ being \textit{strictly} convex. Note that this is not always the case. For example consider a deterministic channel, i.e., $ T = I $, we have
\begin{align}
    \min_{\bv{p} \in \simplex{n}} \left\{ \phi_{\lambda} (\bv{p}) \right\} 
    & = \min_{\bv{p} \in \simplex{n}} \left\{ h_n( \bv{p}) - \lambda h_n(\bv{p}) \right\} \\
    & = \min_{\bv{p} \in \simplex{n}} \left\{ (1-\lambda)h_n( \bv{p}) ) \right\} = 0,
\end{align}
with
\begin{equation}
    \bv{p}^* = 
    \begin{cases}
    \bv{e}_k, & \lambda \neq 1 \\
    \simplex{n}, & \lambda = 1,
    \end{cases}
\end{equation}
and then $\tilde{R}(\eta) = \lambda \eta$, which is not strictly convex.

To conclude, we may restrict the channel from $\rv{Y}$ to $\rv{Z}$ to be a modulo-additive channel without loss of optimality, for regions in which the function $R(\eta)$ is strictly convex. In particular, for every $\eta$ there exists $\lambda$ that solves \eqref{eq:comib_modulo_optimality_tildeR}. A problem arises when a specific $\lambda$ corresponds to two (or more) different values of $\eta$, and when this does holds it implies that a modulo-additive channel is possibly sub-optimal. More explicitly, we have shown that for some $\lambda$, $q(\lambda)$ is obtained by modulo-additive channels, and for each $\lambda$ we obtain a unique $\eta$ which is generated by modulo-additive channels. If $g_T(\eta)$ is not strictly convex, then there exists a set of values of $\eta$ that are generated with channels that are not modulo-additive.

	\subsection{Properties of Hamming Channels}
	\label{appendix:hamming_properties}
	
The entropy of $ (\alpha,n) $-Hamming distribution is given by
\begin{equation}
	h_n(\bv{p}) 
	= -\left(\alpha + \frac{\bar{\alpha}}{n}\right) \log \left(\alpha + \frac{\bar{\alpha}}{n}\right) - \frac{(n-1)\bar{\alpha}}{n} \log \frac{\bar\alpha}{n} .
\end{equation}
Alternatively, if $ h_n(\bv{p}) = \eta $, then $ \alpha $ is the root of
\begin{equation}
	\eta + \left(\alpha + \frac{\bar{\alpha}}{n}\right) \log \left(\alpha + \frac{\bar{\alpha}}{n}\right) + \frac{(n-1)\bar{\alpha}}{n} \log 	\frac{\bar\alpha}{n} = 0 .
\end{equation}

The $ (\alpha,n) $ Hamming channel is defined by a transition matrix which rows are cyclic permutations of the $ (\alpha,n) $ Hamming \acrshort{pmf}, i.e.,
\begin{equation}
	\mat{T} = \mat{T}_{\alpha} = \alpha \mat{I}_n + \frac{(1-\alpha)}{n} \mat{E}_n,
\end{equation}
where $ \mat{I}_n $ is the $ n\times n $ identity matrix and $ \mat{E}_n $ is the all ones $ n \times n $ matrix.

Now assume that $ \rv{V} $ is a regular Hamming with parameter $ \alpha $ and $ \rv{W} $ is a regular Hamming with parameter $ \beta $ then $ \rv{Y} = \rv{V} + \rv{W} $ is a regular Hamming with parameter $ \alpha \cdot \beta $. This is true since we can represent the transition matrix from $ \rv{W} $ to $ \rv{Y} $ as 
\begin{equation}
	\mat{T} = \alpha \mat{I} + \frac{\bar{\alpha}}{n} \mat{E},
\end{equation}
and so
\begin{equation} \label{eq:comib_output_Hamming_full_support}
	\bv{q} = \mat{T}\bv{p} = \alpha \bv{p} + \bar{\alpha} \bv{u} = \alpha \beta \bv{e} +\alpha \bar{\beta} \bv{u} + \bar{\alpha} \bv{u} =\alpha \beta \bv{e} + \overline{\alpha \beta} \bv{u}.
\end{equation}

Next assume that $ \rv{V} $ is a regular Hamming with parameter $ \alpha $ and $ \rv{W} $ is a negative Hamming with parameters $ (\beta, n, k) $ then $ \rv{Y} = \rv{V} + \rv{W} $ has the following \acrshort{pmf}
\begin{equation}
	\bv{q} = \mat{T}\bv{p} = \alpha \bv{p} + \bar{\alpha} \bv{u}_n = [\alpha \beta \cdot \bv{e}_{k}+ \alpha \bar{\beta} \bv{u}_{k},\boldsymbol{0}_{n-k}] + \bar\alpha \bv{u}_n .
\end{equation}
Note that for $ n=k $, i.e., the \acrshort{pmf} of $ \rv{W} $ has a full support, then the \acrshort{pmf} of $ \rv{Y} $ is also $ (\alpha \cdot \beta,n) $ Hamming. However, when $ 1<k < n $, then the resulting \acrshort{pmf} has no specific structure. 
The output entropy for the latter is given by
\begin{multline} \label{eq:oblib_entropy_of_posneg_Hamming_mixture}
	h_n(\bv{q}) = \minus \left(\alpha \beta + \frac{\alpha \bar{\beta}}{k} + \frac{\bar{\alpha}}{n} \right) \log  \left(\alpha \beta + \frac{\alpha \bar{\beta}}{k} + \frac{\bar{\alpha}}{n} \right) \\
	\minus  (k\minus1) \left( \frac{\alpha \bar{\beta}}{k} + \frac{\bar{\alpha}}{n} \right)  \log \left( \frac{\alpha \bar{\beta}}{k} + \frac{\bar{\alpha}}{n} \right) 
	\minus  (n\minus k) \left(  \frac{\bar{\alpha}}{n} \right)  \log \left(  \frac{\bar{\alpha}}{n} \right).
\end{multline}

If $ \bv{p} $ is a negative Hamming distribution with parameters $ (\beta,n,k) $, then its entropy is given by
\begin{equation}
	h_n(\bv{p}) 
	= -\left(\beta + \frac{\bar{\beta}}{k}\right) \log \left(\beta + \frac{\bar{\beta}}{k}\right) - \frac{(k-1)\bar{\beta}}{k} \log \frac{\bar\beta}{k} .
\end{equation}

    \subsection{Alternating Algorithm Proof}
	\label{appendix:altenating_algorithm_proof}
	\begin{proof}[Proof of Lagrangian minimization]
	Since
	\begin{align}
		I(\rv{X};\rv{Z})
		&= \sum_{x \in \set{X},z \in \set{Z}} \mathsf{P}_{\rv{X}\rv{Z}} (x,z) \log \frac{ \mathsf{P}_{\rv{X}\rv{Z}} (x,z)}{ \mathsf{P}_{\rv{X}} (x) \mathsf{P}_{\rv{Z}}(z)} \\
		&= \sum_{\substack{x \in \set{X} \\ y \in \set{Y} \\ z \in \set{Z}}}  \mathsf{P}_{\rv{X} \rv{Y}\rv{Z}} (x,y,z) \log \frac{ \sum_{y' \in \set{Y}} \mathsf{P}_{\rv{X} \rv{Y} \rv{Z}} (x,y',z)}{ \mathsf{P}_{\rv{X}} (x) \mathsf{P}_{\rv{Z}}(z)} \\
		&= \sum_{\substack{x \in \set{X} \\ y \in \set{Y} \\ z \in \set{Z}}}  \mathsf{P}_{\rv{X} \rv{Y}} (x,y) \mathsf{P}_{\rv{Z}|\rv{Y}} (z|y)\log \frac{ \sum_{y' \in \set{Y}} \mathsf{P}_{\rv{X} \rv{Y}} (x,y') \mathsf{P}_{\rv{Z}|\rv{Y}} (z|y')}{ \mathsf{P}_{\rv{X}} (x) \mathsf{P}_{\rv{Z}}(z)},
	\end{align}
	then
	\begin{align}
		\frac{\partial I(\rv{X};\rv{Z})}{\partial \mathsf{P}_{\rv{X} \rv{Y}} (x,y)}
		&= \sum_{z \in \set{Z}} \mathsf{P}_{\rv{Z}|\rv{Y}} (z|y)\log \frac{  \mathsf{P}_{\rv{X} \rv{Z}} (x,z) }{ \mathsf{P}_{\rv{X}} (x) \mathsf{P}_{\rv{Z}}(z)}
		+ \sum_{ z \in \set{Z}}  \frac{\mathsf{P}_{\rv{X} \rv{Z}} (x,z) }{ \mathsf{P}_{\rv{X} \rv{Z}} (x,z)} \cdot  \mathsf{P}_{\rv{Z}|\rv{Y}} (z|y) \\
		&= \sum_{z \in \set{Z}} \mathsf{P}_{\rv{Z}|\rv{Y}} (z|y)  \log \frac{  \mathsf{P}_{\rv{Z} | \rv{X}} (z|x)}{ \mathsf{P}_{\rv{Z}}(z)} + 1 \\
		&= \sum_{z \in \set{Z}} \mathsf{P}_{\rv{Z}|\rv{Y}} (z|y)  \log \frac{  \mathsf{P}_{\rv{Z} | \rv{X}} (z|x) \cdot \mathsf{P}_{\rv{Z}|\rv{Y}} (z|y)}{ \mathsf{P}_{\rv{Z}}(z) \cdot \mathsf{P}_{\rv{Z}|\rv{Y}} (z|y)} + 1 \\
		&= I(y,\rv{Z}) - D\left( \mathsf{P}_{\rv{Z}|\rv{Y}} (\cdot|y) || \mathsf{P}_{\rv{Z}|\rv{X}} (\cdot|x) \right) + 1.
	\end{align}
	Similarly, since
	\begin{equation}
		I(\rv{X};\rv{Y})
		= \sum_{x \in \set{X}, y \in \set{Y}} \mathsf{P}_{\rv{X}\rv{Y}} (x,y) \log \frac{ \mathsf{P}_{\rv{X}\rv{Y}} (x,y)}{ \mathsf{P}_{\rv{X}} (x) \mathsf{P}_{\rv{Y}}(y)},
	\end{equation}
	then
	\begin{equation}
		\frac{\partial I(\rv{X};\rv{Y})}{\partial \mathsf{P}_{\rv{X} \rv{Y}} (x,y)}
		=  \log \frac{ \mathsf{P}_{\rv{X}\rv{Y}} (x,y)}{ \mathsf{P}_{\rv{X}} (x) \mathsf{P}_{\rv{Y}}(y)}  + 1.
	\end{equation}
	Thus, stationarity implies
	\begin{align}
		&0 = \frac{\partial \mathcal{L}_{\min} }{\partial \mathsf{P}_{\rv{X} \rv{Y}} (x,y)}  \\
		&= I(y,\rv{Z}) - D\left( \mathsf{P}_{\rv{Z}|\rv{Y}} (\cdot|y) || \mathsf{P}_{\rv{Z}|\rv{X}} (\cdot|x) \right) + 1
		- \lambda_1 \left(\log \frac{ \mathsf{P}_{\rv{X}\rv{Y}} (x,y)}{ \mathsf{P}_{\rv{X}} (x) \mathsf{P}_{\rv{Y}}(y)}  + 1\right) + \mu_x + \nu_y = 0.
	\end{align}
	Since $ I(y,\rv{Z}) $ is a function of $ y $ only, it can be absorbed in $ \nu_y $ along with $ 1 $ and $ \lambda_1 $, i.e.,
	$ \tilde{\nu}_y  \triangleq \nu_y +1 + I(y,\rv{Z}) - \lambda_1 $. Therefore we obtain
	\begin{equation}
		\mathsf{P}_{\rv{X}\rv{Y}} (x,y) = \mathsf{P}_{\rv{X}} (x) \mathsf{P}_{\rv{Y}} (y) \cdot e^{- \frac{1}{\lambda_1} \left(   D\left( \mathsf{P}_{\rv{Z}|\rv{Y}} (\cdot|y) || \mathsf{P}_{\rv{Z}|\rv{X}} (\cdot|x)   - \mu_x - \tilde{\nu}_y \right)  \right) },
	\end{equation}
	which can be further simplified to the following form:
	\begin{equation}
		\mathsf{P}_{\rv{X}\rv{Y}} (x,y) =  \frac{\mathsf{P}_{\rv{X}} (x) \mathsf{P}_{\rv{Y}}(y) e^{- \beta_1    D\left( \mathsf{P}_{\rv{Z}|\rv{Y}} (\cdot|y) || \mathsf{P}_{\rv{Z}|\rv{X}} (\cdot|x) \right) }}{Z(x,y,\beta_1)},
	\end{equation}
	where $ \beta_1 \triangleq \frac{1}{\lambda_1} $ and $ Z(x,y,\beta_1) $ is the normalization constant, that assures correct marginalization of the joint \acrshort{pmf}, i.e.,
	\begin{equation}
		\sum_{x \in \set{X}} \mathsf{P}_{\rv{X}\rv{Y}} (x,y) = \mathsf{P}_{\rv{Y}} (y) \qquad
		\sum_{y \in \set{Y}} \mathsf{P}_{\rv{X}\rv{Y}} (x,y) = \mathsf{P}_{\rv{X}} (x).
	\end{equation}
	Utilizing Bayes' law, the conditional distribution $ \mathsf{P}_{\rv{Z}|\rv{X}} (z|x) $ is given by
	\begin{equation}
		\mathsf{P}_{\rv{Z}|\rv{X}} (z|x) = \frac{1}{\mathsf{P}_{\rv{X}}(x)} \sum_{y \in \set{Y}} \mathsf{P}_{\rv{Z}|\rv{Y}} (z|y) \mathsf{P}_{\rv{X} \rv{Y}} (x,y).
	\end{equation}
\end{proof}

\begin{proof}[Proof of Lagrangian maximization]
    Consider the Lagrangian of the maximization problem:
    \begin{equation} \label{eq:comib_lagrangian_maximization}
        \mathcal{L}_{\max}(\pmf{\rv{Z}|\rv{Y}},\lambda) = -I(\rv{X};\rv{Z}) + \lambda(I(\rv{Y};\rv{Z}) - C_2).
    \end{equation}
	Since
	\begin{align}
		I(\rv{X};\rv{Z})
		&= \sum_{x \in \set{X},z \in \set{Z}} \mathsf{P}_{\rv{X}\rv{Z}} (x,z) \log \frac{ \mathsf{P}_{\rv{X}\rv{Z}} (x,z)}{ \mathsf{P}_{\rv{X}} (x) \mathsf{P}_{\rv{Z}}(z)} \\
		&= \sum_{\substack{x \in \set{X} \\ y \in \set{Y} \\ z \in \set{Z}}}  \mathsf{P}_{\rv{X} \rv{Y}\rv{Z}} (x,y,z) \log \frac{ \sum_{y' \in \set{Y}} \mathsf{P}_{\rv{X} \rv{Y} \rv{Z}} (x,y',z)}{ \mathsf{P}_{\rv{X}} (x) \mathsf{P}_{\rv{Z}}(z)} \\
		&= \sum_{\substack{x \in \set{X} \\ y \in \set{Y} \\ z \in \set{Z}}}  \mathsf{P}_{\rv{X} \rv{Y}} (x,y) \mathsf{P}_{\rv{Z}|\rv{Y}} (z|y)\log \frac{ \sum_{y' \in \set{Y}} \mathsf{P}_{\rv{X} \rv{Y}} (x,y') \mathsf{P}_{\rv{Z}|\rv{Y}} (z|y')}{ \mathsf{P}_{\rv{X}} (x) \mathsf{P}_{\rv{Z}}(z)},
	\end{align}
	then
	\begin{align}
		\frac{\partial I(\rv{X};\rv{Z})}{\partial \mathsf{P}_{\rv{Z}| \rv{Y}} (z|y)}
		&= \sum_{x \in \set{X}}  \mathsf{P}_{\rv{X} \rv{Y}} (x,y) \log \frac{  \mathsf{P}_{\rv{X}| \rv{Z}} (x|z) }{ \mathsf{P}_{\rv{X}} (x) } 
		+ \sum_{x \in \set{X}}  \frac{\mathsf{P}_{\rv{X} \rv{Z}} (x,z)}{\mathsf{P}_{\rv{X} \rv{Z}} (x,z)} \mathsf{P}_{\rv{X}\rv{Y}} (x,y) 
		\minus \sum_{x \in \set{X}}  \frac{\mathsf{P}_{\rv{X} \rv{Z}} (x,z)}{\mathsf{P}_{\rv{Z}} (z)} \mathsf{P}_{\rv{Y}} (y) \\
		&= \mkern-3mu \mathsf{P}_{\rv{Y}} (y)  \mkern-3mu  \sum_{x \in \set{X}}  \mkern-3mu   \mathsf{P}_{\rv{X}| \rv{Y}} (x|y)  \mkern-3mu  \log  \mkern-3mu  \frac{  \mathsf{P}_{\rv{X}| \rv{Z}} (x|z) }{ \mathsf{P}_{\rv{X}} (x) } \cdot \frac{\mathsf{P}_{\rv{X}| \rv{Y}} (x|y)}{\mathsf{P}_{\rv{X}| \rv{Y}} (x|y)} \\
		&= \mathsf{P}_{\rv{Y}} (y) \left( I(y,\rv{X}) \minus D(\mathsf{P}_{\rv{X}| \rv{Y}} (\cdot|y) || \mathsf{P}_{\rv{X}| \rv{Z}} (\cdot|z))\right).
	\end{align}
	Similarly, since
	\begin{equation}
		I(\rv{Y};\rv{Z})
		= \sum_{ y \in \set{Y}, z \in \set{Z}} \mathsf{P}_{\rv{Y}\rv{Z}} (y,z) \log \frac{ \mathsf{P}_{\rv{Y}\rv{Z}} (y,z)}{ \mathsf{P}_{\rv{Y}} (y) \mathsf{P}_{\rv{Z}}(z)},
	\end{equation}
	then
	\begin{align}
		\frac{\partial I(\rv{Y};\rv{Z})}{\partial \mathsf{P}_{\rv{Z}| \rv{Y}} (z|y)}
		&=  \mathsf{P}_{\rv{Y}}(y)  \log \frac{ \mathsf{P}_{\rv{Z}|\rv{Y}} (z|y)}{ \mathsf{P}_{\rv{Z}}(z)}
		+ \sum_{ y \in \set{Y}} \frac{\mathsf{P}_{\rv{Y}\rv{Z}} (y,z)}{\mathsf{P}_{\rv{Y}\rv{Z}} (y,z)} \mathsf{P}_{\rv{Y}} (y) 
		 - \sum_{ y \in \set{Y}} \frac{\mathsf{P}_{\rv{Y}\rv{Z}} (y,z)}{ \mathsf{P}_{\rv{Z}}(z)} \mathsf{P}_{\rv{Y}} (y)  \\
		&=  \mathsf{P}_{\rv{Y}}(y)  \log \frac{ \mathsf{P}_{\rv{Z}|\rv{Y}} (z|y)}{ \mathsf{P}_{\rv{Z}}(z)}.
	\end{align}
	Thus, stationarity implies 
	\begin{align}
		0 &=\frac{\partial \mathcal{L}_{\max}}{\partial \mathsf{P}_{\rv{Z}| \rv{Y}} (z|y)} \\
		&=   \mathsf{P}_{\rv{Y}}(   y  )   \left[  \minus  
		I   (   y  ,  \rv{X} )   +    D  (  \mathsf{P}_{\rv{X}| \rv{Y}}  (  \cdot  |  y )    ||    \mathsf{P}_{\rv{X}| \rv{Z}}  (  \cdot  |  z ))
 	+   \lambda_2 \log \frac{ \mathsf{P}_{\rv{Z}|\rv{Y}} (z|y)}{ \mathsf{P}_{\rv{Z}}(z)}
		 +   \theta_y 
		\right] .
	\end{align}
	Since $ I(y,\rv{X}) $ is a function of $ y $ only, it can be absorbed in $ \theta_y $  , i.e.,
	$ \tilde{\theta}_y  \triangleq \theta_y  + I(y,\rv{X}) $. Therefore we obtain
	\begin{equation}
		\mathsf{P}_{\rv{Z}|\rv{Y}} (z|y) = \mathsf{P}_{\rv{Z}} (z)  \cdot e^{ - \frac{1}{\lambda_1} \left(   D\left( \mathsf{P}_{\rv{X}|\rv{Y}} (\cdot|y) || \mathsf{P}_{\rv{X}|\rv{Z}} (\cdot|z) \right)   - \tilde{\theta}_y \right)}.
	\end{equation}
	The last equation can be further simplified to the following form:
	\begin{equation}
		\mathsf{P}_{\rv{Z}|\rv{Y}} (z|y) = \frac{\mathsf{P}_{\rv{Z}} (z)}{Z(y,\beta_2)}  \cdot 
		e^{ -\beta_2   D\left( \mathsf{P}_{\rv{X}|\rv{Y}} (\cdot|y) || \mathsf{P}_{\rv{X}|\rv{Z}} (\cdot|z) \right) },
	\end{equation}
	where $ \beta_2 \triangleq \frac{1}{\lambda_2} $ and $ Z(y,\beta_2) $ is the normalization constant.
	
	The conditional distribution $ \mathsf{P}_{\rv{X}|\rv{Z}} (x|z) $ is given by
	\begin{equation}
		\mathsf{P}_{\rv{X}|\rv{Z}} (x|z) = \frac{1}{\mathsf{P}_{\rv{Z}}(z)} \sum_{y \in \set{Y}} \mathsf{P}_{\rv{Z}|\rv{Y}} (z|y) \mathsf{P}_{\rv{X} \rv{Y}} (x,y).
	\end{equation}
\end{proof}


	
	\subsection{Proof of Alternating Algorithm for Modulo Additive Channels}
	\label{appendix:modulo_altenating_algorithm_proof}
	Consider the Lagrangian
\begin{equation} 
    L^{\varphi}(\bv{p},\lambda_1) = -h(T_v \bv{p}) + \lambda_1(h(\bv{p})-\eta_1).
\end{equation}
Its stationary point satisfies,
\begin{equation}
    \nabla_{\bv{p}} L = T_v^T \log (T \bv{p}) - \lambda \log \bv{p} + (\mu + 1-\lambda) \bv{e} = \bv{0}.
\end{equation}
Therefore, $\bv{p}_{w}^*$ satisfies the following equation:
\begin{equation}
    \bv{p}_{w}^* = \frac{e^{\beta_1 T_v^T \log \bv{q}_w}}{Z(\beta_1)},
\end{equation}
where $\beta_1 \triangleq \frac{1}{\lambda_1}$, 
    $\bv{q}_w \triangleq T_v^T \bv{p}_{w}^*$,
and $Z_1(\beta_1)$ is the normalization (partition) function.

Due to similarity of Lagrangians the the proof is similar and is omitted due to space limitations.

\bibliographystyle{IEEEtran}
	
	\bibliography{oblib}

\begin{thebibliography}{10}
\providecommand{\url}[1]{#1}
\csname url@samestyle\endcsname
\providecommand{\newblock}{\relax}
\providecommand{\bibinfo}[2]{#2}
\providecommand{\BIBentrySTDinterwordspacing}{\spaceskip=0pt\relax}
\providecommand{\BIBentryALTinterwordstretchfactor}{4}
\providecommand{\BIBentryALTinterwordspacing}{\spaceskip=\fontdimen2\font plus
\BIBentryALTinterwordstretchfactor\fontdimen3\font minus
  \fontdimen4\font\relax}
\providecommand{\BIBforeignlanguage}[2]{{%
\expandafter\ifx\csname l@#1\endcsname\relax
\typeout{** WARNING: IEEEtran.bst: No hyphenation pattern has been}%
\typeout{** loaded for the language `#1'. Using the pattern for}%
\typeout{** the default language instead.}%
\else
\language=\csname l@#1\endcsname
\fi
#2}}
\providecommand{\BIBdecl}{\relax}
\BIBdecl

\bibitem{Tishby1999}
N.~Tishby, F.~C.~N. Pereira, and W.~Bialek, ``The information bottleneck
  method,'' in \emph{Proc. 37th Annu. Allerton Conf. Commun. Control Comput.},
  Sep. 1999, p. 368–377.

\bibitem{Tishby2015}
N.~{Tishby} and N.~{Zaslavsky}, ``Deep learning and the information bottleneck
  principle,'' in \emph{Proc. IEEE Inf. Theory Workshop (ITW)}, Jerusalem,
  Israel, Apr. 2015, pp. 1--5.

\bibitem{fontana1981universal}
R.~Fontana, ``On universal coding for classes of composite and remote sources
  with memory (corresp.),'' \emph{IEEE Transactions on Information Theory},
  vol.~27, no.~6, pp. 784--786, 1981.

\bibitem{Dembo2003}
A.~Dembo and T.~Weissman, ``The minimax distortion redundancy in noisy source
  coding,'' \emph{IEEE Transactions on Information Theory}, vol.~49, no.~11,
  pp. 3020--3030, 2003.

\bibitem{Weissman2004}
T.~Weissman, ``Universally attainable error exponents for rate-distortion
  coding of noisy sources,'' \emph{IEEE Transactions on Information Theory},
  vol.~50, no.~6, pp. 1229--1246, 2004.

\bibitem{Dobrushin1962}
R.~{Dobrushin} and B.~{Tsybakov}, ``Information transmission with additional
  noise,'' \emph{IRE Trans. Inf. Theory}, vol.~8, no.~5, pp. 293--304, Sep.
  1962.

\bibitem{Wolf1970}
J.~Wolf and J.~Ziv, ``Transmission of noisy information to a noisy receiver
  with minimum distortion,'' \emph{IEEE Trans. Inf. Theory}, vol.~16, pp.
  406--411, Jul. 1970.

\bibitem{Lapidoth1997}
A.~Lapidoth, ``On the role of mismatch in rate distortion theory,'' \emph{IEEE
  Transactions on Information Theory}, vol.~43, no.~1, pp. 38--47, 1997.

\bibitem{Aguerri2019}
I.~Estella~Aguerri, A.~Zaidi, G.~Caire, and S.~Shamai~Shitz, ``On the capacity
  of cloud radio access networks with oblivious relaying,'' \emph{IEEE Trans.
  Inform. Theory}, vol.~65, no.~7, pp. 4575--4596, Jul. 2019.

\bibitem{Makhdoumi2014}
A.~Makhdoumi, S.~Salamatian, N.~Fawaz, and M.~Médard, ``From the information
  bottleneck to the privacy funnel,'' in \emph{Proc. IEEE Inf. Theory Workshop
  (ITW)}, Nov. 2014, pp. 501--505.

\bibitem{Painsky2020}
\BIBentryALTinterwordspacing
A.~Painsky, M.~Feder, and N.~Tishby, ``Nonlinear canonical correlation
  analysis:a compressed representation approach,'' \emph{Entropy}, vol.~22,
  no.~2, 2020. [Online]. Available:
  \url{https://www.mdpi.com/1099-4300/22/2/208}
\BIBentrySTDinterwordspacing

\bibitem{Shamir2010}
\BIBentryALTinterwordspacing
O.~Shamir, S.~Sabato, and N.~Tishby, ``Learning and generalization with the
  information bottleneck,'' \emph{Theoretical Computer Science}, vol. 411,
  no.~29, pp. 2696--2711, 2010, algorithmic Learning Theory (ALT 2008).
  [Online]. Available:
  \url{https://www.sciencedirect.com/science/article/pii/S030439751000201X}
\BIBentrySTDinterwordspacing

\bibitem{McLachlan2008}
G.~J. McLachlan, \emph{\BIBforeignlanguage{eng}{The EM algorithm and
  extensions}}, 2nd~ed., ser. Wiley series in probability and statistics.\hskip
  1em plus 0.5em minus 0.4em\relax Hoboken, N.J: Wiley-Interscience, Jun. 2008.

\bibitem{Amari2000}
S.-i. Amari and H.~Nagaoka, \emph{Methods of information geometry}.\hskip 1em
  plus 0.5em minus 0.4em\relax American Mathematical Soc., 2000, vol. 191.

\bibitem{Levy2009}
B.~C. Levy, ``Robust hypothesis testing with a relative entropy tolerance,''
  \emph{IEEE Trans. Inf. Theory}, vol.~55, no.~1, pp. 413--421, Jan. 2009.

\bibitem{Dytso2019}
A.~Dytso, M.~Fauß, A.~M. Zoubir, and H.~V. Poor, ``{MMSE} bounds for additive
  noise channels under {Kullback–Leibler} divergence constraints on the input
  distribution,'' \emph{IEEE Trans. Signal Process.}, vol.~67, no.~24, pp.
  6352--6367, Dec. 2019.

\bibitem{Cao2020}
W.~Cao, A.~Dytso, M.~Fauß, G.~Feng, and H.~V. Poor, ``Robust power allocation
  for parallel {G}aussian channels with approximately {G}aussian input
  distributions,'' \emph{IEEE Trans. Wireless Commun.}, vol.~19, no.~6, pp.
  3685--3699, Jun. 2020.

\bibitem{Witsenhausen1975}
H.~S. Witsenhausen and A.~D. Wyner, ``A conditional entropy bound for a pair of
  discrete random variables,'' \emph{IEEE Trans. Inf. Theory}, vol.~21, no.~5,
  pp. 493--501, Sep. 1975.

\bibitem{Gacs1973}
P.~G\'{a}cs and J.~K\"{o}rner, ``Common information is far less than mutual
  information,'' \emph{Probl. Contr. Inform. Theory}, vol.~2, no.~2, pp.
  149--162, 1973.

\bibitem{PinCalmon2017}
F.~du~Pin~Calmon, A.~Makhdoumi, M.~Médard, M.~Varia, M.~Christiansen, and
  K.~R. Duffy, ``Principal inertia components and applications,'' \emph{IEEE
  Trans. Inf. Theory}, vol.~63, no.~8, pp. 5011--5038, Aug. 2017.

\bibitem{Shamai2021}
S.~Shamai, ``The information bottleneck: A unified information theoretic
  view,'' National Conference on Communications (NCC2021), Jul. 2021, plenary
  Address.

\bibitem{Wyner1976}
A.~Wyner and J.~Ziv, ``The rate-distortion function for source coding with side
  information at the decoder,'' \emph{IEEE Trans. Inform. Theory}, vol.~22,
  no.~1, pp. 1--10, Jan. 1976.

\bibitem{Zaidi2020}
A.~Zaidi, I.~E. Aguerri, and S.~S. (Shitz), ``On the information bottleneck
  problems: Models, connections, applications and information theoretic
  views,'' \emph{Entropy}, vol.~22, no.~2, p. 151, Feb. 2020.

\bibitem{Wyner1973}
A.~Wyner and J.~Ziv, ``A theorem on the entropy of certain binary sequences and
  applications {I},'' \emph{IEEE Trans. Inf. Theory}, vol.~19, pp. 769--772,
  Nov. 1973.

\bibitem{Sutskover2005}
I.~{Sutskover}, S.~{Shamai}, and J.~{Ziv}, ``Extremes of information
  combining,'' \emph{IEEE Trans. Inf. Theory}, vol.~51, no.~4, pp. 1313--1325,
  Apr. 2005.

\bibitem{Chechik2005}
G.~Chechik, A.~Globerson, N.~Tishby, and Y.~Weiss, ``Information bottleneck for
  {G}aussian variables,'' \emph{J. Mach. Learn. Res.}, vol.~6, pp. 165--188,
  Dec. 2005.

\bibitem{Dembo1991}
A.~Dembo, T.~Cover, and J.~Thomas, ``Information theoretic inequalities,''
  \emph{IEEE Trans. Inf. Theory}, vol.~37, no.~6, pp. 1501--1518, Nov. 1991.

\bibitem{Guo2013}
D.~Guo, S.~{Shamai (Shitz)}, and S.~Verdú, ``The interplay between information
  and estimation measures,'' \emph{Found. Trends Signal Process.}, vol.~6,
  no.~4, pp. 243--429, 2012.

\bibitem{Bustin2013}
R.~Bustin, M.~Payaro, D.~P. Palomar, and S.~Shamai~(Shitz), ``On {MMSE}
  crossing properties and implications in parallel vector {Gaussian}
  channels,'' \emph{IEEE Trans. Inf. Theory}, vol.~59, no.~2, pp. 818--844,
  Feb. 2013.

\bibitem{Blahut1972}
R.~Blahut, ``Computation of channel capacity and rate-distortion functions,''
  \emph{IEEE Transactions on Information Theory}, vol.~18, no.~4, pp. 460--473,
  Jul. 1972.

\bibitem{Arimoto1972}
S.~Arimoto, ``An algorithm for computing the capacity of arbitrary discrete
  memoryless channels,'' \emph{IEEE Trans. Inf. Theory}, vol.~18, pp. 14--20,
  Jan. 1972.

\bibitem{Hassanpour2017}
S.~Hassanpour, D.~Wuebben, and A.~Dekorsy, ``Overview and investigation of
  algorithms for the information bottleneck method,'' in \emph{Proc. 11th Int.
  ITG Conf. Syst., Commun. Coding (SCC)}, Feb. 2017, pp. 1--6.

\bibitem{Aguerri2021}
I.~{Estella-Aguerri} and A.~{Zaidi}, ``Distributed variational representation
  learning,'' \emph{IEEE Trans. Pattern Anal. Mach. Intell.}, vol.~43, no.~1,
  pp. 120--138, Jan. 2021.

\bibitem{Slonim2002}
N.~Slonim, ``The information bottleneck: Theory and applications,'' Ph.D.
  dissertation, Hebrew University of Jerusalem, Jerusalem, Israel, 2002.

\bibitem{Steinberg2009}
Y.~{Steinberg}, ``Coding and common reconstruction,'' \emph{IEEE Trans. Inf.
  Theory}, vol.~55, no.~11, pp. 4995--5010, Nov. 2009.

\bibitem{Land2006}
I.~Land and J.~Huber, ``Information combining,'' \emph{Found. Trends Commun.
  Inf. Theory}, vol.~3, no.~3, pp. 227--330, Nov. 2006.

\bibitem{Farajiparvar2018}
P.~{Farajiparvar}, A.~{Beirami}, and M.~{Nokleby}, ``Information bottleneck
  methods for distributed learning,'' in \emph{Proc. 56th Annu. Allerton Conf.
  Commun., Control Comput.}, 2018, pp. 24--31.

\bibitem{Goldfeld2020}
Z.~{Goldfeld} and Y.~{Polyanskiy}, ``The information bottleneck problem and its
  applications in machine {L}earning,'' \emph{IEEE J. Sel. Areas Inf. Theory},
  vol.~1, no.~1, pp. 19--38, May 2020.

\bibitem{Lewandowsky2016}
J.~Lewandowsky, M.~Stark, and G.~Bauch, ``Information bottleneck graphs for
  receiver design,'' in \emph{Proc. IEEE Int. Symp. Inf. Theory}, Barcelona,
  Spain, Jul. 2016, pp. 2888--2892.

\bibitem{Stark2020}
M.~Stark, G.~Bauch, L.~Wang, and R.~D. Wesel, ``Information bottleneck decoding
  of rate-compatible {5G-LDPC} codes,'' in \emph{Proc. IEEE Inter. Conf. on
  Comm. (ICC)}, Jun. 2020, pp. 1--6.

\bibitem{Bhatt2021}
A.~Bhatt, B.~Nazer, O.~Ordentlich, and Y.~Polyanskiy, ``Information-distilling
  quantizers,'' \emph{IEEE Trans. Inf. Theory}, vol.~67, no.~4, pp. 2472--2487,
  2021.

\bibitem{Stark2018}
M.~Stark, A.~Shah, and G.~Bauch, ``Polar code construction using the
  information bottleneck method,'' in \emph{Proc. IEEE Wireless Comm. Netw.
  Conf. Workshops (WCNCW)}.\hskip 1em plus 0.5em minus 0.4em\relax IEEE, 2018,
  pp. 7--12.

\bibitem{Shah2019}
S.~A.~A. Shah, M.~Stark, and G.~Bauch, ``Design of quantized decoders for polar
  codes using the information bottleneck method,'' in \emph{12th International
  ITG Conference on Systems, Communications and Coding 2019 (SCC’2019)},
  2019, pp. 1--6.

\bibitem{Shah2019a}
------, ``Coarsely quantized decoding and construction of polar codes using the
  information bottleneck method,'' \emph{Algorithms}, vol.~12, no.~9, p. 192,
  Sep. 2019.

\bibitem{Kurkoski2017}
B.~M. Kurkoski, ``On the relationship between the {KL} means algorithm and the
  information bottleneck method,'' in \emph{Proc. 11th International ITG
  Conference on Systems, Communications and Coding (SCC)}, Feb. 2017, pp. 1--6.

\bibitem{Pensia2020}
A.~Pensia, V.~Jog, and P.-L. Loh, ``Extracting robust and accurate features via
  a robust information bottleneck,'' \emph{IEEE Journal on Selected Areas in
  Information Theory}, vol.~1, no.~1, pp. 131--144, 2020.

\bibitem{Kazikli2020}
E.~Kaz{\i}kl{\i}, S.~Gezici, and S.~Y{\"u}ksel, ``Quadratic privacy-signaling
  games and the {MMSE} {G}aussian information bottleneck problem,'' \emph{arXiv
  preprint arXiv:2005.05743}, 2020.

\bibitem{Erkip1998}
E.~{Erkip} and T.~M. {Cover}, ``The efficiency of investment information,''
  \emph{IEEE Trans. Inf. Theory}, vol.~44, no.~3, pp. 1026--1040, May 1998.

\bibitem{Alemi2017}
A.~Alemi, I.~Fischer, J.~Dillon, and K.~Murphy,
  ``\BIBforeignlanguage{English}{Deep variational information bottleneck},'' in
  \emph{\BIBforeignlanguage{English}{Proc. Int. Conf. Learn. Represent.
  (ICLR)}}, 2017.

\bibitem{ShwartzZiv2017}
\BIBentryALTinterwordspacing
R.~Shwartz{-}Ziv and N.~Tishby, ``Opening the black box of deep neural networks
  via information,'' \emph{CoRR}, vol. abs/1703.00810, 2017. [Online].
  Available: \url{http://arxiv.org/abs/1703.00810}
\BIBentrySTDinterwordspacing

\bibitem{Gabrie2019}
M.~Gabri{\'{e}}, A.~Manoel, C.~Luneau, J.~Barbier, N.~Macris, F.~Krzakala, and
  L.~Zdeborov{\'{a}}, ``Entropy and mutual information in models of deep neural
  networks,'' \emph{J. Stat. Mech. Theory Exp.}, vol. 2019, no.~12, Dec. 2019.

\bibitem{Goldfeld2018}
\BIBentryALTinterwordspacing
Z.~Goldfeld, E.~van~den Berg, K.~H. Greenewald, I.~Melnyk, N.~Nguyen,
  B.~Kingsbury, and Y.~Polyanskiy, ``Estimating information flow in deep neural
  networks,'' \emph{CoRR}, vol. abs/1810.05728, 2018. [Online]. Available:
  \url{http://arxiv.org/abs/1810.05728}
\BIBentrySTDinterwordspacing

\bibitem{Cheng2018}
H.~Cheng, D.~Lian, S.~Gao, and Y.~Geng, ``Evaluating capability of deep neural
  networks for image classification via information plane,'' in \emph{Proc.
  Eur. Conf. Comput. Vis. (ECCV)}, Munich, Germany, Sep. 2018, pp. 168--182.

\bibitem{Yu2021}
S.~Yu, K.~Wickstrøm, R.~Jenssen, and J.~C. Príncipe, ``Understanding
  convolutional neural networks with information theory: An initial
  exploration,'' \emph{IEEE Trans. Neural Netw. Learn. Syst.}, vol.~32, no.~1,
  pp. 435--442, Jan. 2021.

\bibitem{Boyd2014}
S.~P. Boyd and L.~Vandenberghe, \emph{{Convex Optimization}}.\hskip 1em plus
  0.5em minus 0.4em\relax New York, NY, USA.: Cambridge University Press, 2014.

\bibitem{Bertsekas2002}
D.~P. Bertsekas and J.~N. Tsitsiklis,
  \emph{\BIBforeignlanguage{eng}{Introduction to probability}}.\hskip 1em plus
  0.5em minus 0.4em\relax Belmont, MA: Athena Scientific, 2002.

\bibitem{Witsenhausen1974}
H.~S. Witsenhausen, ``Entropy inequalities for discrete channels,'' \emph{IEEE
  Trans. Inf. Theory}, vol.~20, no.~5, pp. 610--616, Sep. 1974.

\bibitem{Audenaert2006}
K.~M. Audenaert, ``A sharp {F}annes-type inequality for the von {N}eumann
  entropy,'' \emph{J. Phys. A}, vol.~40, pp. 8127--8136, 2007.

\bibitem{Zhang2007}
Z.~Zhang, ``Estimating mutual information via {K}olmogorov distance,''
  \emph{IEEE Trans. Inform. Theory}, vol.~53, no.~9, pp. 3280--3282, Sep. 2007.

\bibitem{dobrushin1956central}
R.~L. Dobrushin, ``Central limit theorem for nonstationary {M}arkov chains.
  {I},'' \emph{Theory of Probability \& Its Applications}, vol.~1, no.~1, pp.
  65--80, 1956.

\bibitem{ho2010interplay}
S.-W. Ho and R.~W. Yeung, ``The interplay between entropy and variational
  distance,'' \emph{IEEE Trans. Inf. Theory}, vol.~56, no.~12, pp. 5906--5929,
  Dec. 2010.

\bibitem{Royden2010}
H.~L. Royden and P.~Fitzpatrick, \emph{Real analysis}, 4th~ed.\hskip 1em plus
  0.5em minus 0.4em\relax Prentice Hall, 2010.

\bibitem{Beck2014}
A.~Beck, \emph{\BIBforeignlanguage{eng}{Introduction to nonlinear optimization
  : theory, algorithms, and applications with MATLAB}}.\hskip 1em plus 0.5em
  minus 0.4em\relax Philadelphia: MOS-SIAM, 2014, vol.~19.

\bibitem{Steiner2021}
A.~Steiner and S.~Shamai~Shitz, ``Broadcast approach for the information
  bottleneck channel,'' \emph{IEEE Transactions on Communications}, vol.~69,
  no.~3, pp. 1595--1604, 2021.

\bibitem{Cover2006}
T.~M. Cover and J.~A. Thomas, \emph{Elements of {I}nformation {T}heory}.\hskip
  1em plus 0.5em minus 0.4em\relax Hoboken, NJ, USA: Wiley, 2006.

\bibitem{Meidlinger2014}
M.~Meidlinger, A.~Winkelbauer, and G.~Matz, ``On the relation between the
  {G}aussian information bottleneck and {MSE}-optimal rate-distortion
  quantization,'' in \emph{Proc. IEEE Workshop on Statistical Signal Processing
  (SSP 2014)}, 2014, pp. 89--92.

\bibitem{Tian2009}
C.~Tian and J.~Chen, ``Remote vector gaussian source coding with decoder side
  information under mutual information and distortion constraints,'' \emph{IEEE
  Transactions on Information Theory}, vol.~55, no.~10, pp. 4676--4680, 2009.

\bibitem{Horn2012}
R.~A. Horn, \emph{\BIBforeignlanguage{eng}{Matrix Analysis}}, 2nd~ed.\hskip 1em
  plus 0.5em minus 0.4em\relax Cambridge: Cambridge University Press, 2012.

\end{thebibliography}
	
\end{document}